\documentclass[%
 reprint,
 superscriptaddress,
 showkeys,
nofootinbib,
 amsmath,amssymb,
 aps,
 prd,
]{revtex4-2}

\newcommand{\apjs}{Astrophys. J. Suppl. Ser.}
\newcommand{\nphysa}{Nucl. Phys. A}
\newcommand{\apjl}{ApJ Lett.}
\newcommand{\mnras}{MNRAS}

\newcommand{\aap}{A\&A}

\usepackage{soul}
\usepackage{float}
\usepackage[caption=false]{subfig} 
\usepackage{gensymb}
\usepackage{graphicx}
\usepackage{dcolumn}
\usepackage{bm}
\usepackage[hypertexnames=true,colorlinks=true,urlcolor=blue,bookmarks=true,citecolor=blue,breaklinks=true]{hyperref}

\usepackage{array}
\usepackage{multirow}
\newcolumntype{C}[1]{>{\centering\arraybackslash}p{#1}}

\begin{document}

\preprint{APS/123-QED}

\title{Assessing the Distance for Probing the Nuclear Equation of State with Supernova Gravitational Waves}

\author{Y. S. Abylkairov}
\email{sultan.abylkairov@nu.edu.kz} 
\affiliation{Energetic Cosmos Laboratory, Nazarbayev University, 010000 Astana, Kazakhstan} 

\author{M. C. Edwards}
\affiliation{Department of Statistics, University of Auckland, Auckland 1010, New Zealand}

\author{A. Ostrikov}
\affiliation{Department of Informatics, University of Hamburg, 22527 Hamburg, Germany}

\author{Y. Tleukhanov}
\affiliation{Department of Physics, Nazarbayev University, 010000 Astana, Kazakhstan}

\author{A. Torres-Forné}
\affiliation{Departamento de Astronomía y Astrofísica, Universitat de València, Dr. Moliner 50, 46100 Burjassot (Valencia), Spain}
\affiliation{Observatori Astronòmic, Universitat de València, Catedrático José Beltrán 2, 46980, Paterna, Spain}

\author{P. Cerdá-Durán}
\affiliation{Departamento de Astronomía y Astrofísica, Universitat de València, Dr. Moliner 50, 46100 Burjassot (Valencia), Spain}
\affiliation{Observatori Astronòmic, Universitat de València, Catedrático José Beltrán 2, 46980, Paterna, Spain}

\author{J. A. Font}
\affiliation{Departamento de Astronomía y Astrofísica, Universitat de València, Dr. Moliner 50, 46100 Burjassot (Valencia), Spain}
\affiliation{Observatori Astronòmic, Universitat de València, Catedrático José Beltrán 2, 46980, Paterna, Spain}

\author{M. J. Szczepańczyk}
\affiliation{Faculty of Physics, University of Warsaw, Ludwika Pasteura 5, 02-093 Warsaw, Poland}

\author{E. Abdikamalov}
\affiliation{Energetic Cosmos Laboratory, Nazarbayev University, 010000 Astana, Kazakhstan}
\affiliation{Department of Physics, Nazarbayev University, 010000 Astana, Kazakhstan}

\date{\today}

\begin{abstract}
Gravitational waves from core-collapse supernovae provide a unique probe of the equation of state (EOS) of high density matter. In this work, we focus on the bounce signal from numerical simulations of rotating supernovae and explore its potential for EOS inference. We employ a support vector machine, previously shown to perform best among tested methods, to classify GW signals simulated for 18 EOS models. For optimally oriented sources, we estimate that the Advanced LIGO A+ detector can probe the EOS for Galactic events, while third-generation observatories such as the Einstein Telescope and Cosmic Explorer can reach substantially farther. For randomly oriented sources, only these next-generation detectors are expected to have sufficient sensitivity. These results represent the potential observational range for probing the nuclear EOS, although, due to the simplifying assumptions adopted, they should be regarded as approximate upper limits.
\end{abstract}


\maketitle

\section{Introduction}
\label{sec:intro}

Gravitational waves (GWs) offer a unique observational window into astrophysical events. Signals from compact binary mergers are now routinely observed \citep{Abac25GWTC4}, yielding key insights into the physics of neutron stars and black holes. Core-collapse supernovae (CCSNe) represent another promising GW source \citep{abbott2020optically, szczepanczyk2023optically, powell22inferring}. Galactic events are within reach of current detectors \citep{gossan16observing, Szczepanczyk21Detecting}, while extragalactic signals may become accessible with future instruments \citep{srivastava19detection} (see \citep{abdikamalov22gravitational, mezzacappa24gravitational} for recent reviews). 

CCSNe mark the end of massive stars, whose iron cores collapse under gravity once electron degeneracy pressure can no longer hold it. The collapse halts at nuclear density, launching a shock wave that soon stalls from energy losses and must be revived to drive the explosion \citep{muller20hydrodynamics}. Successful explosions leave behind neutron stars, while failed ones form black holes \citep{oconnor11, Burrows23Black}.

A small fraction of neutrinos emitted by the nascent proto-neutron star (PNS) is absorbed in the post-shock region, where it heats the material \citep{mueller:12a} and drives convection \citep{herant94inside, burrows95on, janka95first}. This convective activity, together with the standing accretion shock instability (SASI) \citep{blondin03stability, foglizzo06neutrino}, helps revive the stalled shock and trigger explosion. Additionally, convective eddies from the innermost nuclear burning shells can enhance post-shock turbulence when they interact with the shock, further supporting shock expansion \citep{Couch15Three, Mueller17Supernova, Kazeroni20impact, Vartanyan22collapse, Telman24Convective}. This explosion scenario is known as the neutrino mechanism.

In rapidly rotating progenitors, the PNS is born with high rotational kinetic energy. This energy can be transferred to the shock front via magnetic fields, potentially driving powerful explosions known as hypernovae \citep{burrows:07b, moesta:14b}. This process, referred to as the magnetorotational mechanism, may also produce magnetically driven jets \cite{obergaulinger:20, kuroda:20}. If such a jet successfully breaks through the stellar envelope, it can power a long gamma-ray burst  \citep{metzger11protomagnetar}. Even in cases where the jet is ``choked'' before breakout, its energy can still be deposited into the envelope, contributing to an explosion \citep{Pais23choked, Piran19jet_ccsn}. However, such rapidly rotating progenitors are expected to be rare \citep{Heger05Presupernova}. Recently, progenitors with moderate rotation have received increasing attention \citep[e.g.,][]{Takiwaki16Three, Summa18Rotation, Abdikamalov21Impact, Buellet23Effect}, bridging the gap between the extremes of slow and rapid rotation.

Rotation plays a crucial role in shaping both the dynamics and GW emission of CCSNe. In non-rotating or slowly rotating models, convection and SASI generate GWs and also excite oscillations of the proto–neutron star, which produce the dominant GW signal \citep{Murphy09Model, mueller:13, Yakunin15GW,  radice:19gw, radice:19gw, Andresen17Gravitational, Mezzacappa23Core, Vartanyan23Gravitational}. In rotating models, core collapse becomes asymmetric due to centrifugal deformation. The bounce produces a strong, short-duration GW signal, referred to as the rotating bounce signal \citep{Dimmelmeier07, Fuller15SNseismology}. In sufficiently rapidly rotating models, non-axisymmetric instabilities can develop, causing the PNS to deform into a non-axisymmetric shape over multiple rotation periods \cite{Scheidegger08}. This results in sustained GW emission over many cycles, potentially producing a strong and detectable signal \citep{Shibagaki20new}. Additionally, when neutrino emission is anisotropic, the resulting asymmetric energy flux varies with angle and produces a time-dependent quadrupole moment, which in turn generates GWs at lower frequencies \citep{mueller:97, Choi24GW}. Asymmetric shock propagation \citep{mueller:13} and, if present, jets \citep{Birnholtz13GW_jet, Gottlieb23Jetted, Soker23GWJJ} can both contribute additional low-frequency GW emission below $\sim 100$ Hz.

A quark-deconfinement phase transition in the PNS core, if it occurs, may cause a pressure drop, resulting in a mini-collapse to a more compact state. In rapidly rotating models, this can excite quasi-radial and quadrupole modes of the PNS \citep{abdikamalov:09}, while in slowly rotating models, it can couple to the flow asphericities \citep{zha:20, Kuroda22}, enhancing GW emission.

GW signals carry information about their sources \cite{mezzacappa24gravitational}. Once detected, a key objective is to extract the source parameters from the waveform~\citep{mitra23, pastor24, nunes2024deep, Villegas25Parameter}. Because rotation strongly influences the dynamics, the GW signature may allow us to discriminate between a neutrino-driven explosion in slowly rotating progenitors and a magnetorotational explosion in rapidly rotating ones \citep{Logue12Inferring, Powell24Determining}. One can constrain the PNS mass and radius \citep{Bizouard21Inference, Bruel23Inference} as well as rotation \citep{abdikamalov:14, pajkos19}. If a supernova fails to explode, or if the stellar envelope later falls back onto the proto-neutron star, a black hole may form \cite{cerda:13, Pan18Equation, Shibagaki21, Powell25noEMCCSN, Eggenberger25Black}. Such events produce distinct gravitational-wave signatures \cite{Kuroda23Failed}. 

Constraining the high‐density nuclear equation of state (EOS) is another exciting prospect: since the PNS dynamics, and thus its GW signature, vary with the EOS \citep[e.g.,][]{richers:17, schneider19equation, Powell24GW, Villegas25Parameter}, GW observations provide a means to probe and distinguish between competing EOS models \citep{edwards2017, chao22determining, Wolfe23GW, CasallasLagos23Characterizing, Murphy24Dependence}. GW observations of neutron star mergers have already provided constraints on the EOS in the cold, $\beta$-equilibrated regime through tidal deformability measurements \cite{abbott18gw170817}, and future detections of the post-merger signal are expected to probe the hot, supranuclear phase of the remnant \cite{Takami14Constraining, Radice18Long}. CCSNe, in contrast, explore a distinct thermodynamic domain characterized by hot, lepton-rich matter. Consequently, supernovae and mergers probe the EOS under complementary conditions.

In prior work \citep{mitra24, Abylkairov24Evaluating}, we investigated the feasibility of inferring the nuclear EOS from the GW signal produced by rotating core bounce. Building on earlier studies by \cite{edwards2017, chao22determining}, and using the waveform catalog of \citet{richers:17}, we examined how accurately machine-learning methods can identify the EOS under idealized, noise-free conditions. In the present study, we extend this analysis by incorporating the effects of detector noise. Our objective is to obtain a rough estimate of the distance out to which the EOS can be probed. We adopt a simplified noise model, a discrete set of EOS candidates, and other idealized assumptions (see below). Therefore, our results should be viewed as approximate upper limits on the distance horizon and a step toward more comprehensive future studies.
 
The paper is organized as follows. Section \ref{sec:method} outlines our method, Section \ref{sec:result} presents the results. Section \ref{sec:conclusion} summarizes the main conclusions.

\section{Methods} 
\label{sec:method}

\subsection{Data} 
\label{subsec:data}

\begin{table}[t]
\caption{Mapping of numerical labels to EOS models. The first column shows the numerical labels assigned to each EOS. The second column lists the full set of 18 EOS models. Horizontal lines separate EOSs that differ in their treatment of low-density matter.}
\vskip 0.15in
\begin{center}
\begin{small}
\begin{sc}
\begin{tabular}{C{1.4cm} C{3.6cm}}
\toprule
Label & 18 EOS models \\
\hline
1  & BHB$\Lambda$ \cite{bhbeos} \\
2  & BHB$\Lambda \Phi$ \cite{bhbeos} \\
3  & HSDD2 \cite{hempel:10,hempel:12} \\
4  & HSTMA \cite{hempel:10,hempel:12} \\
5  & HSFSG \cite{hempel:10,hempel:12} \\
6  & HSIUF \cite{hempel:10,hempel:12} \\
7  & HSNL3 \cite{hempel:10,hempel:12} \\
8  & HSTM1 \cite{hempel:10,hempel:12} \\
9  & SFHx \cite{steiner:13b} \\
10  & SFHo \cite{steiner:13b} \\
\hline
11  & GShenFSU1.7 \cite{gshen:11b} \\
12  & GShenFSU2.1 \cite{gshen:11b}  \\
13  & GShenNL3 \cite{gshennl3} \\
\hline
14  & HShen \cite{hsheneos1,hsheneos2,hshenheos3} \\
15  & HShenH \cite{hshenheos3} \\
\hline
16  & LS180 \cite{lseos:91} \\
17  & LS220 \cite{lseos:91} \\
18  & LS375 \cite{lseos:91} \\
\hline
\hline
\end{tabular}
\end{sc}
\end{small}
\end{center}
\vskip 0.3in
\label{Table:EOS_GR_ML}
\end{table}

\begin{figure}
\centering
\includegraphics[width=\linewidth]{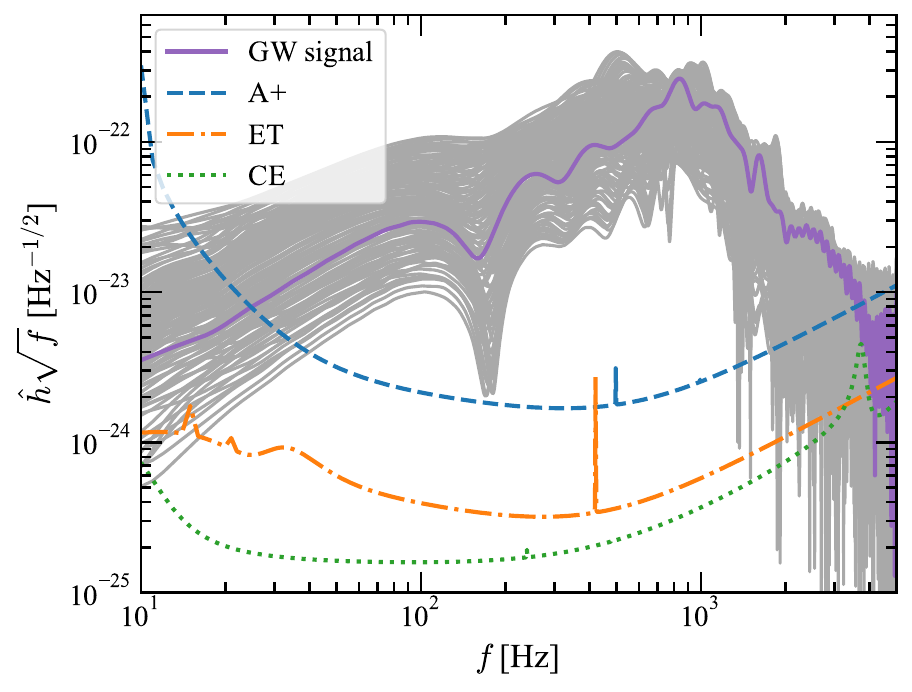}
\caption{Fourier transformed gravitational wave signals for the SFHo EOS at a distance of 10 kpc, scaled by $\sqrt{f}$, for $0.02 < T/|W| < 0.18$ (shown in grey). The purple curve corresponds to the signal at $T/|W| \approx 0.07$. The blue, orange, and green curves represent the sensitivity curves of Advanced LIGO A+, the Einstein Telescope, and the Cosmic Explorer, respectively.}
\label{fig:signal_noise_fft}
\end{figure}

In this work, we use gravitational waveforms from the \citet{richers:17} catalog, which comprises 1824 numerical waveforms\footnote{We exclude 60 waveforms from this analysis due to the absence of core bounce in extreme rotation cases, and an additional 60 waveforms with artificially enhanced or reduced electron capture rates. Waveforms with minor numerical artifacts were cleaned using a low-pass filter \cite{Villegas25Parameter}.} from simulations of a 12 $M_\odot$ progenitor using the general relativistic code {\tt CoCoNuT} \cite{Dimmelmeier02a, dimmelmeier:05MdM}. The simulations use 18 EOS models and follow the evolution up to approximately 50 ms after core bounce. See \cite{richers:17} for the details of the simulations.

These 18 EOS models are listed in Table~\ref{Table:EOS_GR_ML}. Among them, three (LS180, LS220, and LS375) are based on the compressible liquid-drop model \citep{lseos:91}, while the remaining ones are constructed within the relativistic mean-field framework. This group includes BHB$\Lambda$, BHB$\Lambda\Phi$ \cite{bhbeos}, HSDD2, HSTMA, HSFSG, HSIUF, HSNL3, HSTM1 \cite{hempel:10,hempel:12}, SFHx, SFHo \cite{steiner:13b}, GShenFSU1.7, GShenFSU2.1 \cite{gshen:11b}, GShenNL3 \cite{gshennl3}, HShen \cite{hsheneos1,hsheneos2,hshenheos3}, and HShenH \cite{hshenheos3}. Detailed descriptions of these EOSs can be found in \citet{richers:17}. To avoid redundancy, we do not repeat this information here and refer the reader to the original references.

Not all 18 EOS models from this catalog are consistent with modern experimental and astrophysical constraints (see discussion in \cite{richers:17}). Nevertheless, we retain the full set to estimate the maximum distance over which the EOS can be probed. Excluding the inconsistent models and replacing them with more realistic alternatives would yield a more tightly clustered EOS set, making classification more difficult and thus reducing this distance. This consideration is one of the reasons why our results should be regarded as upper limits.

The waveforms were resampled at 16384 Hz to match the sampling rate of LIGO detectors. We restrict our study to the time interval from $-$2 ms to 6 ms, where zero corresponds to the bounce time. This interval was selected because the GW signal prior to $-$2 ms carries negligible energy, while the signal beyond 6 ms is affected by contributions from prompt convection \citep{Raynaud22GW}, which is not accurately modeled in the \cite{richers:17} catalog.

For each EOS, the catalog includes approximately 100 rotational configurations, ranging from slow to rapid rotation \cite{richers:17}. We characterize the rotation at bounce by the parameter $T/|W|$, with $T$ denoting the rotational kinetic energy and $W$ the gravitational binding energy. Our analysis focuses on models with $0.02 < T/|W| < 0.18$. For $T/|W| < 0.02$, the rotation is insufficient to produce significant quadrupole deformation, leading to a weak bounce signal. For $T/|W| > 0.18$, centrifugal effects become prevalent, preventing the core from attaining the high densities where EOS distinctions are strongest. After applying this constraint, our dataset contains 1070 waveforms across 18 EOS models, with approximately 60 waveforms per EOS model on average.

\subsection{Noise}
\label{subsec:noise}

\begin{figure}
\centering
\includegraphics[width= \linewidth]{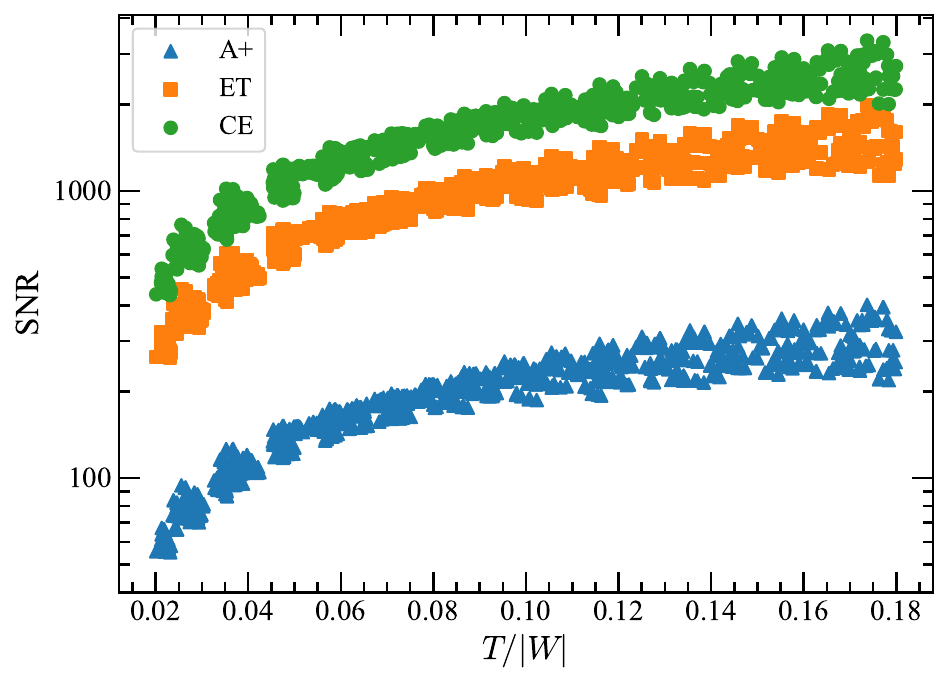}
\caption{Signal-to-noise ratio at 10 kpc as a function of $T/|W|$ for all 18 EOS waveforms with different detector sensitivities: Advanced LIGO A+ (blue), Einstein Telescope (orange), and Cosmic Explorer (green).}
\label{fig:snr_TW_detector}
\end{figure}

The signal-to-noise ratio (SNR) quantifies the strength of a GW signal relative to the background noise in a detector. In our analysis, we consider the Advanced LIGO A+ design sensitivity \cite{barsotti2018a+}. For comparison, we also include the planned sensitivities of the Einstein Telescope (ET) \cite{ET_CE_noise} and Cosmic Explorer (CE) \cite{ET_CE_noise}. Fig.~\ref{fig:signal_noise_fft} shows the sensitivity curves of these detectors together with the Fourier transforms of the GW signals scaled by $\sqrt{f}$, where $f$ is the frequency. We adopt idealized conditions, assuming that the detector arms are optimally oriented relative to the signal. Following \cite{flanagan98}, the matched-filter SNR $\rho$ for a detected GW signal $h$ is defined under the assumption of a perfectly matched template,
\begin{equation}
\rho = \sqrt{\int_0^{\infty}\frac{4\hat{h}^*(f)\hat{h}(f)}{S_n(f)}df},
\end{equation}
where $\hat{h}(f)$ is the Fourier transform of the template waveform and $S_n(f)$ is the one-sided noise spectral density.

Our dataset consists of simulated GW signals combined with detector noise, expressed as:
\begin{equation}
d = h + n,
\end{equation}
where $h$ is the simulated GW signal and $n$ is the detector noise. The latter was generated using the \texttt{PyCBC} library \cite{pycbc} as stationary, zero-mean Gaussian noise colored according to the detector power spectral density, with each signal assigned a unique realization. The GW signals are added directly to this colored noise without additional frequency-dependent weighting. As shown in Fig. \ref{fig:signal_noise_fft}, the detector sensitivity over the relevant frequency range ($\thicksim$100–1000 Hz), where most signal power resides can be approximated as flat, justifying this simplified approach.

At a fixed distance, the GW signal amplitude, and thus the SNR, varies significantly with rotation: rapidly rotating models produce stronger signals, while slowly rotating models yield weaker signals \cite{abdikamalov:14}. This results in substantial variation of the SNR within each EOS class \cite{richers:17}. Fig. \ref{fig:snr_TW_detector} presents the SNR variation with rotation at 10 kpc for different detector sensitivities.

\subsection{Classification algorithm}
\label{subsec:algorithms}

To classify EOS models, we employ the support vector machine (SVM) method, which demonstrated the best classification performance among the classical machine learning and deep learning algorithms tested by \citet{Abylkairov24Evaluating}. We use an SVM with a linear kernel and regularization parameter $C = 10$, optimized for our dataset through grid search cross-validation. For a given SNR, the method is trained on 80$\%$ of the clean dataset and tested on the remaining 20$\%$ with added noise. Prior to training, $z$-normalization was applied to each time frame of the training data. The same normalization parameters were applied to the noisy test set.
To enforce a fixed target SNR across different rotational configurations, we rescale only the noise component, ensuring that each signal achieves the desired SNR. We rescale the noise rather than the signal itself because $z$-normalization relies on preserving the distribution of each time frame; rescaling the signal would distort this distribution. The fixed SNR is obtained by modifying the signal-plus-noise waveform as
\begin{equation}
H = h + n \cdot \frac{\text{SNR}_\mathrm{o}}{\text{SNR}_\mathrm{t}},
\label{eq:snr}
\end{equation}
where $\text{SNR}_\mathrm{o}$ is the original SNR calculated from the injected detector noise and $\text{SNR}_\mathrm{t}$ is the target SNR. This procedure ensures that the intrinsic signal strength distribution is preserved across rotational configurations while maintaining the same effective noise level. Without rescaling the noise, strong signals would be paired with artificially weak noise and weak signals with disproportionately strong noise, biasing the classifier toward recognizing strong signals more easily.

The dynamics during core bounce and the early post-bounce phase are largely axisymmetric \cite[e.g.,][]{ott12correlated}. For axisymmetric sources, the GW strain scales as $h \propto \sin^2\alpha$, where $\alpha$ is the inclination angle between the rotation axis and the line of sight. To evaluate the impact of source orientation, we perform analysis for both optimally and randomly oriented sources. For the latter, we augment the training set with 11 isotropically distributed orientations, following \cite{gossan16observing}.

When evaluating classification accuracy as a function of source distance (Section \ref{sec:result}), we apply SNR-based selection criteria. For optimally oriented sources, we require all signals at a given distance to have SNR $>$ 10, while for general inclination analysis we require SNR $>$ 20. The higher threshold for general inclination provides sufficient margin such that after random scaling by $\mathrm{sin}^2\alpha$, which reduces the SNR, we retain only signals with SNR $>$ 10. This ensures reasonable freedom in inclination angle sampling while maintaining realistic detection thresholds.

We assess EOS classification performance using accuracy metric:
\begin{equation}
\text{Accuracy} = \frac{\text{Number of Correct Predictions}}{\text{Total Number of Predictions}} 
\end{equation}
In our evaluation, the calculations are repeated 100 times, with each run incorporating a random train-test split and a different realization of noise. For random oriented source analysis, each iteration also applies an independent random inclination scaling to each test waveform. This repetition ensures statistical robustness by reducing bias from any single train-test split, noise realization, or orientation sample. We average performance across all iterations and use the standard deviation as the error, making our results independent of these random choices.

\begin{figure}
\centering
\includegraphics[width=\linewidth]{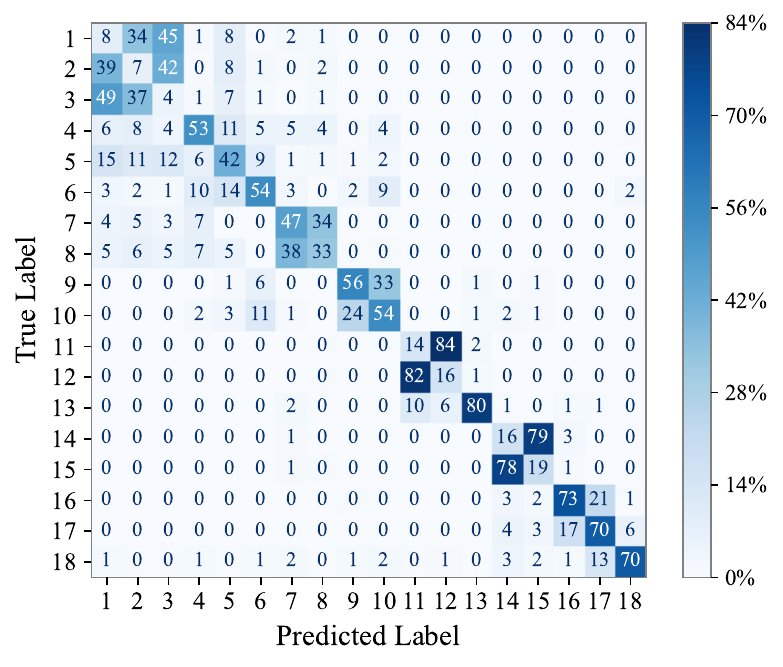}
\caption{Confusion matrix for the classification of 18 EOS models without added noise, using signals with $0.02 < T/|W| < 0.18$. EOS numerical labels correspond to those listed in Table~\ref{Table:EOS_GR_ML}. The overall classification accuracy, averaged over 100 runs, is $39.8\% \pm 3.1\%$.}
\label{fig:18eos_confusion_matrix}
\end{figure}

\begin{figure}
\centering
\includegraphics[width= \linewidth]{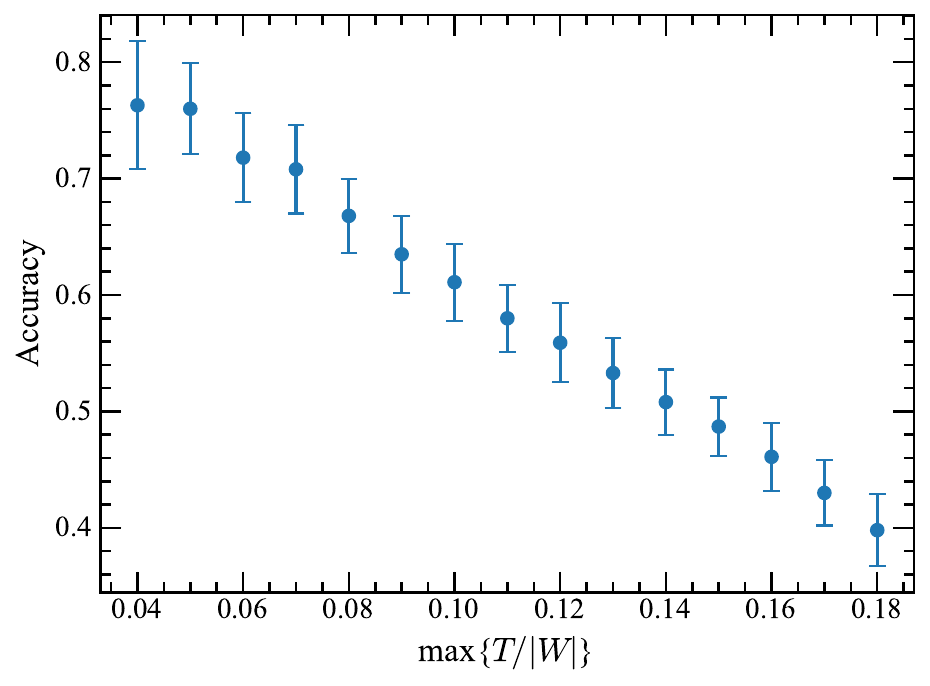}
\caption{Classification accuracy as a function of the maximum $T/|W|$ value included in the dataset, evaluated using clean (noise-free) signals.}
\label{fig:tw_acc}
\end{figure}

\begin{figure*}[t]
    \centering
    \subfloat{{\includegraphics[width=.47\textwidth]{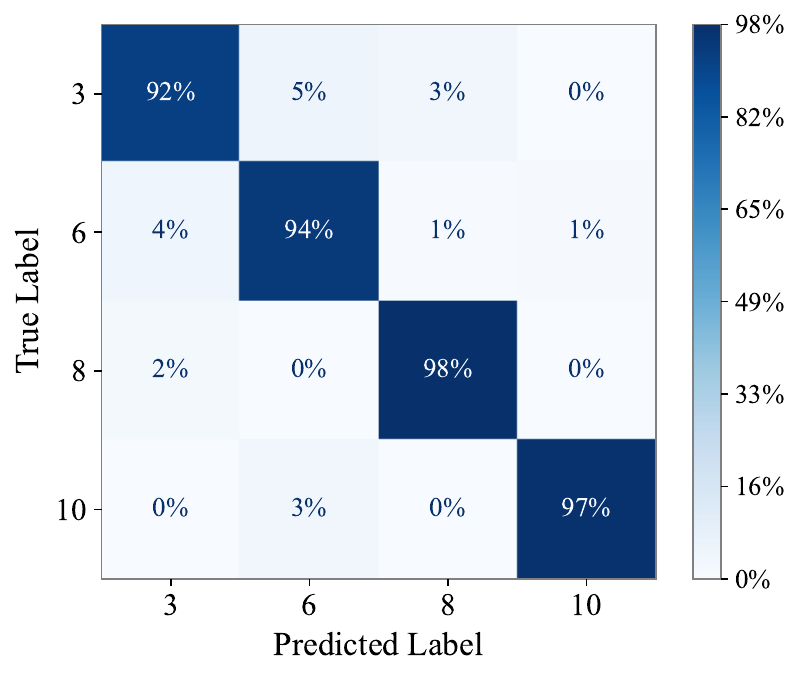} }}
    \hspace{0.03\textwidth}
    \subfloat{{\includegraphics[width=.47\textwidth]{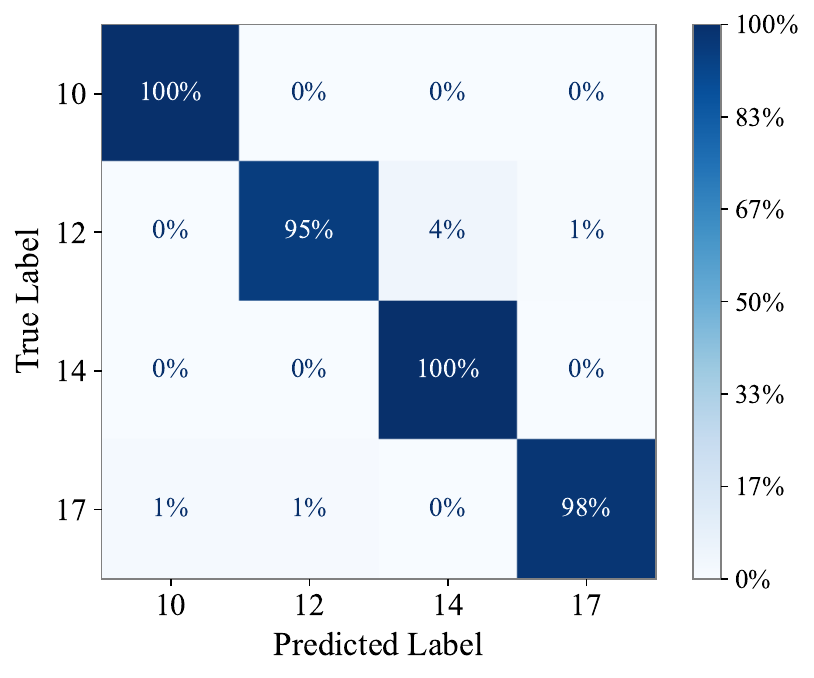} }}
    \caption{Confusion matrices for EOSs with the same (left) and distinct (right) treatments of low-density matter for $T/|W| < 0.10$. The corresponding classification accuracies are $95.6\% \pm 5.5\%$ and $98.3\% \pm 3.1\%$, respectively. The EOS labels are provided in Table~\ref{Table:EOS_GR_ML}.}
    \label{fig:CM_low-density_test}
\end{figure*}

\section{Results} 
\label{sec:result}

\begin{figure*}[t]
    \centering
    \subfloat{{\includegraphics[width=.47\textwidth]{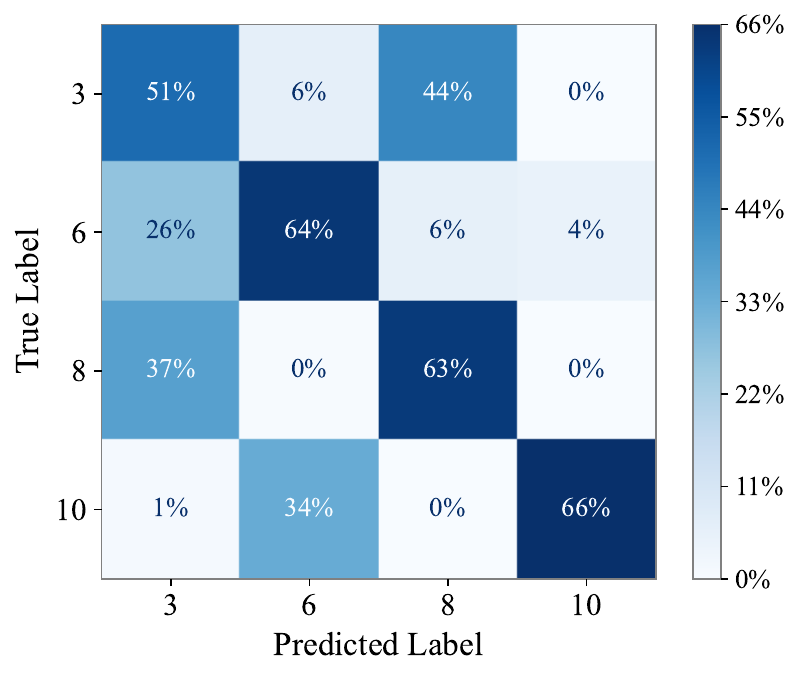} }}
    \hspace{0.03\textwidth}
    \subfloat{{\includegraphics[width=.47\textwidth]{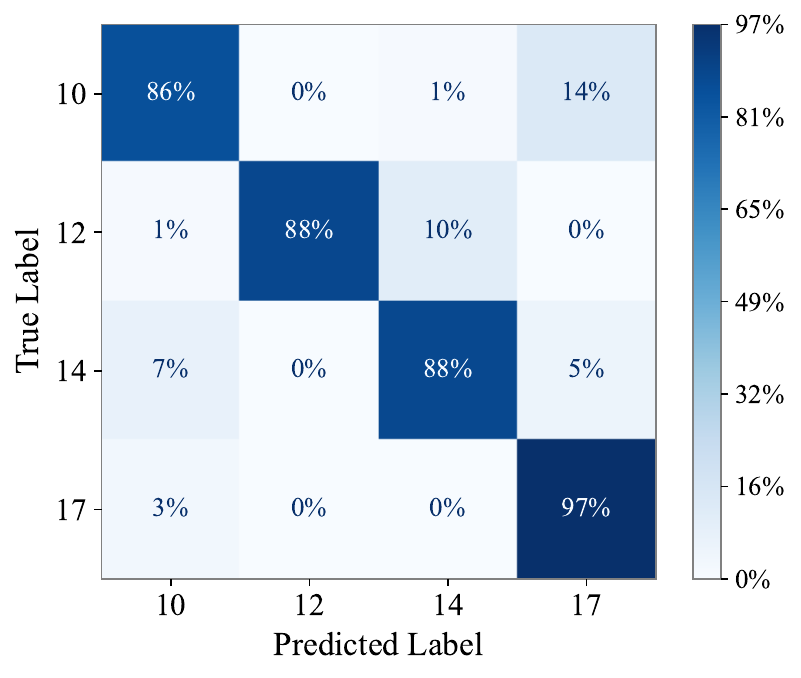} }}
    \caption{Confusion matrices for EOSs with the same (left) and distinct (right) treatments of low-density matter at extreme rotation $0.14 < T/|W| < 0.18$. The corresponding classification accuracies are $61.1\% \pm 4.5\%$ and $90.3\% \pm 3.1\%$, respectively. The EOS labels are provided in Table~\ref{Table:EOS_GR_ML}.}
    \label{fig:CM_low-density_test2}
\end{figure*}

\begin{figure}
\centering
\includegraphics[width= \linewidth]{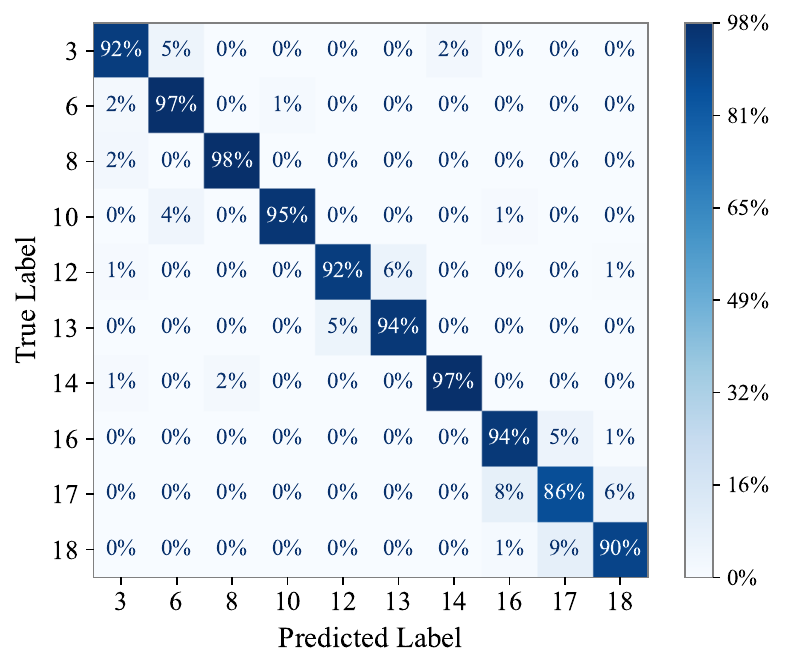}
\caption{Confusion matrix for the reduced set of 10 EOS models. The matrix is computed using noise-free signals, with $T/|W|$ values ranging from 0.02 to 0.10. The overall classification accuracy improves significantly compared to the 18 EOS case, reaching $93.1\% \pm 4.1\%$. The EOS labels are provided in Table~\ref{Table:EOS_GR_ML}.}
\label{fig:10eos_confusion_matrix}
\end{figure}

\begin{figure}
\centering
\includegraphics[width= \linewidth]{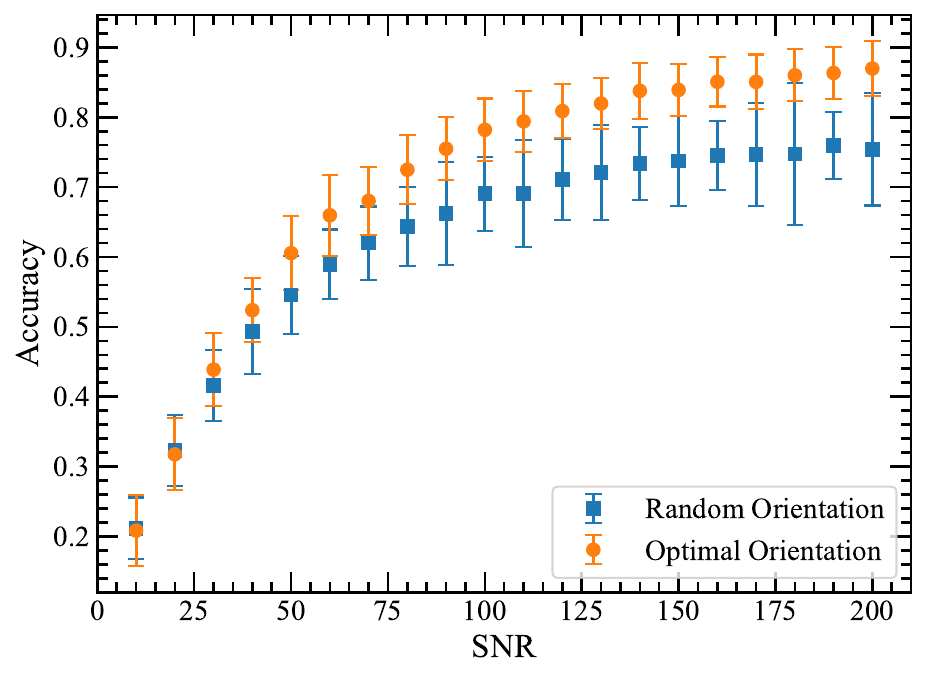}
\caption{EOS classification accuracy as a function of signal-to-noise ratio (SNR) for $T/|W| < 0.10$. For optimal orientation, the accuracy exceeds $70\%$ for SNR $\gtrsim 70$. However, for random source orientation, this is achieved for SNR of $\sim 100$.}
\label{fig:snr_acc}
\end{figure}

\begin{figure*}
\centering
\includegraphics[width= \linewidth]{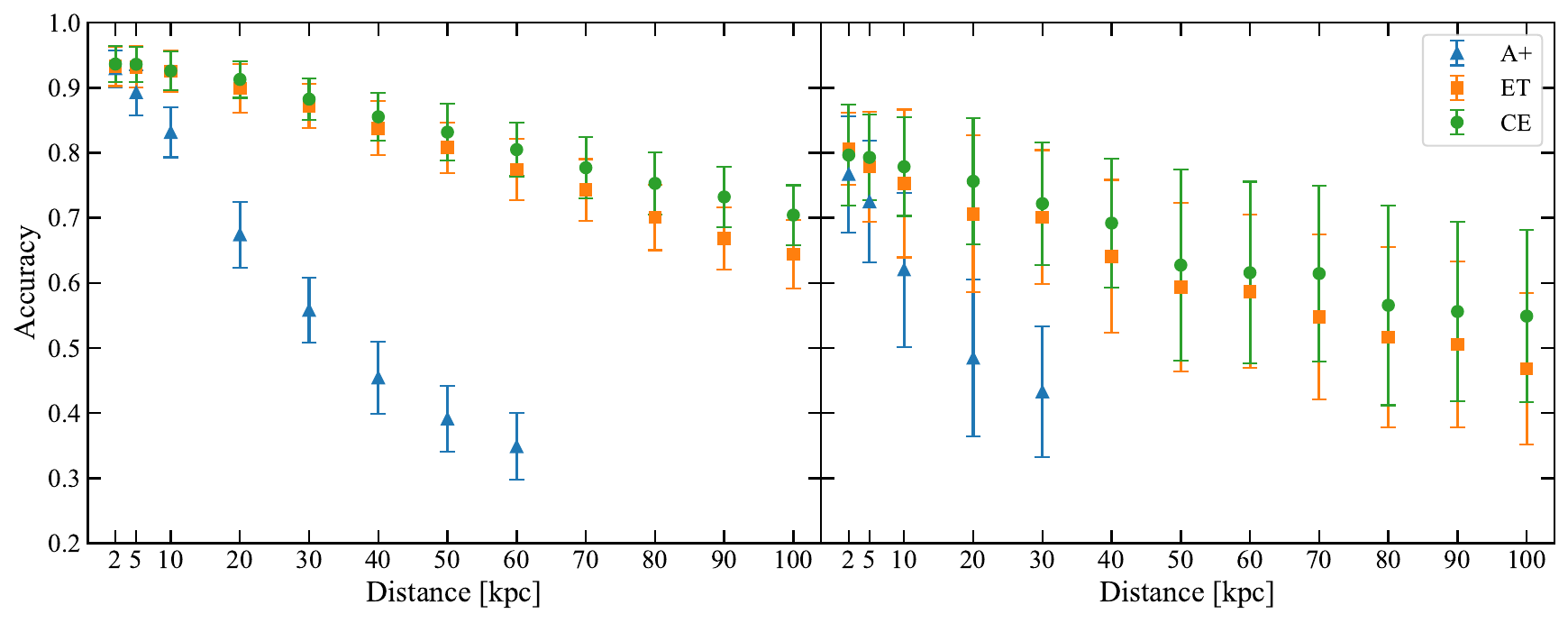}
\caption{Classification accuracy as a function of source distance for different detectors. The left panel corresponds to optimal orientation of the source, while the right panel corresponds to random orientation.}
\label{fig:detector_distance}
\end{figure*}

As a first step, we run our classification algorithm on the entire dataset without adding noise. Fig.~\ref{fig:18eos_confusion_matrix} shows the averaged confusion matrix (in percentage) obtained from 100 runs of the 18 EOS classification on the clean dataset. The EOS numerical labels are decoded in Table~\ref{Table:EOS_GR_ML}, where the first column lists the numerical labels and the second column provides the corresponding EOS model names. The overall classification accuracy is $39.8\% \pm 3.1\%$. This is consistent with the similar calculation in \cite{mitra24} using a convolutional neural network model. In the following, we analyze how these results vary with rotation, EOS family, and source distance.

\subsection{Dependence on Rotation}

In our earlier work \cite{mitra24}, it was shown that for extremely rapid rotation, EOS classification accuracy decreases with increasing $T/|W|$. This decline is attributed to the growing influence of centrifugal support, which prevents the stellar core from reaching the extreme densities where EOS-dependent effects are most pronounced. To further quantify this effect, we evaluate the classification accuracy as a function of the maximum value of $T/|W|$ included in the dataset, hereafter denoted $\max{ \{ T/|W|\} }$. As shown in Fig.~\ref{fig:tw_acc}, the accuracy decreases approximately linearly from $ 76.3\%\pm5.5\% $ at $ \max{ \{ T/|W|\} } = 0.04 $ to $61.1\%\pm3.3\%$ at $\max{ \{ T/|W|\} } = 0.10$, and further to $39.8\%\pm3.1\%$ at $\max{ \{ T/|W|\}} = 0.18$. Reducing $\max{ \{ T/|W|\} } $ below $0.04$ yields a dataset with too few waveforms, which increases the statistical uncertainty. Therefore, we do not consider that limit. 

Based on these results, we restrict our subsequent analysis to the range $0.02 < T/|W| < 0.10$, ensuring a balance between classification performance and dataset size (664 waveforms), while retaining a physically relevant portion of the parameter space. Because rapid rotation is rare, the condition $T/|W| < 0.10$ does not constitute a substantial additional constraint, as progenitors with faster rotation are probably extremely uncommon \cite{Heger05Presupernova}. For a detected signal, the value of $T/|W|$ can be inferred using the method of \citet{pastor24}.

\subsection{Dependence on EOS family}
\label{subsec:EOS family}

As we saw above, the classification accuracy of the full set of 18 EOSs is $61.1\%\pm3.3\%$ for $T/|W| < 0.10$. This relatively low accuracy can be understood by examining Fig.~\ref{fig:18eos_confusion_matrix}, which shows substantial confusion among EOSs that share the same low-density physics\footnote{In Table~\ref{Table:EOS_GR_ML}, different low-density treatments are separated by horizontal lines.}. As noted by \citet{richers:17}, the bounce dynamics, and thus the GW signal, depend on both the low- and high-density treatments. For example, GShenFSU2.1 and GShenFSU1.7 differ only at densities above nuclear saturation. Similarly, HShenH extends HShen by including hyperons. BHB$\Lambda$ extends HSDD2 by including hyperons, while BHB$\Lambda\Phi$ introduces an additional hyperonic interaction. All Hempel-based EOS families (HS, SFH, BHB) share the same treatment of low-density nonuniform matter \cite{richers:17}.

This raises an important question: is our classifier probing the low-density regime, the high-density regime, or both? To address this, we perform the following test. Among the full set of 18 EOSs, there are four distinct treatments of the low-density regime. From this, we select four EOSs with different low-density prescriptions (labels 10, 12, 14, and 17; see Table~\ref{Table:EOS_GR_ML}). We then select another four EOSs that share the same low-density prescription (labels 3, 6, 8, and 10). The resulting classification accuracies are $98.3\%\pm 3.1\%$ and $95.6\%\pm 5.5\%$, respectively. The corresponding confusion matrices are shown in Fig.~\ref{fig:CM_low-density_test}. These results indicate that even for EOSs sharing the same low-density physics, the models can still be distinguished.

To further examine the influence of low-density matter treatment, we perform another test under extreme rotation $0.14 < T/|W| < 0.18$. The resulting confusion matrices are shown in Fig.~\ref{fig:CM_low-density_test2}. For EOSs sharing the same low-density prescription, the overall classification accuracy reduces to $61.1\%\pm4.5\%$, significantly lower than $95.6\%\pm5.5\%$ obtained in the $T/|W| < 0.10$ case. This reduction is expected: at extremely high rotation rates, the core does not reach high densities, where EOS differences are largest, limiting our ability to probe the high-density regime. 

In contrast, for EOSs with different low-density treatments, the accuracy remains high at $90.3\%\pm3.1\%$ even under extremely high rotation, though still somewhat lower than the $98.3\%\pm3.1\%$ value obtained in the $T/|W| < 0.10$ case. In summary, at very rapid rotation the GW signal becomes less sensitive to high-density physics, but remains sensitive to differences in the low-density regime.

One caveat must be noted. Because the current set of EOS models is limited in size, we cannot robustly quantify how much of the classification accuracy arises from variations in the low- versus high-density regions of the EOS. Overcoming this limitation will require a large, parametrized family of EOS models, enabling a systematic study of the respective impacts of the low- and high-density components on the GW signal. We leave this for future work.

In the following analysis, we select ten well distinguishable EOS models (labeled 3, 6, 8, 10, 12, 13, 14, 16, 17, and 18), encompassing both the same and different low-density treatments. The corresponding confusion matrix is shown in Fig.~\ref{fig:10eos_confusion_matrix}. The overall classification accuracy is $93.1\%\pm 4.1\%$, which is significantly higher than that obtained for the full 18 EOS dataset.

\subsection{Dependence on Distance}

We now investigate classification performance under noisy conditions, assuming a detected GW signal. The orange circles in Fig.~\ref{fig:snr_acc} show the classification accuracy as a function of SNR for optimally oriented sources. The accuracy reaches $68.0\%\pm4.8\%$ at SNR = 70 and $87.1 \pm 3.9\%$ at SNR = 200. This is notable performance for a ten-class classification problem. For randomly oriented sources (blue squares in Fig.~\ref{fig:snr_acc}), the accuracy is lower by about $10\%$, reaching $69.1\%\pm5.1\%$ at SNR = 100. We also notice somewhat larger standard deviations for randomly oriented sources, which reflects the variability introduced by averaging over different orientations and corresponding signal amplitudes.

We now evaluate classification performance as a function of distance to the source for different detectors. This is shown in Fig.~\ref{fig:detector_distance}. For optimal orientation, shown on the left panel, Advanced LIGO A+ achieves $\gtrsim 70\%$ classification accuracy up to 20 kpc, encompassing the most probable region for Galactic CCSNe. The next-generation detectors Einstein Telescope and Cosmic Explorer extend the horizon substantially, achieving accuracies above $\sim 90\%$ within 10 kpc, and maintaining $\gtrsim 70\%$ accuracy out to $\sim$80 kpc and $\sim$100 kpc, respectively. 

For randomly oriented sources, the corresponding accuracy values are shown on the right panel of Fig.~\ref{fig:detector_distance}. For Advanced LIGO A+, the accuracy drops below $70\%$ already at a distance of $\sim$10 kpc. For ET and CE, the classification accuracy decreases from $92.4\%\pm2.9\%$ (both detectors) for optimal orientation to $75.0\%\pm11.4\%$ (ET) and $77.5\%\pm7.5\%$ (CE) for sources around 10 kpc. The accuracy drops below $\sim 70\%$ beyond a distance of 30 kpc. Moreover, we observe significant increase in the standard deviation of the accuracy values. This suggests that source orientation plays a significant role in achieving accurate classification. For non-optimal orientations, only next-generation detectors such as ET and CE may enable EOS inference for Galactic sources.

\section{Conclusion} 
\label{sec:conclusion}

In this work, we investigated the feasibility of classifying the nuclear equation of state (EOS) from gravitational wave (GW) signals from rotating during core-collapse supernovae. Using a catalog of 18 EOS models simulated across a broad range of rotational configurations, we quantified the classification performance of a support vector machine classifier. 

We estimated the distances out to which the nuclear EOS can be probed. For optimally oriented sources, Advanced LIGO A+ may enable EOS classification within $\sim$20 kpc. Next-generation detectors, such as the Einstein Telescope (ET) and Cosmic Explorer (CE), substantially extend this reach, allowing classification at distances of up to $\sim$80 kpc and $\sim$100 kpc, respectively. However, this capability is strongly dependent on source orientation. For orientation is not optimal, the classification accuracy decreases considerably. In such cases, only ET and CE are expected to probe the EOS for Galactic events (cf. Section~\ref{sec:result}).

We emphasize that these results represent an optimistic scenario with several limitations. We rely on a limited set of EOS models. Ideally, one would perform a regression analysis on underlying nuclear physics parameters, but this is left to future work. We also focus on bounce signals from rotating progenitors and restrict the analysis to the first $\sim$6 ms after bounce. Such signals are relatively easier to detect, even at moderate SNR, compared to those from non-rotating or slowly rotating stars \cite{Szczepanczyk21Detecting}. Extending the framework to longer timescales remains computationally prohibitive, as producing sufficiently large sets of long-duration simulations for machine-learning analyses remains computationally demanding with present simulation codes. Moreover, accurately determining the bounce time in real GW data may be difficult \cite{gossan16observing}, limiting the ability to align signals within the $-2$ to $6$ ms window used in our analysis. Also, rapid rotation is expected to be rare \citep{Heger05Presupernova}, and it remains uncertain how our results extrapolate to more slowly rotating models. Furthermore, our dataset is restricted to waveforms from a single progenitor mass. In realistic cases, uncertainties in the progenitor mass may impose additional constraints, an aspect we leave for future investigation. 

Despite these caveats, our study provides an approximate estimate of EOS distinguishability across detector sensitivities, rotation rates, and noise levels. This work is a step toward more realistic studies integrating detector modeling, improved machine learning, and multi-messenger data to probe dense matter.

\begin{acknowledgments}

We thank Evan O'Connor for useful discussions. This research was funded by the Science Committee of the Ministry of Science and Higher Education of the Republic of Kazakhstan (Grant No. AP26103591) and partially supported by the Nazarbayev University Faculty Development Competitive Research Grant Program (no. 040225FD4713).  MCE kindly acknowledges support by the Marsden Grant No. MFP-UOA2131 from New Zealand Government funding, administered by the Royal Society Te Apārangi. PCD, ATF and JAF are supported by the Spanish Agencia Estatal de Investigaci\'on (grants PID2021-125485NB-C21 and PID2024-159689NB-C21) funded by MCIN/AEI/10.13039/501100011033 and ERDF A way of making Europe, and by the Generalitat Valenciana (grant CIPROM/2022/49). M.S. acknowledges Polish National Science Centre Grants No. UMO-2023/49/B/ST9/02777 and No. UMO-2024/03/1/ST9/00005, and the Polish National Agency for Academic Exchange within Polish Returns Programme Grant No. BPN/PPO/2023/1/0001.
\end{acknowledgments}

\bibliography{gw_sn}

\begin{thebibliography}{101}%
\makeatletter
\providecommand \@ifxundefined [1]{%
 \@ifx{#1\undefined}
}%
\providecommand \@ifnum [1]{%
 \ifnum #1\expandafter \@firstoftwo
 \else \expandafter \@secondoftwo
 \fi
}%
\providecommand \@ifx [1]{%
 \ifx #1\expandafter \@firstoftwo
 \else \expandafter \@secondoftwo
 \fi
}%
\providecommand \natexlab [1]{#1}%
\providecommand \enquote  [1]{``#1''}%
\providecommand \bibnamefont  [1]{#1}%
\providecommand \bibfnamefont [1]{#1}%
\providecommand \citenamefont [1]{#1}%
\providecommand \href@noop [0]{\@secondoftwo}%
\providecommand \href [0]{\begingroup \@sanitize@url \@href}%
\providecommand \@href[1]{\@@startlink{#1}\@@href}%
\providecommand \@@href[1]{\endgroup#1\@@endlink}%
\providecommand \@sanitize@url [0]{\catcode `\\12\catcode `\$12\catcode `\&12\catcode `\#12\catcode `\^12\catcode `\_12\catcode `\%12\relax}%
\providecommand \@@startlink[1]{}%
\providecommand \@@endlink[0]{}%
\providecommand \url  [0]{\begingroup\@sanitize@url \@url }%
\providecommand \@url [1]{\endgroup\@href {#1}{\urlprefix }}%
\providecommand \urlprefix  [0]{URL }%
\providecommand \Eprint [0]{\href }%
\providecommand \doibase [0]{https://doi.org/}%
\providecommand \selectlanguage [0]{\@gobble}%
\providecommand \bibinfo  [0]{\@secondoftwo}%
\providecommand \bibfield  [0]{\@secondoftwo}%
\providecommand \translation [1]{[#1]}%
\providecommand \BibitemOpen [0]{}%
\providecommand \bibitemStop [0]{}%
\providecommand \bibitemNoStop [0]{.\EOS\space}%
\providecommand \EOS [0]{\spacefactor3000\relax}%
\providecommand \BibitemShut  [1]{\csname bibitem#1\endcsname}%
\let\auto@bib@innerbib\@empty
\bibitem [{\citenamefont {{The LIGO Scientific Collaboration}}\ \emph {et~al.}(2025)\citenamefont {{The LIGO Scientific Collaboration}}, \citenamefont {{the Virgo Collaboration}}, \citenamefont {{the KAGRA Collaboration}}, \citenamefont {{Abac}}, \citenamefont {{Abouelfettouh}}, \citenamefont {{Acernese}},\ and\ \citenamefont {{Ackley}}}]{Abac25GWTC4}%
  \BibitemOpen
  \bibfield  {author} {\bibinfo {author} {\bibnamefont {{The LIGO Scientific Collaboration}}}, \bibinfo {author} {\bibnamefont {{the Virgo Collaboration}}}, \bibinfo {author} {\bibnamefont {{the KAGRA Collaboration}}}, \bibinfo {author} {\bibfnamefont {A.~G.}\ \bibnamefont {{Abac}}}, \bibinfo {author} {\bibfnamefont {I.}~\bibnamefont {{Abouelfettouh}}}, \bibinfo {author} {\bibfnamefont {F.}~\bibnamefont {{Acernese}}},\ and\ \bibinfo {author} {\bibfnamefont {K.~e.~a.}\ \bibnamefont {{Ackley}}},\ }\bibfield  {title} {\bibinfo {title} {{GWTC-4.0: Updating the Gravitational-Wave Transient Catalog with Observations from the First Part of the Fourth LIGO-Virgo-KAGRA Observing Run}},\ }\href {https://doi.org/10.48550/arXiv.2508.18082} {\bibfield  {journal} {\bibinfo  {journal} {arXiv e-prints}\ ,\ \bibinfo {eid} {arXiv:2508.18082}} (\bibinfo {year} {2025})},\ \Eprint {https://arxiv.org/abs/2508.18082} {arXiv:2508.18082 [gr-qc]} \BibitemShut {NoStop}%
\bibitem [{\citenamefont {Abbott}\ \emph {et~al.}(2020)\citenamefont {Abbott}, \citenamefont {Abbott}, \citenamefont {Abbott}, \citenamefont {Abraham},\ and\ \citenamefont {Acernese}}]{abbott2020optically}%
  \BibitemOpen
  \bibfield  {author} {\bibinfo {author} {\bibfnamefont {B.~P.}\ \bibnamefont {Abbott}}, \bibinfo {author} {\bibfnamefont {R.}~\bibnamefont {Abbott}}, \bibinfo {author} {\bibfnamefont {T.~D.}\ \bibnamefont {Abbott}}, \bibinfo {author} {\bibfnamefont {S.}~\bibnamefont {Abraham}},\ and\ \bibinfo {author} {\bibfnamefont {F.}~\bibnamefont {Acernese}} (\bibinfo {collaboration} {LIGO Scientific Collaboration and Virgo Collaboration and ASAS-SN Collaboration and DLT40 Collaboration}),\ }\bibfield  {title} {\bibinfo {title} {Optically targeted search for gravitational waves emitted by core-collapse supernovae during the first and second observing runs of advanced ligo and advanced virgo},\ }\href {https://doi.org/10.1103/PhysRevD.101.084002} {\bibfield  {journal} {\bibinfo  {journal} {Phys. Rev. D}\ }\textbf {\bibinfo {volume} {101}},\ \bibinfo {pages} {084002} (\bibinfo {year} {2020})}\BibitemShut {NoStop}%
\bibitem [{\citenamefont {Szczepa\ifmmode~\acute{n}\else \'{n}\fi{}czyk}\ \emph {et~al.}(2024)\citenamefont {Szczepa\ifmmode~\acute{n}\else \'{n}\fi{}czyk}, \citenamefont {Zheng}, \citenamefont {Antelis}, \citenamefont {Benjamin}, \citenamefont {Bizouard}, \citenamefont {Casallas-Lagos}, \citenamefont {Cerd\'a-Dur\'an}, \citenamefont {Davis}, \citenamefont {Gondek-Rosi\ifmmode~\acute{n}\else \'{n}\fi{}ska}, \citenamefont {Klimenko}, \citenamefont {Moreno}, \citenamefont {Obergaulinger}, \citenamefont {Powell}, \citenamefont {Ramirez}, \citenamefont {Ratto}, \citenamefont {Richardson}, \citenamefont {Rijal}, \citenamefont {Stuver}, \citenamefont {Szewczyk}, \citenamefont {Vedovato}, \citenamefont {Zanolin}, \citenamefont {Bartos}, \citenamefont {Bhaumik}, \citenamefont {Bulik}, \citenamefont {Drago}, \citenamefont {Font}, \citenamefont {De~Colle}, \citenamefont {Garc\'{\i}a-Bellido}, \citenamefont {Gayathri}, \citenamefont {Hughey}, \citenamefont {Mitselmakher}, \citenamefont {Mishra}, \citenamefont
  {Mukherjee}, \citenamefont {Nguyen}, \citenamefont {Chan}, \citenamefont {Di~Palma}, \citenamefont {Piotrzkowski},\ and\ \citenamefont {Singh}}]{szczepanczyk2023optically}%
  \BibitemOpen
  \bibfield  {author} {\bibinfo {author} {\bibfnamefont {M.~J.}\ \bibnamefont {Szczepa\ifmmode~\acute{n}\else \'{n}\fi{}czyk}}, \bibinfo {author} {\bibfnamefont {Y.}~\bibnamefont {Zheng}}, \bibinfo {author} {\bibfnamefont {J.~M.}\ \bibnamefont {Antelis}}, \bibinfo {author} {\bibfnamefont {M.}~\bibnamefont {Benjamin}}, \bibinfo {author} {\bibfnamefont {M.-A.}\ \bibnamefont {Bizouard}}, \bibinfo {author} {\bibfnamefont {A.}~\bibnamefont {Casallas-Lagos}}, \bibinfo {author} {\bibfnamefont {P.}~\bibnamefont {Cerd\'a-Dur\'an}}, \bibinfo {author} {\bibfnamefont {D.}~\bibnamefont {Davis}}, \bibinfo {author} {\bibfnamefont {D.}~\bibnamefont {Gondek-Rosi\ifmmode~\acute{n}\else \'{n}\fi{}ska}}, \bibinfo {author} {\bibfnamefont {S.}~\bibnamefont {Klimenko}}, \bibinfo {author} {\bibfnamefont {C.}~\bibnamefont {Moreno}}, \bibinfo {author} {\bibfnamefont {M.}~\bibnamefont {Obergaulinger}}, \bibinfo {author} {\bibfnamefont {J.}~\bibnamefont {Powell}}, \bibinfo {author} {\bibfnamefont {D.}~\bibnamefont {Ramirez}}, \bibinfo
  {author} {\bibfnamefont {B.}~\bibnamefont {Ratto}}, \bibinfo {author} {\bibfnamefont {C.}~\bibnamefont {Richardson}}, \bibinfo {author} {\bibfnamefont {A.}~\bibnamefont {Rijal}}, \bibinfo {author} {\bibfnamefont {A.~L.}\ \bibnamefont {Stuver}}, \bibinfo {author} {\bibfnamefont {P.}~\bibnamefont {Szewczyk}}, \bibinfo {author} {\bibfnamefont {G.}~\bibnamefont {Vedovato}}, \bibinfo {author} {\bibfnamefont {M.}~\bibnamefont {Zanolin}}, \bibinfo {author} {\bibfnamefont {I.}~\bibnamefont {Bartos}}, \bibinfo {author} {\bibfnamefont {S.}~\bibnamefont {Bhaumik}}, \bibinfo {author} {\bibfnamefont {T.}~\bibnamefont {Bulik}}, \bibinfo {author} {\bibfnamefont {M.}~\bibnamefont {Drago}}, \bibinfo {author} {\bibfnamefont {J.~A.}\ \bibnamefont {Font}}, \bibinfo {author} {\bibfnamefont {F.}~\bibnamefont {De~Colle}}, \bibinfo {author} {\bibfnamefont {J.}~\bibnamefont {Garc\'{\i}a-Bellido}}, \bibinfo {author} {\bibfnamefont {V.}~\bibnamefont {Gayathri}}, \bibinfo {author} {\bibfnamefont {B.}~\bibnamefont {Hughey}}, \bibinfo
  {author} {\bibfnamefont {G.}~\bibnamefont {Mitselmakher}}, \bibinfo {author} {\bibfnamefont {T.}~\bibnamefont {Mishra}}, \bibinfo {author} {\bibfnamefont {S.}~\bibnamefont {Mukherjee}}, \bibinfo {author} {\bibfnamefont {Q.~L.}\ \bibnamefont {Nguyen}}, \bibinfo {author} {\bibfnamefont {M.~L.}\ \bibnamefont {Chan}}, \bibinfo {author} {\bibfnamefont {I.}~\bibnamefont {Di~Palma}}, \bibinfo {author} {\bibfnamefont {B.~J.}\ \bibnamefont {Piotrzkowski}},\ and\ \bibinfo {author} {\bibfnamefont {N.}~\bibnamefont {Singh}},\ }\bibfield  {title} {\bibinfo {title} {Optically targeted search for gravitational waves emitted by core-collapse supernovae during the third observing run of advanced ligo and advanced virgo},\ }\href {https://doi.org/10.1103/PhysRevD.110.042007} {\bibfield  {journal} {\bibinfo  {journal} {Phys. Rev. D}\ }\textbf {\bibinfo {volume} {110}},\ \bibinfo {pages} {042007} (\bibinfo {year} {2024})}\BibitemShut {NoStop}%
\bibitem [{\citenamefont {{Powell}}\ and\ \citenamefont {{M{\"u}ller}}(2022)}]{powell22inferring}%
  \BibitemOpen
  \bibfield  {author} {\bibinfo {author} {\bibfnamefont {J.}~\bibnamefont {{Powell}}}\ and\ \bibinfo {author} {\bibfnamefont {B.}~\bibnamefont {{M{\"u}ller}}},\ }\bibfield  {title} {\bibinfo {title} {{Inferring astrophysical parameters of core-collapse supernovae from their gravitational-wave emission}},\ }\href {https://doi.org/10.1103/PhysRevD.105.063018} {\bibfield  {journal} {\bibinfo  {journal} {\prd}\ }\textbf {\bibinfo {volume} {105}},\ \bibinfo {eid} {063018} (\bibinfo {year} {2022})},\ \Eprint {https://arxiv.org/abs/2201.01397} {arXiv:2201.01397 [astro-ph.HE]} \BibitemShut {NoStop}%
\bibitem [{\citenamefont {{Gossan}}\ \emph {et~al.}(2016)\citenamefont {{Gossan}}, \citenamefont {{Sutton}}, \citenamefont {{Stuver}}, \citenamefont {{Zanolin}}, \citenamefont {{Gill}},\ and\ \citenamefont {{Ott}}}]{gossan16observing}%
  \BibitemOpen
  \bibfield  {author} {\bibinfo {author} {\bibfnamefont {S.~E.}\ \bibnamefont {{Gossan}}}, \bibinfo {author} {\bibfnamefont {P.}~\bibnamefont {{Sutton}}}, \bibinfo {author} {\bibfnamefont {A.}~\bibnamefont {{Stuver}}}, \bibinfo {author} {\bibfnamefont {M.}~\bibnamefont {{Zanolin}}}, \bibinfo {author} {\bibfnamefont {K.}~\bibnamefont {{Gill}}},\ and\ \bibinfo {author} {\bibfnamefont {C.~D.}\ \bibnamefont {{Ott}}},\ }\bibfield  {title} {\bibinfo {title} {{Observing gravitational waves from core-collapse supernovae in the advanced detector era}},\ }\href {https://doi.org/10.1103/PhysRevD.93.042002} {\bibfield  {journal} {\bibinfo  {journal} {\prd}\ }\textbf {\bibinfo {volume} {93}},\ \bibinfo {eid} {042002} (\bibinfo {year} {2016})},\ \Eprint {https://arxiv.org/abs/1511.02836} {arXiv:1511.02836 [astro-ph.HE]} \BibitemShut {NoStop}%
\bibitem [{\citenamefont {{Szczepa{\'n}czyk}}\ \emph {et~al.}(2021)\citenamefont {{Szczepa{\'n}czyk}}, \citenamefont {{Antelis}}, \citenamefont {{Benjamin}}, \citenamefont {{Cavagli{\`a}}}, \citenamefont {{Gondek-Rosi{\'n}ska}}, \citenamefont {{Hansen}}, \citenamefont {{Klimenko}}, \citenamefont {{Morales}}, \citenamefont {{Moreno}}, \citenamefont {{Mukherjee}}, \citenamefont {{Nurbek}}, \citenamefont {{Powell}}, \citenamefont {{Singh}}, \citenamefont {{Sitmukhambetov}}, \citenamefont {{Szewczyk}}, \citenamefont {{Valdez}}, \citenamefont {{Vedovato}}, \citenamefont {{Westhouse}}, \citenamefont {{Zanolin}},\ and\ \citenamefont {{Zheng}}}]{Szczepanczyk21Detecting}%
  \BibitemOpen
  \bibfield  {author} {\bibinfo {author} {\bibfnamefont {M.~J.}\ \bibnamefont {{Szczepa{\'n}czyk}}}, \bibinfo {author} {\bibfnamefont {J.~M.}\ \bibnamefont {{Antelis}}}, \bibinfo {author} {\bibfnamefont {M.}~\bibnamefont {{Benjamin}}}, \bibinfo {author} {\bibfnamefont {M.}~\bibnamefont {{Cavagli{\`a}}}}, \bibinfo {author} {\bibfnamefont {D.}~\bibnamefont {{Gondek-Rosi{\'n}ska}}}, \bibinfo {author} {\bibfnamefont {T.}~\bibnamefont {{Hansen}}}, \bibinfo {author} {\bibfnamefont {S.}~\bibnamefont {{Klimenko}}}, \bibinfo {author} {\bibfnamefont {M.~D.}\ \bibnamefont {{Morales}}}, \bibinfo {author} {\bibfnamefont {C.}~\bibnamefont {{Moreno}}}, \bibinfo {author} {\bibfnamefont {S.}~\bibnamefont {{Mukherjee}}}, \bibinfo {author} {\bibfnamefont {G.}~\bibnamefont {{Nurbek}}}, \bibinfo {author} {\bibfnamefont {J.}~\bibnamefont {{Powell}}}, \bibinfo {author} {\bibfnamefont {N.}~\bibnamefont {{Singh}}}, \bibinfo {author} {\bibfnamefont {S.}~\bibnamefont {{Sitmukhambetov}}}, \bibinfo {author} {\bibfnamefont
  {P.}~\bibnamefont {{Szewczyk}}}, \bibinfo {author} {\bibfnamefont {O.}~\bibnamefont {{Valdez}}}, \bibinfo {author} {\bibfnamefont {G.}~\bibnamefont {{Vedovato}}}, \bibinfo {author} {\bibfnamefont {J.}~\bibnamefont {{Westhouse}}}, \bibinfo {author} {\bibfnamefont {M.}~\bibnamefont {{Zanolin}}},\ and\ \bibinfo {author} {\bibfnamefont {Y.}~\bibnamefont {{Zheng}}},\ }\bibfield  {title} {\bibinfo {title} {{Detecting and reconstructing gravitational waves from the next galactic core-collapse supernova in the advanced detector era}},\ }\href {https://doi.org/10.1103/PhysRevD.104.102002} {\bibfield  {journal} {\bibinfo  {journal} {\prd}\ }\textbf {\bibinfo {volume} {104}},\ \bibinfo {eid} {102002} (\bibinfo {year} {2021})},\ \Eprint {https://arxiv.org/abs/2104.06462} {arXiv:2104.06462 [astro-ph.HE]} \BibitemShut {NoStop}%
\bibitem [{\citenamefont {{Srivastava}}\ \emph {et~al.}(2019)\citenamefont {{Srivastava}}, \citenamefont {{Ballmer}}, \citenamefont {{Brown}}, \citenamefont {{Afle}}, \citenamefont {{Burrows}}, \citenamefont {{Radice}},\ and\ \citenamefont {{Vartanyan}}}]{srivastava19detection}%
  \BibitemOpen
  \bibfield  {author} {\bibinfo {author} {\bibfnamefont {V.}~\bibnamefont {{Srivastava}}}, \bibinfo {author} {\bibfnamefont {S.}~\bibnamefont {{Ballmer}}}, \bibinfo {author} {\bibfnamefont {D.~A.}\ \bibnamefont {{Brown}}}, \bibinfo {author} {\bibfnamefont {C.}~\bibnamefont {{Afle}}}, \bibinfo {author} {\bibfnamefont {A.}~\bibnamefont {{Burrows}}}, \bibinfo {author} {\bibfnamefont {D.}~\bibnamefont {{Radice}}},\ and\ \bibinfo {author} {\bibfnamefont {D.}~\bibnamefont {{Vartanyan}}},\ }\bibfield  {title} {\bibinfo {title} {{Detection prospects of core-collapse supernovae with supernova-optimized third-generation gravitational-wave detectors}},\ }\href {https://doi.org/10.1103/PhysRevD.100.043026} {\bibfield  {journal} {\bibinfo  {journal} {\prd}\ }\textbf {\bibinfo {volume} {100}},\ \bibinfo {eid} {043026} (\bibinfo {year} {2019})},\ \Eprint {https://arxiv.org/abs/1906.00084} {arXiv:1906.00084 [gr-qc]} \BibitemShut {NoStop}%
\bibitem [{\citenamefont {{Abdikamalov}}\ \emph {et~al.}(2022)\citenamefont {{Abdikamalov}}, \citenamefont {{Pagliaroli}},\ and\ \citenamefont {{Radice}}}]{abdikamalov22gravitational}%
  \BibitemOpen
  \bibfield  {author} {\bibinfo {author} {\bibfnamefont {E.}~\bibnamefont {{Abdikamalov}}}, \bibinfo {author} {\bibfnamefont {G.}~\bibnamefont {{Pagliaroli}}},\ and\ \bibinfo {author} {\bibfnamefont {D.}~\bibnamefont {{Radice}}},\ }\bibfield  {title} {\bibinfo {title} {{Gravitational Waves from Core-Collapse Supernovae}},\ }in\ \href {https://doi.org/10.1007/978-981-15-4702-7_21-1} {\emph {\bibinfo {booktitle} {Handbook of Gravitational Wave Astronomy}}},\ \bibinfo {editor} {edited by\ \bibinfo {editor} {\bibfnamefont {C.}~\bibnamefont {{Bambi}}}, \bibinfo {editor} {\bibfnamefont {S.}~\bibnamefont {{Katsanevas}}},\ and\ \bibinfo {editor} {\bibfnamefont {K.~D.}\ \bibnamefont {{Kokkotas}}}}\ (\bibinfo {year} {2022})\ p.~\bibinfo {pages} {21}\BibitemShut {NoStop}%
\bibitem [{\citenamefont {{Mezzacappa}}\ and\ \citenamefont {{Zanolin}}(2024)}]{mezzacappa24gravitational}%
  \BibitemOpen
  \bibfield  {author} {\bibinfo {author} {\bibfnamefont {A.}~\bibnamefont {{Mezzacappa}}}\ and\ \bibinfo {author} {\bibfnamefont {M.}~\bibnamefont {{Zanolin}}},\ }\bibfield  {title} {\bibinfo {title} {{Gravitational Waves from Neutrino-Driven Core Collapse Supernovae: Predictions, Detection, and Parameter Estimation}},\ }\href {https://doi.org/10.48550/arXiv.2401.11635} {\bibfield  {journal} {\bibinfo  {journal} {arXiv e-prints}\ ,\ \bibinfo {eid} {arXiv:2401.11635}} (\bibinfo {year} {2024})},\ \Eprint {https://arxiv.org/abs/2401.11635} {arXiv:2401.11635 [astro-ph.HE]} \BibitemShut {NoStop}%
\bibitem [{\citenamefont {{M{\"u}ller}}(2020)}]{muller20hydrodynamics}%
  \BibitemOpen
  \bibfield  {author} {\bibinfo {author} {\bibfnamefont {B.}~\bibnamefont {{M{\"u}ller}}},\ }\bibfield  {title} {\bibinfo {title} {{Hydrodynamics of core-collapse supernovae and their progenitors}},\ }\href {https://doi.org/10.1007/s41115-020-0008-5} {\bibfield  {journal} {\bibinfo  {journal} {Living Reviews in Computational Astrophysics}\ }\textbf {\bibinfo {volume} {6}},\ \bibinfo {eid} {3} (\bibinfo {year} {2020})},\ \Eprint {https://arxiv.org/abs/2006.05083} {arXiv:2006.05083 [astro-ph.SR]} \BibitemShut {NoStop}%
\bibitem [{\citenamefont {{O'Connor}}\ and\ \citenamefont {{Ott}}(2011)}]{oconnor11}%
  \BibitemOpen
  \bibfield  {author} {\bibinfo {author} {\bibfnamefont {E.}~\bibnamefont {{O'Connor}}}\ and\ \bibinfo {author} {\bibfnamefont {C.~D.}\ \bibnamefont {{Ott}}},\ }\bibfield  {title} {\bibinfo {title} {{Black Hole Formation in Failing Core-Collapse Supernovae}},\ }\href {https://doi.org/10.1088/0004-637X/730/2/70} {\bibfield  {journal} {\bibinfo  {journal} {\apj}\ }\textbf {\bibinfo {volume} {730}},\ \bibinfo {eid} {70} (\bibinfo {year} {2011})},\ \Eprint {https://arxiv.org/abs/1010.5550} {arXiv:1010.5550 [astro-ph.HE]} \BibitemShut {NoStop}%
\bibitem [{\citenamefont {{Burrows}}\ \emph {et~al.}(2023)\citenamefont {{Burrows}}, \citenamefont {{Vartanyan}},\ and\ \citenamefont {{Wang}}}]{Burrows23Black}%
  \BibitemOpen
  \bibfield  {author} {\bibinfo {author} {\bibfnamefont {A.}~\bibnamefont {{Burrows}}}, \bibinfo {author} {\bibfnamefont {D.}~\bibnamefont {{Vartanyan}}},\ and\ \bibinfo {author} {\bibfnamefont {T.}~\bibnamefont {{Wang}}},\ }\bibfield  {title} {\bibinfo {title} {{Black Hole Formation Accompanied by the Supernova Explosion of a 40 M $_{{\ensuremath{\odot}}}$ Progenitor Star}},\ }\href {https://doi.org/10.3847/1538-4357/acfc1c} {\bibfield  {journal} {\bibinfo  {journal} {\apj}\ }\textbf {\bibinfo {volume} {957}},\ \bibinfo {eid} {68} (\bibinfo {year} {2023})},\ \Eprint {https://arxiv.org/abs/2308.05798} {arXiv:2308.05798 [astro-ph.SR]} \BibitemShut {NoStop}%
\bibitem [{\citenamefont {{M{\"u}ller}}\ \emph {et~al.}(2012)\citenamefont {{M{\"u}ller}}, \citenamefont {{Janka}},\ and\ \citenamefont {{Marek}}}]{mueller:12a}%
  \BibitemOpen
  \bibfield  {author} {\bibinfo {author} {\bibfnamefont {B.}~\bibnamefont {{M{\"u}ller}}}, \bibinfo {author} {\bibfnamefont {H.-T.}\ \bibnamefont {{Janka}}},\ and\ \bibinfo {author} {\bibfnamefont {A.}~\bibnamefont {{Marek}}},\ }\bibfield  {title} {\bibinfo {title} {{A New Multi-dimensional General Relativistic Neutrino Hydrodynamics Code for Core-collapse Supernovae. II. Relativistic Explosion Models of Core-collapse Supernovae}},\ }\href {https://doi.org/10.1088/0004-637X/756/1/84} {\bibfield  {journal} {\bibinfo  {journal} {\apj}\ }\textbf {\bibinfo {volume} {756}},\ \bibinfo {eid} {84} (\bibinfo {year} {2012})},\ \Eprint {https://arxiv.org/abs/1202.0815} {arXiv:1202.0815 [astro-ph.SR]} \BibitemShut {NoStop}%
\bibitem [{\citenamefont {{Herant}}\ \emph {et~al.}(1994)\citenamefont {{Herant}}, \citenamefont {{Benz}}, \citenamefont {{Hix}}, \citenamefont {{Fryer}},\ and\ \citenamefont {{Colgate}}}]{herant94inside}%
  \BibitemOpen
  \bibfield  {author} {\bibinfo {author} {\bibfnamefont {M.}~\bibnamefont {{Herant}}}, \bibinfo {author} {\bibfnamefont {W.}~\bibnamefont {{Benz}}}, \bibinfo {author} {\bibfnamefont {W.~R.}\ \bibnamefont {{Hix}}}, \bibinfo {author} {\bibfnamefont {C.~L.}\ \bibnamefont {{Fryer}}},\ and\ \bibinfo {author} {\bibfnamefont {S.~A.}\ \bibnamefont {{Colgate}}},\ }\bibfield  {title} {\bibinfo {title} {{Inside the Supernova: A Powerful Convective Engine}},\ }\href {https://doi.org/10.1086/174817} {\bibfield  {journal} {\bibinfo  {journal} {\apj}\ }\textbf {\bibinfo {volume} {435}},\ \bibinfo {pages} {339} (\bibinfo {year} {1994})},\ \Eprint {https://arxiv.org/abs/astro-ph/9404024} {arXiv:astro-ph/9404024 [astro-ph]} \BibitemShut {NoStop}%
\bibitem [{\citenamefont {{Burrows}}\ \emph {et~al.}(1995)\citenamefont {{Burrows}}, \citenamefont {{Hayes}},\ and\ \citenamefont {{Fryxell}}}]{burrows95on}%
  \BibitemOpen
  \bibfield  {author} {\bibinfo {author} {\bibfnamefont {A.}~\bibnamefont {{Burrows}}}, \bibinfo {author} {\bibfnamefont {J.}~\bibnamefont {{Hayes}}},\ and\ \bibinfo {author} {\bibfnamefont {B.~A.}\ \bibnamefont {{Fryxell}}},\ }\bibfield  {title} {\bibinfo {title} {{On the Nature of Core-Collapse Supernova Explosions}},\ }\href {https://doi.org/10.1086/176188} {\bibfield  {journal} {\bibinfo  {journal} {\apj}\ }\textbf {\bibinfo {volume} {450}},\ \bibinfo {pages} {830} (\bibinfo {year} {1995})},\ \Eprint {https://arxiv.org/abs/astro-ph/9506061} {arXiv:astro-ph/9506061 [astro-ph]} \BibitemShut {NoStop}%
\bibitem [{\citenamefont {{Janka}}\ and\ \citenamefont {{Mueller}}(1995)}]{janka95first}%
  \BibitemOpen
  \bibfield  {author} {\bibinfo {author} {\bibfnamefont {H.~T.}\ \bibnamefont {{Janka}}}\ and\ \bibinfo {author} {\bibfnamefont {E.}~\bibnamefont {{Mueller}}},\ }\bibfield  {title} {\bibinfo {title} {{The First Second of a Type II Supernova: Convection, Accretion, and Shock Propagation}},\ }\href {https://doi.org/10.1086/309604} {\bibfield  {journal} {\bibinfo  {journal} {Astrophys. J.}\ }\textbf {\bibinfo {volume} {448}},\ \bibinfo {pages} {L109} (\bibinfo {year} {1995})},\ \Eprint {https://arxiv.org/abs/astro-ph/9503015} {arXiv:astro-ph/9503015 [astro-ph]} \BibitemShut {NoStop}%
\bibitem [{\citenamefont {{Blondin}}\ \emph {et~al.}(2003)\citenamefont {{Blondin}}, \citenamefont {{Mezzacappa}},\ and\ \citenamefont {{DeMarino}}}]{blondin03stability}%
  \BibitemOpen
  \bibfield  {author} {\bibinfo {author} {\bibfnamefont {J.~M.}\ \bibnamefont {{Blondin}}}, \bibinfo {author} {\bibfnamefont {A.}~\bibnamefont {{Mezzacappa}}},\ and\ \bibinfo {author} {\bibfnamefont {C.}~\bibnamefont {{DeMarino}}},\ }\bibfield  {title} {\bibinfo {title} {{Stability of Standing Accretion Shocks, with an Eye toward Core-Collapse Supernovae}},\ }\href {https://doi.org/10.1086/345812} {\bibfield  {journal} {\bibinfo  {journal} {\apj}\ }\textbf {\bibinfo {volume} {584}},\ \bibinfo {pages} {971} (\bibinfo {year} {2003})},\ \Eprint {https://arxiv.org/abs/astro-ph/0210634} {arXiv:astro-ph/0210634 [astro-ph]} \BibitemShut {NoStop}%
\bibitem [{\citenamefont {{Foglizzo}}\ \emph {et~al.}(2006)\citenamefont {{Foglizzo}}, \citenamefont {{Scheck}},\ and\ \citenamefont {{Janka}}}]{foglizzo06neutrino}%
  \BibitemOpen
  \bibfield  {author} {\bibinfo {author} {\bibfnamefont {T.}~\bibnamefont {{Foglizzo}}}, \bibinfo {author} {\bibfnamefont {L.}~\bibnamefont {{Scheck}}},\ and\ \bibinfo {author} {\bibfnamefont {H.~T.}\ \bibnamefont {{Janka}}},\ }\bibfield  {title} {\bibinfo {title} {{Neutrino-driven Convection versus Advection in Core-Collapse Supernovae}},\ }\href {https://doi.org/10.1086/508443} {\bibfield  {journal} {\bibinfo  {journal} {\apj}\ }\textbf {\bibinfo {volume} {652}},\ \bibinfo {pages} {1436} (\bibinfo {year} {2006})},\ \Eprint {https://arxiv.org/abs/astro-ph/0507636} {arXiv:astro-ph/0507636 [astro-ph]} \BibitemShut {NoStop}%
\bibitem [{\citenamefont {{Couch}}\ \emph {et~al.}(2015)\citenamefont {{Couch}}, \citenamefont {{Chatzopoulos}}, \citenamefont {{Arnett}},\ and\ \citenamefont {{Timmes}}}]{Couch15Three}%
  \BibitemOpen
  \bibfield  {author} {\bibinfo {author} {\bibfnamefont {S.~M.}\ \bibnamefont {{Couch}}}, \bibinfo {author} {\bibfnamefont {E.}~\bibnamefont {{Chatzopoulos}}}, \bibinfo {author} {\bibfnamefont {W.~D.}\ \bibnamefont {{Arnett}}},\ and\ \bibinfo {author} {\bibfnamefont {F.~X.}\ \bibnamefont {{Timmes}}},\ }\bibfield  {title} {\bibinfo {title} {{The Three-dimensional Evolution to Core Collapse of a Massive Star}},\ }\href {https://doi.org/10.1088/2041-8205/808/1/L21} {\bibfield  {journal} {\bibinfo  {journal} {\apjl}\ }\textbf {\bibinfo {volume} {808}},\ \bibinfo {eid} {L21} (\bibinfo {year} {2015})},\ \Eprint {https://arxiv.org/abs/1503.02199} {arXiv:1503.02199 [astro-ph.HE]} \BibitemShut {NoStop}%
\bibitem [{\citenamefont {{M{\"u}ller}}\ \emph {et~al.}(2017)\citenamefont {{M{\"u}ller}}, \citenamefont {{Melson}}, \citenamefont {{Heger}},\ and\ \citenamefont {{Janka}}}]{Mueller17Supernova}%
  \BibitemOpen
  \bibfield  {author} {\bibinfo {author} {\bibfnamefont {B.}~\bibnamefont {{M{\"u}ller}}}, \bibinfo {author} {\bibfnamefont {T.}~\bibnamefont {{Melson}}}, \bibinfo {author} {\bibfnamefont {A.}~\bibnamefont {{Heger}}},\ and\ \bibinfo {author} {\bibfnamefont {H.-T.}\ \bibnamefont {{Janka}}},\ }\bibfield  {title} {\bibinfo {title} {{Supernova simulations from a 3D progenitor model - Impact of perturbations and evolution of explosion properties}},\ }\href {https://doi.org/10.1093/mnras/stx1962} {\bibfield  {journal} {\bibinfo  {journal} {\mnras}\ }\textbf {\bibinfo {volume} {472}},\ \bibinfo {pages} {491} (\bibinfo {year} {2017})},\ \Eprint {https://arxiv.org/abs/1705.00620} {arXiv:1705.00620 [astro-ph.SR]} \BibitemShut {NoStop}%
\bibitem [{\citenamefont {{Kazeroni}}\ and\ \citenamefont {{Abdikamalov}}(2020)}]{Kazeroni20impact}%
  \BibitemOpen
  \bibfield  {author} {\bibinfo {author} {\bibfnamefont {R.}~\bibnamefont {{Kazeroni}}}\ and\ \bibinfo {author} {\bibfnamefont {E.}~\bibnamefont {{Abdikamalov}}},\ }\bibfield  {title} {\bibinfo {title} {{The impact of progenitor asymmetries on the neutrino-driven convection in core-collapse supernovae}},\ }\href {https://doi.org/10.1093/mnras/staa944} {\bibfield  {journal} {\bibinfo  {journal} {\mnras}\ }\textbf {\bibinfo {volume} {494}},\ \bibinfo {pages} {5360} (\bibinfo {year} {2020})},\ \Eprint {https://arxiv.org/abs/1911.08819} {arXiv:1911.08819 [astro-ph.SR]} \BibitemShut {NoStop}%
\bibitem [{\citenamefont {{Vartanyan}}\ \emph {et~al.}(2022)\citenamefont {{Vartanyan}}, \citenamefont {{Coleman}},\ and\ \citenamefont {{Burrows}}}]{Vartanyan22collapse}%
  \BibitemOpen
  \bibfield  {author} {\bibinfo {author} {\bibfnamefont {D.}~\bibnamefont {{Vartanyan}}}, \bibinfo {author} {\bibfnamefont {M.~S.~B.}\ \bibnamefont {{Coleman}}},\ and\ \bibinfo {author} {\bibfnamefont {A.}~\bibnamefont {{Burrows}}},\ }\bibfield  {title} {\bibinfo {title} {{The collapse and three-dimensional explosion of three-dimensional massive-star supernova progenitor models}},\ }\href {https://doi.org/10.1093/mnras/stab3702} {\bibfield  {journal} {\bibinfo  {journal} {\mnras}\ }\textbf {\bibinfo {volume} {510}},\ \bibinfo {pages} {4689} (\bibinfo {year} {2022})},\ \Eprint {https://arxiv.org/abs/2109.10920} {arXiv:2109.10920 [astro-ph.SR]} \BibitemShut {NoStop}%
\bibitem [{\citenamefont {{Telman}}\ \emph {et~al.}(2024)\citenamefont {{Telman}}, \citenamefont {{Abdikamalov}},\ and\ \citenamefont {{Foglizzo}}}]{Telman24Convective}%
  \BibitemOpen
  \bibfield  {author} {\bibinfo {author} {\bibfnamefont {Y.}~\bibnamefont {{Telman}}}, \bibinfo {author} {\bibfnamefont {E.}~\bibnamefont {{Abdikamalov}}},\ and\ \bibinfo {author} {\bibfnamefont {T.}~\bibnamefont {{Foglizzo}}},\ }\bibfield  {title} {\bibinfo {title} {{Convective vortices in collapsing stars}},\ }\href {https://doi.org/10.1093/mnras/stae2448} {\bibfield  {journal} {\bibinfo  {journal} {\mnras}\ }\textbf {\bibinfo {volume} {535}},\ \bibinfo {pages} {1388} (\bibinfo {year} {2024})},\ \Eprint {https://arxiv.org/abs/2409.17737} {arXiv:2409.17737 [astro-ph.SR]} \BibitemShut {NoStop}%
\bibitem [{\citenamefont {{Burrows}}\ \emph {et~al.}(2007)\citenamefont {{Burrows}}, \citenamefont {{Dessart}}, \citenamefont {{Livne}}, \citenamefont {{Ott}},\ and\ \citenamefont {{Murphy}}}]{burrows:07b}%
  \BibitemOpen
  \bibfield  {author} {\bibinfo {author} {\bibfnamefont {A.}~\bibnamefont {{Burrows}}}, \bibinfo {author} {\bibfnamefont {L.}~\bibnamefont {{Dessart}}}, \bibinfo {author} {\bibfnamefont {E.}~\bibnamefont {{Livne}}}, \bibinfo {author} {\bibfnamefont {C.~D.}\ \bibnamefont {{Ott}}},\ and\ \bibinfo {author} {\bibfnamefont {J.}~\bibnamefont {{Murphy}}},\ }\bibfield  {title} {\bibinfo {title} {{Simulations of Magnetically Driven Supernova and Hypernova Explosions in the Context of Rapid Rotation}},\ }\href {https://doi.org/10.1086/519161} {\bibfield  {journal} {\bibinfo  {journal} {\apj}\ }\textbf {\bibinfo {volume} {664}},\ \bibinfo {pages} {416} (\bibinfo {year} {2007})},\ \Eprint {https://arxiv.org/abs/astro-ph/0702539} {arXiv:astro-ph/0702539} \BibitemShut {NoStop}%
\bibitem [{\citenamefont {{M{\"o}sta}}\ \emph {et~al.}(2014)\citenamefont {{M{\"o}sta}}, \citenamefont {{Richers}}, \citenamefont {{Ott}}, \citenamefont {{Haas}}, \citenamefont {{Piro}}, \citenamefont {{Boydstun}}, \citenamefont {{Abdikamalov}}, \citenamefont {{Reisswig}},\ and\ \citenamefont {{Schnetter}}}]{moesta:14b}%
  \BibitemOpen
  \bibfield  {author} {\bibinfo {author} {\bibfnamefont {P.}~\bibnamefont {{M{\"o}sta}}}, \bibinfo {author} {\bibfnamefont {S.}~\bibnamefont {{Richers}}}, \bibinfo {author} {\bibfnamefont {C.~D.}\ \bibnamefont {{Ott}}}, \bibinfo {author} {\bibfnamefont {R.}~\bibnamefont {{Haas}}}, \bibinfo {author} {\bibfnamefont {A.~L.}\ \bibnamefont {{Piro}}}, \bibinfo {author} {\bibfnamefont {K.}~\bibnamefont {{Boydstun}}}, \bibinfo {author} {\bibfnamefont {E.}~\bibnamefont {{Abdikamalov}}}, \bibinfo {author} {\bibfnamefont {C.}~\bibnamefont {{Reisswig}}},\ and\ \bibinfo {author} {\bibfnamefont {E.}~\bibnamefont {{Schnetter}}},\ }\bibfield  {title} {\bibinfo {title} {{Magnetorotational Core-collapse Supernovae in Three Dimensions}},\ }\href {https://doi.org/10.1088/2041-8205/785/2/L29} {\bibfield  {journal} {\bibinfo  {journal} {ApJL}\ }\textbf {\bibinfo {volume} {785}},\ \bibinfo {eid} {L29} (\bibinfo {year} {2014})},\ \Eprint {https://arxiv.org/abs/1403.1230} {arXiv:1403.1230 [astro-ph.HE]} \BibitemShut {NoStop}%
\bibitem [{\citenamefont {{Obergaulinger}}\ and\ \citenamefont {{Aloy}}(2020)}]{obergaulinger:20}%
  \BibitemOpen
  \bibfield  {author} {\bibinfo {author} {\bibfnamefont {M.}~\bibnamefont {{Obergaulinger}}}\ and\ \bibinfo {author} {\bibfnamefont {M.~{\'A}.}\ \bibnamefont {{Aloy}}},\ }\bibfield  {title} {\bibinfo {title} {{Magnetorotational core collapse of possible GRB progenitors - I. Explosion mechanisms}},\ }\href {https://doi.org/10.1093/mnras/staa096} {\bibfield  {journal} {\bibinfo  {journal} {MNRAS}\ }\textbf {\bibinfo {volume} {492}},\ \bibinfo {pages} {4613} (\bibinfo {year} {2020})},\ \Eprint {https://arxiv.org/abs/1909.01105} {arXiv:1909.01105 [astro-ph.HE]} \BibitemShut {NoStop}%
\bibitem [{\citenamefont {{Kuroda}}\ \emph {et~al.}(2020)\citenamefont {{Kuroda}}, \citenamefont {{Arcones}}, \citenamefont {{Takiwaki}},\ and\ \citenamefont {{Kotake}}}]{kuroda:20}%
  \BibitemOpen
  \bibfield  {author} {\bibinfo {author} {\bibfnamefont {T.}~\bibnamefont {{Kuroda}}}, \bibinfo {author} {\bibfnamefont {A.}~\bibnamefont {{Arcones}}}, \bibinfo {author} {\bibfnamefont {T.}~\bibnamefont {{Takiwaki}}},\ and\ \bibinfo {author} {\bibfnamefont {K.}~\bibnamefont {{Kotake}}},\ }\bibfield  {title} {\bibinfo {title} {{Magnetorotational Explosion of a Massive Star Supported by Neutrino Heating in General Relativistic Three-dimensional Simulations}},\ }\href {https://doi.org/10.3847/1538-4357/ab9308} {\bibfield  {journal} {\bibinfo  {journal} {\apj}\ }\textbf {\bibinfo {volume} {896}},\ \bibinfo {eid} {102} (\bibinfo {year} {2020})},\ \Eprint {https://arxiv.org/abs/2003.02004} {arXiv:2003.02004 [astro-ph.HE]} \BibitemShut {NoStop}%
\bibitem [{\citenamefont {{Metzger}}\ \emph {et~al.}(2011)\citenamefont {{Metzger}}, \citenamefont {{Giannios}}, \citenamefont {{Thompson}}, \citenamefont {{Bucciantini}},\ and\ \citenamefont {{Quataert}}}]{metzger11protomagnetar}%
  \BibitemOpen
  \bibfield  {author} {\bibinfo {author} {\bibfnamefont {B.~D.}\ \bibnamefont {{Metzger}}}, \bibinfo {author} {\bibfnamefont {D.}~\bibnamefont {{Giannios}}}, \bibinfo {author} {\bibfnamefont {T.~A.}\ \bibnamefont {{Thompson}}}, \bibinfo {author} {\bibfnamefont {N.}~\bibnamefont {{Bucciantini}}},\ and\ \bibinfo {author} {\bibfnamefont {E.}~\bibnamefont {{Quataert}}},\ }\bibfield  {title} {\bibinfo {title} {{The protomagnetar model for gamma-ray bursts}},\ }\href {https://doi.org/10.1111/j.1365-2966.2011.18280.x} {\bibfield  {journal} {\bibinfo  {journal} {MNRAS}\ }\textbf {\bibinfo {volume} {413}},\ \bibinfo {pages} {2031} (\bibinfo {year} {2011})},\ \Eprint {https://arxiv.org/abs/1012.0001} {arXiv:1012.0001 [astro-ph.HE]} \BibitemShut {NoStop}%
\bibitem [{\citenamefont {{Pais}}\ \emph {et~al.}(2023)\citenamefont {{Pais}}, \citenamefont {{Piran}},\ and\ \citenamefont {{Nakar}}}]{Pais23choked}%
  \BibitemOpen
  \bibfield  {author} {\bibinfo {author} {\bibfnamefont {M.}~\bibnamefont {{Pais}}}, \bibinfo {author} {\bibfnamefont {T.}~\bibnamefont {{Piran}}},\ and\ \bibinfo {author} {\bibfnamefont {E.}~\bibnamefont {{Nakar}}},\ }\bibfield  {title} {\bibinfo {title} {{The velocity distribution of outflows driven by choked jets in stellar envelopes}},\ }\href {https://doi.org/10.1093/mnras/stac3640} {\bibfield  {journal} {\bibinfo  {journal} {\mnras}\ }\textbf {\bibinfo {volume} {519}},\ \bibinfo {pages} {1941} (\bibinfo {year} {2023})},\ \Eprint {https://arxiv.org/abs/2208.14459} {arXiv:2208.14459 [astro-ph.HE]} \BibitemShut {NoStop}%
\bibitem [{\citenamefont {{Piran}}\ \emph {et~al.}(2019)\citenamefont {{Piran}}, \citenamefont {{Nakar}}, \citenamefont {{Mazzali}},\ and\ \citenamefont {{Pian}}}]{Piran19jet_ccsn}%
  \BibitemOpen
  \bibfield  {author} {\bibinfo {author} {\bibfnamefont {T.}~\bibnamefont {{Piran}}}, \bibinfo {author} {\bibfnamefont {E.}~\bibnamefont {{Nakar}}}, \bibinfo {author} {\bibfnamefont {P.}~\bibnamefont {{Mazzali}}},\ and\ \bibinfo {author} {\bibfnamefont {E.}~\bibnamefont {{Pian}}},\ }\bibfield  {title} {\bibinfo {title} {{Relativistic Jets in Core-collapse Supernovae}},\ }\href {https://doi.org/10.3847/2041-8213/aaffce} {\bibfield  {journal} {\bibinfo  {journal} {\apjl}\ }\textbf {\bibinfo {volume} {871}},\ \bibinfo {eid} {L25} (\bibinfo {year} {2019})}\BibitemShut {NoStop}%
\bibitem [{\citenamefont {{Heger}}\ \emph {et~al.}(2005)\citenamefont {{Heger}}, \citenamefont {{Woosley}},\ and\ \citenamefont {{Spruit}}}]{Heger05Presupernova}%
  \BibitemOpen
  \bibfield  {author} {\bibinfo {author} {\bibfnamefont {A.}~\bibnamefont {{Heger}}}, \bibinfo {author} {\bibfnamefont {S.~E.}\ \bibnamefont {{Woosley}}},\ and\ \bibinfo {author} {\bibfnamefont {H.~C.}\ \bibnamefont {{Spruit}}},\ }\bibfield  {title} {\bibinfo {title} {{Presupernova Evolution of Differentially Rotating Massive Stars Including Magnetic Fields}},\ }\href {https://doi.org/10.1086/429868} {\bibfield  {journal} {\bibinfo  {journal} {\apj}\ }\textbf {\bibinfo {volume} {626}},\ \bibinfo {pages} {350} (\bibinfo {year} {2005})},\ \Eprint {https://arxiv.org/abs/astro-ph/0409422} {arXiv:astro-ph/0409422 [astro-ph]} \BibitemShut {NoStop}%
\bibitem [{\citenamefont {{Takiwaki}}\ \emph {et~al.}(2016)\citenamefont {{Takiwaki}}, \citenamefont {{Kotake}},\ and\ \citenamefont {{Suwa}}}]{Takiwaki16Three}%
  \BibitemOpen
  \bibfield  {author} {\bibinfo {author} {\bibfnamefont {T.}~\bibnamefont {{Takiwaki}}}, \bibinfo {author} {\bibfnamefont {K.}~\bibnamefont {{Kotake}}},\ and\ \bibinfo {author} {\bibfnamefont {Y.}~\bibnamefont {{Suwa}}},\ }\bibfield  {title} {\bibinfo {title} {{Three-dimensional simulations of rapidly rotating core-collapse supernovae: finding a neutrino-powered explosion aided by non-axisymmetric flows}},\ }\href {https://doi.org/10.1093/mnrasl/slw105} {\bibfield  {journal} {\bibinfo  {journal} {\mnras}\ }\textbf {\bibinfo {volume} {461}},\ \bibinfo {pages} {L112} (\bibinfo {year} {2016})},\ \Eprint {https://arxiv.org/abs/1602.06759} {arXiv:1602.06759 [astro-ph.HE]} \BibitemShut {NoStop}%
\bibitem [{\citenamefont {{Summa}}\ \emph {et~al.}(2018)\citenamefont {{Summa}}, \citenamefont {{Janka}}, \citenamefont {{Melson}},\ and\ \citenamefont {{Marek}}}]{Summa18Rotation}%
  \BibitemOpen
  \bibfield  {author} {\bibinfo {author} {\bibfnamefont {A.}~\bibnamefont {{Summa}}}, \bibinfo {author} {\bibfnamefont {H.-T.}\ \bibnamefont {{Janka}}}, \bibinfo {author} {\bibfnamefont {T.}~\bibnamefont {{Melson}}},\ and\ \bibinfo {author} {\bibfnamefont {A.}~\bibnamefont {{Marek}}},\ }\bibfield  {title} {\bibinfo {title} {{Rotation-supported Neutrino-driven Supernova Explosions in Three Dimensions and the Critical Luminosity Condition}},\ }\href {https://doi.org/10.3847/1538-4357/aa9ce8} {\bibfield  {journal} {\bibinfo  {journal} {\apj}\ }\textbf {\bibinfo {volume} {852}},\ \bibinfo {eid} {28} (\bibinfo {year} {2018})},\ \Eprint {https://arxiv.org/abs/1708.04154} {arXiv:1708.04154 [astro-ph.HE]} \BibitemShut {NoStop}%
\bibitem [{\citenamefont {{Abdikamalov}}\ \emph {et~al.}(2021)\citenamefont {{Abdikamalov}}, \citenamefont {{Foglizzo}},\ and\ \citenamefont {{Mukazhanov}}}]{Abdikamalov21Impact}%
  \BibitemOpen
  \bibfield  {author} {\bibinfo {author} {\bibfnamefont {E.}~\bibnamefont {{Abdikamalov}}}, \bibinfo {author} {\bibfnamefont {T.}~\bibnamefont {{Foglizzo}}},\ and\ \bibinfo {author} {\bibfnamefont {O.}~\bibnamefont {{Mukazhanov}}},\ }\bibfield  {title} {\bibinfo {title} {{Impact of rotation on the evolution of convective vortices in collapsing stars}},\ }\href {https://doi.org/10.1093/mnras/stab715} {\bibfield  {journal} {\bibinfo  {journal} {\mnras}\ }\textbf {\bibinfo {volume} {503}},\ \bibinfo {pages} {3617} (\bibinfo {year} {2021})},\ \Eprint {https://arxiv.org/abs/2012.06710} {arXiv:2012.06710 [astro-ph.SR]} \BibitemShut {NoStop}%
\bibitem [{\citenamefont {{Buellet}}\ \emph {et~al.}(2023)\citenamefont {{Buellet}}, \citenamefont {{Foglizzo}}, \citenamefont {{Guilet}},\ and\ \citenamefont {{Abdikamalov}}}]{Buellet23Effect}%
  \BibitemOpen
  \bibfield  {author} {\bibinfo {author} {\bibfnamefont {A.~C.}\ \bibnamefont {{Buellet}}}, \bibinfo {author} {\bibfnamefont {T.}~\bibnamefont {{Foglizzo}}}, \bibinfo {author} {\bibfnamefont {J.}~\bibnamefont {{Guilet}}},\ and\ \bibinfo {author} {\bibfnamefont {E.}~\bibnamefont {{Abdikamalov}}},\ }\bibfield  {title} {\bibinfo {title} {{Effect of stellar rotation on the development of post-shock instabilities during core-collapse supernovae}},\ }\href {https://doi.org/10.1051/0004-6361/202245799} {\bibfield  {journal} {\bibinfo  {journal} {\aap}\ }\textbf {\bibinfo {volume} {674}},\ \bibinfo {eid} {A205} (\bibinfo {year} {2023})},\ \Eprint {https://arxiv.org/abs/2301.01962} {arXiv:2301.01962 [astro-ph.HE]} \BibitemShut {NoStop}%
\bibitem [{\citenamefont {{Murphy}}\ \emph {et~al.}(2009)\citenamefont {{Murphy}}, \citenamefont {{Ott}},\ and\ \citenamefont {{Burrows}}}]{Murphy09Model}%
  \BibitemOpen
  \bibfield  {author} {\bibinfo {author} {\bibfnamefont {J.~W.}\ \bibnamefont {{Murphy}}}, \bibinfo {author} {\bibfnamefont {C.~D.}\ \bibnamefont {{Ott}}},\ and\ \bibinfo {author} {\bibfnamefont {A.}~\bibnamefont {{Burrows}}},\ }\bibfield  {title} {\bibinfo {title} {{A Model for Gravitational Wave Emission from Neutrino-Driven Core-Collapse Supernovae}},\ }\href {https://doi.org/10.1088/0004-637X/707/2/1173} {\bibfield  {journal} {\bibinfo  {journal} {\apj}\ }\textbf {\bibinfo {volume} {707}},\ \bibinfo {pages} {1173} (\bibinfo {year} {2009})},\ \Eprint {https://arxiv.org/abs/0907.4762} {arXiv:0907.4762 [astro-ph.SR]} \BibitemShut {NoStop}%
\bibitem [{\citenamefont {{M{\"u}ller}}\ \emph {et~al.}(2013)\citenamefont {{M{\"u}ller}}, \citenamefont {{Janka}},\ and\ \citenamefont {{Marek}}}]{mueller:13}%
  \BibitemOpen
  \bibfield  {author} {\bibinfo {author} {\bibfnamefont {B.}~\bibnamefont {{M{\"u}ller}}}, \bibinfo {author} {\bibfnamefont {H.-T.}\ \bibnamefont {{Janka}}},\ and\ \bibinfo {author} {\bibfnamefont {A.}~\bibnamefont {{Marek}}},\ }\bibfield  {title} {\bibinfo {title} {{A New Multi-dimensional General Relativistic Neutrino Hydrodynamics Code of Core-collapse Supernovae. III. Gravitational Wave Signals from Supernova Explosion Models}},\ }\href {https://doi.org/10.1088/0004-637X/766/1/43} {\bibfield  {journal} {\bibinfo  {journal} {\apj}\ }\textbf {\bibinfo {volume} {766}},\ \bibinfo {eid} {43} (\bibinfo {year} {2013})},\ \Eprint {https://arxiv.org/abs/1210.6984} {arXiv:1210.6984 [astro-ph.SR]} \BibitemShut {NoStop}%
\bibitem [{\citenamefont {{Yakunin}}\ \emph {et~al.}(2015)\citenamefont {{Yakunin}}, \citenamefont {{Mezzacappa}}, \citenamefont {{Marronetti}}, \citenamefont {{Yoshida}}, \citenamefont {{Bruenn}}, \citenamefont {{Hix}}, \citenamefont {{Lentz}}, \citenamefont {{Bronson Messer}}, \citenamefont {{Harris}}, \citenamefont {{Endeve}}, \citenamefont {{Blondin}},\ and\ \citenamefont {{Lingerfelt}}}]{Yakunin15GW}%
  \BibitemOpen
  \bibfield  {author} {\bibinfo {author} {\bibfnamefont {K.~N.}\ \bibnamefont {{Yakunin}}}, \bibinfo {author} {\bibfnamefont {A.}~\bibnamefont {{Mezzacappa}}}, \bibinfo {author} {\bibfnamefont {P.}~\bibnamefont {{Marronetti}}}, \bibinfo {author} {\bibfnamefont {S.}~\bibnamefont {{Yoshida}}}, \bibinfo {author} {\bibfnamefont {S.~W.}\ \bibnamefont {{Bruenn}}}, \bibinfo {author} {\bibfnamefont {W.~R.}\ \bibnamefont {{Hix}}}, \bibinfo {author} {\bibfnamefont {E.~J.}\ \bibnamefont {{Lentz}}}, \bibinfo {author} {\bibfnamefont {O.~E.}\ \bibnamefont {{Bronson Messer}}}, \bibinfo {author} {\bibfnamefont {J.~A.}\ \bibnamefont {{Harris}}}, \bibinfo {author} {\bibfnamefont {E.}~\bibnamefont {{Endeve}}}, \bibinfo {author} {\bibfnamefont {J.~M.}\ \bibnamefont {{Blondin}}},\ and\ \bibinfo {author} {\bibfnamefont {E.~J.}\ \bibnamefont {{Lingerfelt}}},\ }\bibfield  {title} {\bibinfo {title} {{Gravitational wave signatures of ab initio two-dimensional core collapse supernova explosion models for 12 -25
  M$_{{\ensuremath{\odot}}}$ stars}},\ }\href {https://doi.org/10.1103/PhysRevD.92.084040} {\bibfield  {journal} {\bibinfo  {journal} {\prd}\ }\textbf {\bibinfo {volume} {92}},\ \bibinfo {eid} {084040} (\bibinfo {year} {2015})},\ \Eprint {https://arxiv.org/abs/1505.05824} {arXiv:1505.05824 [astro-ph.HE]} \BibitemShut {NoStop}%
\bibitem [{\citenamefont {{Radice}}\ \emph {et~al.}(2019)\citenamefont {{Radice}}, \citenamefont {{Morozova}}, \citenamefont {{Burrows}}, \citenamefont {{Vartanyan}},\ and\ \citenamefont {{Nagakura}}}]{radice:19gw}%
  \BibitemOpen
  \bibfield  {author} {\bibinfo {author} {\bibfnamefont {D.}~\bibnamefont {{Radice}}}, \bibinfo {author} {\bibfnamefont {V.}~\bibnamefont {{Morozova}}}, \bibinfo {author} {\bibfnamefont {A.}~\bibnamefont {{Burrows}}}, \bibinfo {author} {\bibfnamefont {D.}~\bibnamefont {{Vartanyan}}},\ and\ \bibinfo {author} {\bibfnamefont {H.}~\bibnamefont {{Nagakura}}},\ }\bibfield  {title} {\bibinfo {title} {{Characterizing the Gravitational Wave Signal from Core-collapse Supernovae}},\ }\href {https://doi.org/10.3847/2041-8213/ab191a} {\bibfield  {journal} {\bibinfo  {journal} {ApJL}\ }\textbf {\bibinfo {volume} {876}},\ \bibinfo {eid} {L9} (\bibinfo {year} {2019})},\ \Eprint {https://arxiv.org/abs/1812.07703} {arXiv:1812.07703 [astro-ph.HE]} \BibitemShut {NoStop}%
\bibitem [{\citenamefont {{Andresen}}\ \emph {et~al.}(2017)\citenamefont {{Andresen}}, \citenamefont {{M{\"u}ller}}, \citenamefont {{M{\"u}ller}},\ and\ \citenamefont {{Janka}}}]{Andresen17Gravitational}%
  \BibitemOpen
  \bibfield  {author} {\bibinfo {author} {\bibfnamefont {H.}~\bibnamefont {{Andresen}}}, \bibinfo {author} {\bibfnamefont {B.}~\bibnamefont {{M{\"u}ller}}}, \bibinfo {author} {\bibfnamefont {E.}~\bibnamefont {{M{\"u}ller}}},\ and\ \bibinfo {author} {\bibfnamefont {H.~T.}\ \bibnamefont {{Janka}}},\ }\bibfield  {title} {\bibinfo {title} {{Gravitational wave signals from 3D neutrino hydrodynamics simulations of core-collapse supernovae}},\ }\href {https://doi.org/10.1093/mnras/stx618} {\bibfield  {journal} {\bibinfo  {journal} {MNRAS}\ }\textbf {\bibinfo {volume} {468}},\ \bibinfo {pages} {2032} (\bibinfo {year} {2017})},\ \Eprint {https://arxiv.org/abs/1607.05199} {arXiv:1607.05199 [astro-ph.HE]} \BibitemShut {NoStop}%
\bibitem [{\citenamefont {{Mezzacappa}}\ \emph {et~al.}(2023)\citenamefont {{Mezzacappa}}, \citenamefont {{Marronetti}}, \citenamefont {{Landfield}}, \citenamefont {{Lentz}}, \citenamefont {{Murphy}}, \citenamefont {{Raphael Hix}}, \citenamefont {{Harris}}, \citenamefont {{Bruenn}}, \citenamefont {{Blondin}}, \citenamefont {{Bronson Messer}}, \citenamefont {{Casanova}},\ and\ \citenamefont {{Kronzer}}}]{Mezzacappa23Core}%
  \BibitemOpen
  \bibfield  {author} {\bibinfo {author} {\bibfnamefont {A.}~\bibnamefont {{Mezzacappa}}}, \bibinfo {author} {\bibfnamefont {P.}~\bibnamefont {{Marronetti}}}, \bibinfo {author} {\bibfnamefont {R.~E.}\ \bibnamefont {{Landfield}}}, \bibinfo {author} {\bibfnamefont {E.~J.}\ \bibnamefont {{Lentz}}}, \bibinfo {author} {\bibfnamefont {R.~D.}\ \bibnamefont {{Murphy}}}, \bibinfo {author} {\bibfnamefont {W.}~\bibnamefont {{Raphael Hix}}}, \bibinfo {author} {\bibfnamefont {J.~A.}\ \bibnamefont {{Harris}}}, \bibinfo {author} {\bibfnamefont {S.~W.}\ \bibnamefont {{Bruenn}}}, \bibinfo {author} {\bibfnamefont {J.~M.}\ \bibnamefont {{Blondin}}}, \bibinfo {author} {\bibfnamefont {O.~E.}\ \bibnamefont {{Bronson Messer}}}, \bibinfo {author} {\bibfnamefont {J.}~\bibnamefont {{Casanova}}},\ and\ \bibinfo {author} {\bibfnamefont {L.~L.}\ \bibnamefont {{Kronzer}}},\ }\bibfield  {title} {\bibinfo {title} {{Core collapse supernova gravitational wave emission for progenitors of 9.6, 15, and 25M{\ensuremath{\odot}}}},\ }\href
  {https://doi.org/10.1103/PhysRevD.107.043008} {\bibfield  {journal} {\bibinfo  {journal} {\prd}\ }\textbf {\bibinfo {volume} {107}},\ \bibinfo {eid} {043008} (\bibinfo {year} {2023})},\ \Eprint {https://arxiv.org/abs/2208.10643} {arXiv:2208.10643 [astro-ph.SR]} \BibitemShut {NoStop}%
\bibitem [{\citenamefont {{Vartanyan}}\ \emph {et~al.}(2023)\citenamefont {{Vartanyan}}, \citenamefont {{Burrows}}, \citenamefont {{Wang}}, \citenamefont {{Coleman}},\ and\ \citenamefont {{White}}}]{Vartanyan23Gravitational}%
  \BibitemOpen
  \bibfield  {author} {\bibinfo {author} {\bibfnamefont {D.}~\bibnamefont {{Vartanyan}}}, \bibinfo {author} {\bibfnamefont {A.}~\bibnamefont {{Burrows}}}, \bibinfo {author} {\bibfnamefont {T.}~\bibnamefont {{Wang}}}, \bibinfo {author} {\bibfnamefont {M.~S.~B.}\ \bibnamefont {{Coleman}}},\ and\ \bibinfo {author} {\bibfnamefont {C.~J.}\ \bibnamefont {{White}}},\ }\bibfield  {title} {\bibinfo {title} {{Gravitational-wave signature of core-collapse supernovae}},\ }\href {https://doi.org/10.1103/PhysRevD.107.103015} {\bibfield  {journal} {\bibinfo  {journal} {\prd}\ }\textbf {\bibinfo {volume} {107}},\ \bibinfo {eid} {103015} (\bibinfo {year} {2023})},\ \Eprint {https://arxiv.org/abs/2302.07092} {arXiv:2302.07092 [astro-ph.HE]} \BibitemShut {NoStop}%
\bibitem [{\citenamefont {{Dimmelmeier}}\ \emph {et~al.}(2007)\citenamefont {{Dimmelmeier}}, \citenamefont {{Ott}}, \citenamefont {{Janka}}, \citenamefont {{Marek}},\ and\ \citenamefont {{M{\"u}ller}}}]{Dimmelmeier07}%
  \BibitemOpen
  \bibfield  {author} {\bibinfo {author} {\bibfnamefont {H.}~\bibnamefont {{Dimmelmeier}}}, \bibinfo {author} {\bibfnamefont {C.~D.}\ \bibnamefont {{Ott}}}, \bibinfo {author} {\bibfnamefont {H.~T.}\ \bibnamefont {{Janka}}}, \bibinfo {author} {\bibfnamefont {A.}~\bibnamefont {{Marek}}},\ and\ \bibinfo {author} {\bibfnamefont {E.}~\bibnamefont {{M{\"u}ller}}},\ }\bibfield  {title} {\bibinfo {title} {{Generic Gravitational-Wave Signals from the Collapse of Rotating Stellar Cores}},\ }\href {https://doi.org/10.1103/PhysRevLett.98.251101} {\bibfield  {journal} {\bibinfo  {journal} {\prl}\ }\textbf {\bibinfo {volume} {98}},\ \bibinfo {eid} {251101} (\bibinfo {year} {2007})},\ \Eprint {https://arxiv.org/abs/astro-ph/0702305} {arXiv:astro-ph/0702305 [astro-ph]} \BibitemShut {NoStop}%
\bibitem [{\citenamefont {{Fuller}}\ \emph {et~al.}(2015)\citenamefont {{Fuller}}, \citenamefont {{Klion}}, \citenamefont {{Abdikamalov}},\ and\ \citenamefont {{Ott}}}]{Fuller15SNseismology}%
  \BibitemOpen
  \bibfield  {author} {\bibinfo {author} {\bibfnamefont {J.}~\bibnamefont {{Fuller}}}, \bibinfo {author} {\bibfnamefont {H.}~\bibnamefont {{Klion}}}, \bibinfo {author} {\bibfnamefont {E.}~\bibnamefont {{Abdikamalov}}},\ and\ \bibinfo {author} {\bibfnamefont {C.~D.}\ \bibnamefont {{Ott}}},\ }\bibfield  {title} {\bibinfo {title} {{Supernova seismology: gravitational wave signatures of rapidly rotating core collapse}},\ }\href {https://doi.org/10.1093/mnras/stv698} {\bibfield  {journal} {\bibinfo  {journal} {\mnras}\ }\textbf {\bibinfo {volume} {450}},\ \bibinfo {pages} {414} (\bibinfo {year} {2015})},\ \Eprint {https://arxiv.org/abs/1501.06951} {arXiv:1501.06951 [astro-ph.HE]} \BibitemShut {NoStop}%
\bibitem [{\citenamefont {{Scheidegger}}\ \emph {et~al.}(2008)\citenamefont {{Scheidegger}}, \citenamefont {{Fischer}}, \citenamefont {{Whitehouse}},\ and\ \citenamefont {{Liebend{\"o}rfer}}}]{Scheidegger08}%
  \BibitemOpen
  \bibfield  {author} {\bibinfo {author} {\bibfnamefont {S.}~\bibnamefont {{Scheidegger}}}, \bibinfo {author} {\bibfnamefont {T.}~\bibnamefont {{Fischer}}}, \bibinfo {author} {\bibfnamefont {S.~C.}\ \bibnamefont {{Whitehouse}}},\ and\ \bibinfo {author} {\bibfnamefont {M.}~\bibnamefont {{Liebend{\"o}rfer}}},\ }\bibfield  {title} {\bibinfo {title} {{Gravitational waves from 3D MHD core collapse simulations}},\ }\href {https://doi.org/10.1051/0004-6361:20078577} {\bibfield  {journal} {\bibinfo  {journal} {\aap}\ }\textbf {\bibinfo {volume} {490}},\ \bibinfo {pages} {231} (\bibinfo {year} {2008})},\ \Eprint {https://arxiv.org/abs/0709.0168} {arXiv:0709.0168 [astro-ph]} \BibitemShut {NoStop}%
\bibitem [{\citenamefont {{Shibagaki}}\ \emph {et~al.}(2020)\citenamefont {{Shibagaki}}, \citenamefont {{Kuroda}}, \citenamefont {{Kotake}},\ and\ \citenamefont {{Takiwaki}}}]{Shibagaki20new}%
  \BibitemOpen
  \bibfield  {author} {\bibinfo {author} {\bibfnamefont {S.}~\bibnamefont {{Shibagaki}}}, \bibinfo {author} {\bibfnamefont {T.}~\bibnamefont {{Kuroda}}}, \bibinfo {author} {\bibfnamefont {K.}~\bibnamefont {{Kotake}}},\ and\ \bibinfo {author} {\bibfnamefont {T.}~\bibnamefont {{Takiwaki}}},\ }\bibfield  {title} {\bibinfo {title} {{A new gravitational-wave signature of low-T/|W| instability in rapidly rotating stellar core collapse}},\ }\href {https://doi.org/10.1093/mnrasl/slaa021} {\bibfield  {journal} {\bibinfo  {journal} {\mnras}\ }\textbf {\bibinfo {volume} {493}},\ \bibinfo {pages} {L138} (\bibinfo {year} {2020})},\ \Eprint {https://arxiv.org/abs/1909.09730} {arXiv:1909.09730 [astro-ph.HE]} \BibitemShut {NoStop}%
\bibitem [{\citenamefont {{Mueller}}\ and\ \citenamefont {{Janka}}(1997)}]{mueller:97}%
  \BibitemOpen
  \bibfield  {author} {\bibinfo {author} {\bibfnamefont {E.}~\bibnamefont {{Mueller}}}\ and\ \bibinfo {author} {\bibfnamefont {H.~T.}\ \bibnamefont {{Janka}}},\ }\bibfield  {title} {\bibinfo {title} {{Gravitational radiation from convective instabilities in Type II supernova explosions.}},\ }\href@noop {} {\bibfield  {journal} {\bibinfo  {journal} {\aap}\ }\textbf {\bibinfo {volume} {317}},\ \bibinfo {pages} {140} (\bibinfo {year} {1997})}\BibitemShut {NoStop}%
\bibitem [{\citenamefont {{Choi}}\ \emph {et~al.}(2024)\citenamefont {{Choi}}, \citenamefont {{Burrows}},\ and\ \citenamefont {{Vartanyan}}}]{Choi24GW}%
  \BibitemOpen
  \bibfield  {author} {\bibinfo {author} {\bibfnamefont {L.}~\bibnamefont {{Choi}}}, \bibinfo {author} {\bibfnamefont {A.}~\bibnamefont {{Burrows}}},\ and\ \bibinfo {author} {\bibfnamefont {D.}~\bibnamefont {{Vartanyan}}},\ }\bibfield  {title} {\bibinfo {title} {{Gravitational-wave and Gravitational-wave Memory Signatures of Core-collapse Supernovae}},\ }\href {https://doi.org/10.3847/1538-4357/ad74f8} {\bibfield  {journal} {\bibinfo  {journal} {\apj}\ }\textbf {\bibinfo {volume} {975}},\ \bibinfo {eid} {12} (\bibinfo {year} {2024})},\ \Eprint {https://arxiv.org/abs/2408.01525} {arXiv:2408.01525 [astro-ph.HE]} \BibitemShut {NoStop}%
\bibitem [{\citenamefont {{Birnholtz}}\ and\ \citenamefont {{Piran}}(2013)}]{Birnholtz13GW_jet}%
  \BibitemOpen
  \bibfield  {author} {\bibinfo {author} {\bibfnamefont {O.}~\bibnamefont {{Birnholtz}}}\ and\ \bibinfo {author} {\bibfnamefont {T.}~\bibnamefont {{Piran}}},\ }\bibfield  {title} {\bibinfo {title} {{Gravitational wave memory from gamma ray bursts' jets}},\ }\href {https://doi.org/10.1103/PhysRevD.87.123007} {\bibfield  {journal} {\bibinfo  {journal} {\prd}\ }\textbf {\bibinfo {volume} {87}},\ \bibinfo {eid} {123007} (\bibinfo {year} {2013})},\ \Eprint {https://arxiv.org/abs/1302.5713} {arXiv:1302.5713 [astro-ph.HE]} \BibitemShut {NoStop}%
\bibitem [{\citenamefont {{Gottlieb}}\ \emph {et~al.}(2023)\citenamefont {{Gottlieb}}, \citenamefont {{Nagakura}}, \citenamefont {{Tchekhovskoy}}, \citenamefont {{Natarajan}}, \citenamefont {{Ramirez-Ruiz}}, \citenamefont {{Banagiri}}, \citenamefont {{Jacquemin-Ide}}, \citenamefont {{Kaaz}},\ and\ \citenamefont {{Kalogera}}}]{Gottlieb23Jetted}%
  \BibitemOpen
  \bibfield  {author} {\bibinfo {author} {\bibfnamefont {O.}~\bibnamefont {{Gottlieb}}}, \bibinfo {author} {\bibfnamefont {H.}~\bibnamefont {{Nagakura}}}, \bibinfo {author} {\bibfnamefont {A.}~\bibnamefont {{Tchekhovskoy}}}, \bibinfo {author} {\bibfnamefont {P.}~\bibnamefont {{Natarajan}}}, \bibinfo {author} {\bibfnamefont {E.}~\bibnamefont {{Ramirez-Ruiz}}}, \bibinfo {author} {\bibfnamefont {S.}~\bibnamefont {{Banagiri}}}, \bibinfo {author} {\bibfnamefont {J.}~\bibnamefont {{Jacquemin-Ide}}}, \bibinfo {author} {\bibfnamefont {N.}~\bibnamefont {{Kaaz}}},\ and\ \bibinfo {author} {\bibfnamefont {V.}~\bibnamefont {{Kalogera}}},\ }\bibfield  {title} {\bibinfo {title} {{Jetted and Turbulent Stellar Deaths: New LVK-detectable Gravitational-wave Sources}},\ }\href {https://doi.org/10.3847/2041-8213/ace03a} {\bibfield  {journal} {\bibinfo  {journal} {\apjl}\ }\textbf {\bibinfo {volume} {951}},\ \bibinfo {eid} {L30} (\bibinfo {year} {2023})},\ \Eprint {https://arxiv.org/abs/2209.09256} {arXiv:2209.09256 [astro-ph.HE]}
  \BibitemShut {NoStop}%
\bibitem [{\citenamefont {{Soker}}(2023)}]{Soker23GWJJ}%
  \BibitemOpen
  \bibfield  {author} {\bibinfo {author} {\bibfnamefont {N.}~\bibnamefont {{Soker}}},\ }\bibfield  {title} {\bibinfo {title} {{Predicting Gravitational Waves from Jittering-jets-driven Core Collapse Supernovae}},\ }\href {https://doi.org/10.1088/1674-4527/ad013e} {\bibfield  {journal} {\bibinfo  {journal} {Research in Astronomy and Astrophysics}\ }\textbf {\bibinfo {volume} {23}},\ \bibinfo {eid} {121001} (\bibinfo {year} {2023})},\ \Eprint {https://arxiv.org/abs/2308.04329} {arXiv:2308.04329 [astro-ph.HE]} \BibitemShut {NoStop}%
\bibitem [{\citenamefont {{Abdikamalov}}\ \emph {et~al.}(2009)\citenamefont {{Abdikamalov}}, \citenamefont {{Dimmelmeier}}, \citenamefont {{Rezzolla}},\ and\ \citenamefont {{Miller}}}]{abdikamalov:09}%
  \BibitemOpen
  \bibfield  {author} {\bibinfo {author} {\bibfnamefont {E.~B.}\ \bibnamefont {{Abdikamalov}}}, \bibinfo {author} {\bibfnamefont {H.}~\bibnamefont {{Dimmelmeier}}}, \bibinfo {author} {\bibfnamefont {L.}~\bibnamefont {{Rezzolla}}},\ and\ \bibinfo {author} {\bibfnamefont {J.~C.}\ \bibnamefont {{Miller}}},\ }\bibfield  {title} {\bibinfo {title} {{Relativistic simulations of the phase-transition-induced collapse of neutron stars}},\ }\href {https://doi.org/10.1111/j.1365-2966.2008.14056.x} {\bibfield  {journal} {\bibinfo  {journal} {\mnras}\ }\textbf {\bibinfo {volume} {392}},\ \bibinfo {pages} {52} (\bibinfo {year} {2009})},\ \Eprint {https://arxiv.org/abs/0806.1700} {arXiv:0806.1700 [astro-ph]} \BibitemShut {NoStop}%
\bibitem [{\citenamefont {{Zha}}\ \emph {et~al.}(2020)\citenamefont {{Zha}}, \citenamefont {{O'Connor}}, \citenamefont {{Chu}}, \citenamefont {{Lin}},\ and\ \citenamefont {{Couch}}}]{zha:20}%
  \BibitemOpen
  \bibfield  {author} {\bibinfo {author} {\bibfnamefont {S.}~\bibnamefont {{Zha}}}, \bibinfo {author} {\bibfnamefont {E.~P.}\ \bibnamefont {{O'Connor}}}, \bibinfo {author} {\bibfnamefont {M.-c.}\ \bibnamefont {{Chu}}}, \bibinfo {author} {\bibfnamefont {L.-M.}\ \bibnamefont {{Lin}}},\ and\ \bibinfo {author} {\bibfnamefont {S.~M.}\ \bibnamefont {{Couch}}},\ }\bibfield  {title} {\bibinfo {title} {{Gravitational-wave Signature of a First-order Quantum Chromodynamics Phase Transition in Core-Collapse Supernovae}},\ }\href {https://doi.org/10.1103/PhysRevLett.125.051102} {\bibfield  {journal} {\bibinfo  {journal} {\prl}\ }\textbf {\bibinfo {volume} {125}},\ \bibinfo {eid} {051102} (\bibinfo {year} {2020})},\ \Eprint {https://arxiv.org/abs/2007.04716} {arXiv:2007.04716 [astro-ph.HE]} \BibitemShut {NoStop}%
\bibitem [{\citenamefont {{Kuroda}}\ \emph {et~al.}(2022)\citenamefont {{Kuroda}}, \citenamefont {{Fischer}}, \citenamefont {{Takiwaki}},\ and\ \citenamefont {{Kotake}}}]{Kuroda22}%
  \BibitemOpen
  \bibfield  {author} {\bibinfo {author} {\bibfnamefont {T.}~\bibnamefont {{Kuroda}}}, \bibinfo {author} {\bibfnamefont {T.}~\bibnamefont {{Fischer}}}, \bibinfo {author} {\bibfnamefont {T.}~\bibnamefont {{Takiwaki}}},\ and\ \bibinfo {author} {\bibfnamefont {K.}~\bibnamefont {{Kotake}}},\ }\bibfield  {title} {\bibinfo {title} {{Core-collapse Supernova Simulations and the Formation of Neutron Stars, Hybrid Stars, and Black Holes}},\ }\href {https://doi.org/10.3847/1538-4357/ac31a8} {\bibfield  {journal} {\bibinfo  {journal} {\apj}\ }\textbf {\bibinfo {volume} {924}},\ \bibinfo {eid} {38} (\bibinfo {year} {2022})},\ \Eprint {https://arxiv.org/abs/2109.01508} {arXiv:2109.01508 [astro-ph.HE]} \BibitemShut {NoStop}%
\bibitem [{\citenamefont {{Mitra}}\ \emph {et~al.}(2023)\citenamefont {{Mitra}}, \citenamefont {{Shukirgaliyev}}, \citenamefont {{Abylkairov}},\ and\ \citenamefont {{Abdikamalov}}}]{mitra23}%
  \BibitemOpen
  \bibfield  {author} {\bibinfo {author} {\bibfnamefont {A.}~\bibnamefont {{Mitra}}}, \bibinfo {author} {\bibfnamefont {B.}~\bibnamefont {{Shukirgaliyev}}}, \bibinfo {author} {\bibfnamefont {Y.~S.}\ \bibnamefont {{Abylkairov}}},\ and\ \bibinfo {author} {\bibfnamefont {E.}~\bibnamefont {{Abdikamalov}}},\ }\bibfield  {title} {\bibinfo {title} {{Exploring supernova gravitational waves with machine learning}},\ }\href {https://doi.org/10.1093/mnras/stad169} {\bibfield  {journal} {\bibinfo  {journal} {MNRAS}\ }\textbf {\bibinfo {volume} {520}},\ \bibinfo {pages} {2473} (\bibinfo {year} {2023})},\ \Eprint {https://arxiv.org/abs/2209.14542} {arXiv:2209.14542 [astro-ph.HE]} \BibitemShut {NoStop}%
\bibitem [{\citenamefont {{Pastor-Marcos}}\ \emph {et~al.}(2024)\citenamefont {{Pastor-Marcos}}, \citenamefont {{Cerd{\'a}-Dur{\'a}n}}, \citenamefont {{Walker}}, \citenamefont {{Torres-Forn{\'e}}}, \citenamefont {{Abdikamalov}}, \citenamefont {{Richers}},\ and\ \citenamefont {{Font}}}]{pastor24}%
  \BibitemOpen
  \bibfield  {author} {\bibinfo {author} {\bibfnamefont {C.}~\bibnamefont {{Pastor-Marcos}}}, \bibinfo {author} {\bibfnamefont {P.}~\bibnamefont {{Cerd{\'a}-Dur{\'a}n}}}, \bibinfo {author} {\bibfnamefont {D.}~\bibnamefont {{Walker}}}, \bibinfo {author} {\bibfnamefont {A.}~\bibnamefont {{Torres-Forn{\'e}}}}, \bibinfo {author} {\bibfnamefont {E.}~\bibnamefont {{Abdikamalov}}}, \bibinfo {author} {\bibfnamefont {S.}~\bibnamefont {{Richers}}},\ and\ \bibinfo {author} {\bibfnamefont {J.~A.}\ \bibnamefont {{Font}}},\ }\bibfield  {title} {\bibinfo {title} {{Bayesian inference from gravitational waves in fast-rotating, core-collapse supernovae}},\ }\href {https://doi.org/10.1103/PhysRevD.109.063028} {\bibfield  {journal} {\bibinfo  {journal} {\prd}\ }\textbf {\bibinfo {volume} {109}},\ \bibinfo {eid} {063028} (\bibinfo {year} {2024})},\ \Eprint {https://arxiv.org/abs/2308.03456} {arXiv:2308.03456 [astro-ph.HE]} \BibitemShut {NoStop}%
\bibitem [{\citenamefont {{Nunes}}\ \emph {et~al.}(2024)\citenamefont {{Nunes}}, \citenamefont {{Escrig}}, \citenamefont {{Freitas}}, \citenamefont {{Font}}, \citenamefont {{Fernandes}}, \citenamefont {{Onofre}},\ and\ \citenamefont {{Torres-Forn{\'e}}}}]{nunes2024deep}%
  \BibitemOpen
  \bibfield  {author} {\bibinfo {author} {\bibfnamefont {S.}~\bibnamefont {{Nunes}}}, \bibinfo {author} {\bibfnamefont {G.}~\bibnamefont {{Escrig}}}, \bibinfo {author} {\bibfnamefont {O.~G.}\ \bibnamefont {{Freitas}}}, \bibinfo {author} {\bibfnamefont {J.~A.}\ \bibnamefont {{Font}}}, \bibinfo {author} {\bibfnamefont {T.}~\bibnamefont {{Fernandes}}}, \bibinfo {author} {\bibfnamefont {A.}~\bibnamefont {{Onofre}}},\ and\ \bibinfo {author} {\bibfnamefont {A.}~\bibnamefont {{Torres-Forn{\'e}}}},\ }\bibfield  {title} {\bibinfo {title} {{Deep-learning classification and parameter inference of rotational core-collapse supernovae}},\ }\href {https://doi.org/10.1103/PhysRevD.110.064037} {\bibfield  {journal} {\bibinfo  {journal} {\prd}\ }\textbf {\bibinfo {volume} {110}},\ \bibinfo {eid} {064037} (\bibinfo {year} {2024})},\ \Eprint {https://arxiv.org/abs/2403.04938} {arXiv:2403.04938 [astro-ph.HE]} \BibitemShut {NoStop}%
\bibitem [{\citenamefont {{Villegas}}\ \emph {et~al.}(2025)\citenamefont {{Villegas}}, \citenamefont {{Moreno}}, \citenamefont {{Pajkos}}, \citenamefont {{Zanolin}},\ and\ \citenamefont {{Antelis}}}]{Villegas25Parameter}%
  \BibitemOpen
  \bibfield  {author} {\bibinfo {author} {\bibfnamefont {L.~O.}\ \bibnamefont {{Villegas}}}, \bibinfo {author} {\bibfnamefont {C.}~\bibnamefont {{Moreno}}}, \bibinfo {author} {\bibfnamefont {M.~A.}\ \bibnamefont {{Pajkos}}}, \bibinfo {author} {\bibfnamefont {M.}~\bibnamefont {{Zanolin}}},\ and\ \bibinfo {author} {\bibfnamefont {J.~M.}\ \bibnamefont {{Antelis}}},\ }\bibfield  {title} {\bibinfo {title} {{Parameter estimation from the core-bounce phase of rotating core collapse supernovae in real interferometer noise}},\ }\href {https://doi.org/10.1088/1361-6382/add235} {\bibfield  {journal} {\bibinfo  {journal} {Classical and Quantum Gravity}\ }\textbf {\bibinfo {volume} {42}},\ \bibinfo {eid} {115001} (\bibinfo {year} {2025})}\BibitemShut {NoStop}%
\bibitem [{\citenamefont {{Logue}}\ \emph {et~al.}(2012)\citenamefont {{Logue}}, \citenamefont {{Ott}}, \citenamefont {{Heng}}, \citenamefont {{Kalmus}},\ and\ \citenamefont {{Scargill}}}]{Logue12Inferring}%
  \BibitemOpen
  \bibfield  {author} {\bibinfo {author} {\bibfnamefont {J.}~\bibnamefont {{Logue}}}, \bibinfo {author} {\bibfnamefont {C.~D.}\ \bibnamefont {{Ott}}}, \bibinfo {author} {\bibfnamefont {I.~S.}\ \bibnamefont {{Heng}}}, \bibinfo {author} {\bibfnamefont {P.}~\bibnamefont {{Kalmus}}},\ and\ \bibinfo {author} {\bibfnamefont {J.~H.~C.}\ \bibnamefont {{Scargill}}},\ }\bibfield  {title} {\bibinfo {title} {{Inferring core-collapse supernova physics with gravitational waves}},\ }\href {https://doi.org/10.1103/PhysRevD.86.044023} {\bibfield  {journal} {\bibinfo  {journal} {\prd}\ }\textbf {\bibinfo {volume} {86}},\ \bibinfo {eid} {044023} (\bibinfo {year} {2012})},\ \Eprint {https://arxiv.org/abs/1202.3256} {arXiv:1202.3256 [gr-qc]} \BibitemShut {NoStop}%
\bibitem [{\citenamefont {{Powell}}\ \emph {et~al.}(2024)\citenamefont {{Powell}}, \citenamefont {{Iess}}, \citenamefont {{Llorens-Monteagudo}}, \citenamefont {{Obergaulinger}}, \citenamefont {{M{\"u}ller}}, \citenamefont {{Torres-Forn{\'e}}}, \citenamefont {{Cuoco}},\ and\ \citenamefont {{Font}}}]{Powell24Determining}%
  \BibitemOpen
  \bibfield  {author} {\bibinfo {author} {\bibfnamefont {J.}~\bibnamefont {{Powell}}}, \bibinfo {author} {\bibfnamefont {A.}~\bibnamefont {{Iess}}}, \bibinfo {author} {\bibfnamefont {M.}~\bibnamefont {{Llorens-Monteagudo}}}, \bibinfo {author} {\bibfnamefont {M.}~\bibnamefont {{Obergaulinger}}}, \bibinfo {author} {\bibfnamefont {B.}~\bibnamefont {{M{\"u}ller}}}, \bibinfo {author} {\bibfnamefont {A.}~\bibnamefont {{Torres-Forn{\'e}}}}, \bibinfo {author} {\bibfnamefont {E.}~\bibnamefont {{Cuoco}}},\ and\ \bibinfo {author} {\bibfnamefont {J.~A.}\ \bibnamefont {{Font}}},\ }\bibfield  {title} {\bibinfo {title} {{Determining the core-collapse supernova explosion mechanism with current and future gravitational-wave observatories}},\ }\href {https://doi.org/10.1103/PhysRevD.109.063019} {\bibfield  {journal} {\bibinfo  {journal} {\prd}\ }\textbf {\bibinfo {volume} {109}},\ \bibinfo {eid} {063019} (\bibinfo {year} {2024})},\ \Eprint {https://arxiv.org/abs/2311.18221} {arXiv:2311.18221 [astro-ph.HE]} \BibitemShut
  {NoStop}%
\bibitem [{\citenamefont {{Bizouard}}\ \emph {et~al.}(2021)\citenamefont {{Bizouard}}, \citenamefont {{Maturana-Russel}}, \citenamefont {{Torres-Forn{\'e}}}, \citenamefont {{Obergaulinger}}, \citenamefont {{Cerd{\'a}-Dur{\'a}n}}, \citenamefont {{Christensen}}, \citenamefont {{Font}},\ and\ \citenamefont {{Meyer}}}]{Bizouard21Inference}%
  \BibitemOpen
  \bibfield  {author} {\bibinfo {author} {\bibfnamefont {M.-A.}\ \bibnamefont {{Bizouard}}}, \bibinfo {author} {\bibfnamefont {P.}~\bibnamefont {{Maturana-Russel}}}, \bibinfo {author} {\bibfnamefont {A.}~\bibnamefont {{Torres-Forn{\'e}}}}, \bibinfo {author} {\bibfnamefont {M.}~\bibnamefont {{Obergaulinger}}}, \bibinfo {author} {\bibfnamefont {P.}~\bibnamefont {{Cerd{\'a}-Dur{\'a}n}}}, \bibinfo {author} {\bibfnamefont {N.}~\bibnamefont {{Christensen}}}, \bibinfo {author} {\bibfnamefont {J.~A.}\ \bibnamefont {{Font}}},\ and\ \bibinfo {author} {\bibfnamefont {R.}~\bibnamefont {{Meyer}}},\ }\bibfield  {title} {\bibinfo {title} {{Inference of protoneutron star properties from gravitational-wave data in core-collapse supernovae}},\ }\href {https://doi.org/10.1103/PhysRevD.103.063006} {\bibfield  {journal} {\bibinfo  {journal} {\prd}\ }\textbf {\bibinfo {volume} {103}},\ \bibinfo {eid} {063006} (\bibinfo {year} {2021})},\ \Eprint {https://arxiv.org/abs/2012.00846} {arXiv:2012.00846 [gr-qc]} \BibitemShut {NoStop}%
\bibitem [{\citenamefont {{Bruel}}\ \emph {et~al.}(2023)\citenamefont {{Bruel}}, \citenamefont {{Bizouard}}, \citenamefont {{Obergaulinger}}, \citenamefont {{Maturana-Russel}}, \citenamefont {{Torres-Forn{\'e}}}, \citenamefont {{Cerd{\'a}-Dur{\'a}n}}, \citenamefont {{Christensen}}, \citenamefont {{Font}},\ and\ \citenamefont {{Meyer}}}]{Bruel23Inference}%
  \BibitemOpen
  \bibfield  {author} {\bibinfo {author} {\bibfnamefont {T.}~\bibnamefont {{Bruel}}}, \bibinfo {author} {\bibfnamefont {M.-A.}\ \bibnamefont {{Bizouard}}}, \bibinfo {author} {\bibfnamefont {M.}~\bibnamefont {{Obergaulinger}}}, \bibinfo {author} {\bibfnamefont {P.}~\bibnamefont {{Maturana-Russel}}}, \bibinfo {author} {\bibfnamefont {A.}~\bibnamefont {{Torres-Forn{\'e}}}}, \bibinfo {author} {\bibfnamefont {P.}~\bibnamefont {{Cerd{\'a}-Dur{\'a}n}}}, \bibinfo {author} {\bibfnamefont {N.}~\bibnamefont {{Christensen}}}, \bibinfo {author} {\bibfnamefont {J.~A.}\ \bibnamefont {{Font}}},\ and\ \bibinfo {author} {\bibfnamefont {R.}~\bibnamefont {{Meyer}}},\ }\bibfield  {title} {\bibinfo {title} {{Inference of protoneutron star properties in core-collapse supernovae from a gravitational-wave detector network}},\ }\href {https://doi.org/10.1103/PhysRevD.107.083029} {\bibfield  {journal} {\bibinfo  {journal} {\prd}\ }\textbf {\bibinfo {volume} {107}},\ \bibinfo {eid} {083029} (\bibinfo {year} {2023})},\ \Eprint
  {https://arxiv.org/abs/2301.10019} {arXiv:2301.10019 [astro-ph.HE]} \BibitemShut {NoStop}%
\bibitem [{\citenamefont {{Abdikamalov}}\ \emph {et~al.}(2014)\citenamefont {{Abdikamalov}}, \citenamefont {{Gossan}}, \citenamefont {{DeMaio}},\ and\ \citenamefont {{Ott}}}]{abdikamalov:14}%
  \BibitemOpen
  \bibfield  {author} {\bibinfo {author} {\bibfnamefont {E.}~\bibnamefont {{Abdikamalov}}}, \bibinfo {author} {\bibfnamefont {S.}~\bibnamefont {{Gossan}}}, \bibinfo {author} {\bibfnamefont {A.~M.}\ \bibnamefont {{DeMaio}}},\ and\ \bibinfo {author} {\bibfnamefont {C.~D.}\ \bibnamefont {{Ott}}},\ }\bibfield  {title} {\bibinfo {title} {{Measuring the angular momentum distribution in core-collapse supernova progenitors with gravitational waves}},\ }\href {https://doi.org/10.1103/PhysRevD.90.044001} {\bibfield  {journal} {\bibinfo  {journal} {\prd}\ }\textbf {\bibinfo {volume} {90}},\ \bibinfo {eid} {044001} (\bibinfo {year} {2014})},\ \Eprint {https://arxiv.org/abs/1311.3678} {arXiv:1311.3678 [astro-ph.SR]} \BibitemShut {NoStop}%
\bibitem [{\citenamefont {{Pajkos}}\ \emph {et~al.}(2019)\citenamefont {{Pajkos}}, \citenamefont {{Couch}}, \citenamefont {{Pan}},\ and\ \citenamefont {{O'Connor}}}]{pajkos19}%
  \BibitemOpen
  \bibfield  {author} {\bibinfo {author} {\bibfnamefont {M.~A.}\ \bibnamefont {{Pajkos}}}, \bibinfo {author} {\bibfnamefont {S.~M.}\ \bibnamefont {{Couch}}}, \bibinfo {author} {\bibfnamefont {K.-C.}\ \bibnamefont {{Pan}}},\ and\ \bibinfo {author} {\bibfnamefont {E.~P.}\ \bibnamefont {{O'Connor}}},\ }\bibfield  {title} {\bibinfo {title} {{Features of Accretion-phase Gravitational-wave Emission from Two-dimensional Rotating Core-collapse Supernovae}},\ }\href {https://doi.org/10.3847/1538-4357/ab1de2} {\bibfield  {journal} {\bibinfo  {journal} {\apj}\ }\textbf {\bibinfo {volume} {878}},\ \bibinfo {eid} {13} (\bibinfo {year} {2019})},\ \Eprint {https://arxiv.org/abs/1901.09055} {arXiv:1901.09055 [astro-ph.HE]} \BibitemShut {NoStop}%
\bibitem [{\citenamefont {{Cerd{\'a}-Dur{\'a}n}}\ \emph {et~al.}(2013)\citenamefont {{Cerd{\'a}-Dur{\'a}n}}, \citenamefont {{DeBrye}}, \citenamefont {{Aloy}}, \citenamefont {{Font}},\ and\ \citenamefont {{Obergaulinger}}}]{cerda:13}%
  \BibitemOpen
  \bibfield  {author} {\bibinfo {author} {\bibfnamefont {P.}~\bibnamefont {{Cerd{\'a}-Dur{\'a}n}}}, \bibinfo {author} {\bibfnamefont {N.}~\bibnamefont {{DeBrye}}}, \bibinfo {author} {\bibfnamefont {M.~A.}\ \bibnamefont {{Aloy}}}, \bibinfo {author} {\bibfnamefont {J.~A.}\ \bibnamefont {{Font}}},\ and\ \bibinfo {author} {\bibfnamefont {M.}~\bibnamefont {{Obergaulinger}}},\ }\bibfield  {title} {\bibinfo {title} {{Gravitational Wave Signatures in Black Hole Forming Core Collapse}},\ }\href {https://doi.org/10.1088/2041-8205/779/2/L18} {\bibfield  {journal} {\bibinfo  {journal} {ApJL}\ }\textbf {\bibinfo {volume} {779}},\ \bibinfo {eid} {L18} (\bibinfo {year} {2013})},\ \Eprint {https://arxiv.org/abs/1310.8290} {arXiv:1310.8290 [astro-ph.SR]} \BibitemShut {NoStop}%
\bibitem [{\citenamefont {{Pan}}\ \emph {et~al.}(2018)\citenamefont {{Pan}}, \citenamefont {{Liebend{\"o}rfer}}, \citenamefont {{Couch}},\ and\ \citenamefont {{Thielemann}}}]{Pan18Equation}%
  \BibitemOpen
  \bibfield  {author} {\bibinfo {author} {\bibfnamefont {K.-C.}\ \bibnamefont {{Pan}}}, \bibinfo {author} {\bibfnamefont {M.}~\bibnamefont {{Liebend{\"o}rfer}}}, \bibinfo {author} {\bibfnamefont {S.~M.}\ \bibnamefont {{Couch}}},\ and\ \bibinfo {author} {\bibfnamefont {F.-K.}\ \bibnamefont {{Thielemann}}},\ }\bibfield  {title} {\bibinfo {title} {{Equation of State Dependent Dynamics and Multi-messenger Signals from Stellar-mass Black Hole Formation}},\ }\href {https://doi.org/10.3847/1538-4357/aab71d} {\bibfield  {journal} {\bibinfo  {journal} {\apj}\ }\textbf {\bibinfo {volume} {857}},\ \bibinfo {eid} {13} (\bibinfo {year} {2018})},\ \Eprint {https://arxiv.org/abs/1710.01690} {arXiv:1710.01690 [astro-ph.HE]} \BibitemShut {NoStop}%
\bibitem [{\citenamefont {{Shibagaki}}\ \emph {et~al.}(2021)\citenamefont {{Shibagaki}}, \citenamefont {{Kuroda}}, \citenamefont {{Kotake}},\ and\ \citenamefont {{Takiwaki}}}]{Shibagaki21}%
  \BibitemOpen
  \bibfield  {author} {\bibinfo {author} {\bibfnamefont {S.}~\bibnamefont {{Shibagaki}}}, \bibinfo {author} {\bibfnamefont {T.}~\bibnamefont {{Kuroda}}}, \bibinfo {author} {\bibfnamefont {K.}~\bibnamefont {{Kotake}}},\ and\ \bibinfo {author} {\bibfnamefont {T.}~\bibnamefont {{Takiwaki}}},\ }\bibfield  {title} {\bibinfo {title} {{Characteristic time variability of gravitational-wave and neutrino signals from three-dimensional simulations of non-rotating and rapidly rotating stellar core collapse}},\ }\href {https://doi.org/10.1093/mnras/stab228} {\bibfield  {journal} {\bibinfo  {journal} {\mnras}\ }\textbf {\bibinfo {volume} {502}},\ \bibinfo {pages} {3066} (\bibinfo {year} {2021})},\ \Eprint {https://arxiv.org/abs/2010.03882} {arXiv:2010.03882 [astro-ph.HE]} \BibitemShut {NoStop}%
\bibitem [{\citenamefont {{Powell}}\ and\ \citenamefont {{M{\"u}ller}}(2025)}]{Powell25noEMCCSN}%
  \BibitemOpen
  \bibfield  {author} {\bibinfo {author} {\bibfnamefont {J.}~\bibnamefont {{Powell}}}\ and\ \bibinfo {author} {\bibfnamefont {B.}~\bibnamefont {{M{\"u}ller}}},\ }\bibfield  {title} {\bibinfo {title} {{Gravitational waves from core-collapse supernovae with no electromagnetic counterparts}},\ }\href {https://doi.org/10.48550/arXiv.2506.03581} {\bibfield  {journal} {\bibinfo  {journal} {arXiv e-prints}\ ,\ \bibinfo {eid} {arXiv:2506.03581}} (\bibinfo {year} {2025})},\ \Eprint {https://arxiv.org/abs/2506.03581} {arXiv:2506.03581 [astro-ph.HE]} \BibitemShut {NoStop}%
\bibitem [{\citenamefont {{Eggenberger Andersen}}\ \emph {et~al.}(2025)\citenamefont {{Eggenberger Andersen}}, \citenamefont {{O'Connor}}, \citenamefont {{Andresen}}, \citenamefont {{da Silva Schneider}},\ and\ \citenamefont {{Couch}}}]{Eggenberger25Black}%
  \BibitemOpen
  \bibfield  {author} {\bibinfo {author} {\bibfnamefont {O.}~\bibnamefont {{Eggenberger Andersen}}}, \bibinfo {author} {\bibfnamefont {E.}~\bibnamefont {{O'Connor}}}, \bibinfo {author} {\bibfnamefont {H.}~\bibnamefont {{Andresen}}}, \bibinfo {author} {\bibfnamefont {A.}~\bibnamefont {{da Silva Schneider}}},\ and\ \bibinfo {author} {\bibfnamefont {S.~M.}\ \bibnamefont {{Couch}}},\ }\bibfield  {title} {\bibinfo {title} {{Black Hole Supernovae, Their Equation of State Dependence, and Ejecta Composition}},\ }\href {https://doi.org/10.3847/1538-4357/ada899} {\bibfield  {journal} {\bibinfo  {journal} {\apj}\ }\textbf {\bibinfo {volume} {980}},\ \bibinfo {eid} {53} (\bibinfo {year} {2025})},\ \Eprint {https://arxiv.org/abs/2411.11969} {arXiv:2411.11969 [astro-ph.HE]} \BibitemShut {NoStop}%
\bibitem [{\citenamefont {{Kuroda}}\ and\ \citenamefont {{Shibata}}(2023)}]{Kuroda23Failed}%
  \BibitemOpen
  \bibfield  {author} {\bibinfo {author} {\bibfnamefont {T.}~\bibnamefont {{Kuroda}}}\ and\ \bibinfo {author} {\bibfnamefont {M.}~\bibnamefont {{Shibata}}},\ }\bibfield  {title} {\bibinfo {title} {{Failed supernova simulations beyond black hole formation}},\ }\href {https://doi.org/10.1093/mnras/stad2710} {\bibfield  {journal} {\bibinfo  {journal} {\mnras}\ }\textbf {\bibinfo {volume} {526}},\ \bibinfo {pages} {152} (\bibinfo {year} {2023})},\ \Eprint {https://arxiv.org/abs/2307.06192} {arXiv:2307.06192 [astro-ph.HE]} \BibitemShut {NoStop}%
\bibitem [{\citenamefont {{Richers}}\ \emph {et~al.}(2017)\citenamefont {{Richers}}, \citenamefont {{Ott}}, \citenamefont {{Abdikamalov}}, \citenamefont {{O'Connor}},\ and\ \citenamefont {{Sullivan}}}]{richers:17}%
  \BibitemOpen
  \bibfield  {author} {\bibinfo {author} {\bibfnamefont {S.}~\bibnamefont {{Richers}}}, \bibinfo {author} {\bibfnamefont {C.~D.}\ \bibnamefont {{Ott}}}, \bibinfo {author} {\bibfnamefont {E.}~\bibnamefont {{Abdikamalov}}}, \bibinfo {author} {\bibfnamefont {E.}~\bibnamefont {{O'Connor}}},\ and\ \bibinfo {author} {\bibfnamefont {C.}~\bibnamefont {{Sullivan}}},\ }\bibfield  {title} {\bibinfo {title} {{Equation of state effects on gravitational waves from rotating core collapse}},\ }\href {https://doi.org/10.1103/PhysRevD.95.063019} {\bibfield  {journal} {\bibinfo  {journal} {\prd}\ }\textbf {\bibinfo {volume} {95}},\ \bibinfo {eid} {063019} (\bibinfo {year} {2017})},\ \Eprint {https://arxiv.org/abs/1701.02752} {arXiv:1701.02752 [astro-ph.HE]} \BibitemShut {NoStop}%
\bibitem [{\citenamefont {{Schneider}}\ \emph {et~al.}(2019)\citenamefont {{Schneider}}, \citenamefont {{Roberts}}, \citenamefont {{Ott}},\ and\ \citenamefont {{O'Connor}}}]{schneider19equation}%
  \BibitemOpen
  \bibfield  {author} {\bibinfo {author} {\bibfnamefont {A.~S.}\ \bibnamefont {{Schneider}}}, \bibinfo {author} {\bibfnamefont {L.~F.}\ \bibnamefont {{Roberts}}}, \bibinfo {author} {\bibfnamefont {C.~D.}\ \bibnamefont {{Ott}}},\ and\ \bibinfo {author} {\bibfnamefont {E.}~\bibnamefont {{O'Connor}}},\ }\bibfield  {title} {\bibinfo {title} {{Equation of state effects in the core collapse of a 20 -M$_{{\ensuremath{\odot}}}$ star}},\ }\href {https://doi.org/10.1103/PhysRevC.100.055802} {\bibfield  {journal} {\bibinfo  {journal} {\prc}\ }\textbf {\bibinfo {volume} {100}},\ \bibinfo {eid} {055802} (\bibinfo {year} {2019})},\ \Eprint {https://arxiv.org/abs/1906.02009} {arXiv:1906.02009 [astro-ph.HE]} \BibitemShut {NoStop}%
\bibitem [{\citenamefont {{Powell}}\ and\ \citenamefont {{M{\"u}ller}}(2024)}]{Powell24GW}%
  \BibitemOpen
  \bibfield  {author} {\bibinfo {author} {\bibfnamefont {J.}~\bibnamefont {{Powell}}}\ and\ \bibinfo {author} {\bibfnamefont {B.}~\bibnamefont {{M{\"u}ller}}},\ }\bibfield  {title} {\bibinfo {title} {{The gravitational-wave emission from the explosion of a 15 solar mass star with rotation and magnetic fields}},\ }\href {https://doi.org/10.1093/mnras/stae1731} {\bibfield  {journal} {\bibinfo  {journal} {MNRAS}\ }\textbf {\bibinfo {volume} {532}},\ \bibinfo {pages} {4326} (\bibinfo {year} {2024})},\ \Eprint {https://arxiv.org/abs/2406.09691} {arXiv:2406.09691 [astro-ph.HE]} \BibitemShut {NoStop}%
\bibitem [{\citenamefont {{Edwards}}(2021)}]{edwards2017}%
  \BibitemOpen
  \bibfield  {author} {\bibinfo {author} {\bibfnamefont {M.~C.}\ \bibnamefont {{Edwards}}},\ }\bibfield  {title} {\bibinfo {title} {{Classifying the equation of state from rotating core collapse gravitational waves with deep learning}},\ }\href {https://doi.org/10.1103/PhysRevD.103.024025} {\bibfield  {journal} {\bibinfo  {journal} {\prd}\ }\textbf {\bibinfo {volume} {103}},\ \bibinfo {eid} {024025} (\bibinfo {year} {2021})},\ \Eprint {https://arxiv.org/abs/2009.07367} {arXiv:2009.07367 [astro-ph.IM]} \BibitemShut {NoStop}%
\bibitem [{\citenamefont {{Chao}}\ \emph {et~al.}(2022)\citenamefont {{Chao}}, \citenamefont {{Su}}, \citenamefont {{Chen}}, \citenamefont {{Wang}},\ and\ \citenamefont {{Pan}}}]{chao22determining}%
  \BibitemOpen
  \bibfield  {author} {\bibinfo {author} {\bibfnamefont {Y.-S.}\ \bibnamefont {{Chao}}}, \bibinfo {author} {\bibfnamefont {C.-Z.}\ \bibnamefont {{Su}}}, \bibinfo {author} {\bibfnamefont {T.-Y.}\ \bibnamefont {{Chen}}}, \bibinfo {author} {\bibfnamefont {D.-W.}\ \bibnamefont {{Wang}}},\ and\ \bibinfo {author} {\bibfnamefont {K.-C.}\ \bibnamefont {{Pan}}},\ }\bibfield  {title} {\bibinfo {title} {{Determining the Core Structure and Nuclear Equation of State of Rotating Core-collapse Supernovae with Gravitational Waves by Convolutional Neural Networks}},\ }\href {https://doi.org/10.3847/1538-4357/ac930e} {\bibfield  {journal} {\bibinfo  {journal} {\apj}\ }\textbf {\bibinfo {volume} {939}},\ \bibinfo {eid} {13} (\bibinfo {year} {2022})},\ \Eprint {https://arxiv.org/abs/2209.10089} {arXiv:2209.10089 [astro-ph.HE]} \BibitemShut {NoStop}%
\bibitem [{\citenamefont {{Wolfe}}\ \emph {et~al.}(2023)\citenamefont {{Wolfe}}, \citenamefont {{Fr{\"o}hlich}}, \citenamefont {{Miller}}, \citenamefont {{Torres-Forn{\'e}}},\ and\ \citenamefont {{Cerd{\'a}-Dur{\'a}n}}}]{Wolfe23GW}%
  \BibitemOpen
  \bibfield  {author} {\bibinfo {author} {\bibfnamefont {N.~E.}\ \bibnamefont {{Wolfe}}}, \bibinfo {author} {\bibfnamefont {C.}~\bibnamefont {{Fr{\"o}hlich}}}, \bibinfo {author} {\bibfnamefont {J.~M.}\ \bibnamefont {{Miller}}}, \bibinfo {author} {\bibfnamefont {A.}~\bibnamefont {{Torres-Forn{\'e}}}},\ and\ \bibinfo {author} {\bibfnamefont {P.}~\bibnamefont {{Cerd{\'a}-Dur{\'a}n}}},\ }\bibfield  {title} {\bibinfo {title} {{Gravitational Wave Eigenfrequencies from Neutrino-driven Core-collapse Supernovae}},\ }\href {https://doi.org/10.3847/1538-4357/ace693} {\bibfield  {journal} {\bibinfo  {journal} {\apj}\ }\textbf {\bibinfo {volume} {954}},\ \bibinfo {eid} {161} (\bibinfo {year} {2023})},\ \Eprint {https://arxiv.org/abs/2303.16962} {arXiv:2303.16962 [astro-ph.HE]} \BibitemShut {NoStop}%
\bibitem [{\citenamefont {{Casallas-Lagos}}\ \emph {et~al.}(2023)\citenamefont {{Casallas-Lagos}}, \citenamefont {{Antelis}}, \citenamefont {{Moreno}}, \citenamefont {{Zanolin}}, \citenamefont {{Mezzacappa}},\ and\ \citenamefont {{Szczepa{\'n}czyk}}}]{CasallasLagos23Characterizing}%
  \BibitemOpen
  \bibfield  {author} {\bibinfo {author} {\bibfnamefont {A.}~\bibnamefont {{Casallas-Lagos}}}, \bibinfo {author} {\bibfnamefont {J.~M.}\ \bibnamefont {{Antelis}}}, \bibinfo {author} {\bibfnamefont {C.}~\bibnamefont {{Moreno}}}, \bibinfo {author} {\bibfnamefont {M.}~\bibnamefont {{Zanolin}}}, \bibinfo {author} {\bibfnamefont {A.}~\bibnamefont {{Mezzacappa}}},\ and\ \bibinfo {author} {\bibfnamefont {M.~J.}\ \bibnamefont {{Szczepa{\'n}czyk}}},\ }\bibfield  {title} {\bibinfo {title} {{Characterizing the temporal evolution of the high-frequency gravitational wave emission for a core collapse supernova with laser interferometric data: A neural network approach}},\ }\href {https://doi.org/10.1103/PhysRevD.108.084027} {\bibfield  {journal} {\bibinfo  {journal} {\prd}\ }\textbf {\bibinfo {volume} {108}},\ \bibinfo {eid} {084027} (\bibinfo {year} {2023})}\BibitemShut {NoStop}%
\bibitem [{\citenamefont {{Murphy}}\ \emph {et~al.}(2024)\citenamefont {{Murphy}}, \citenamefont {{Casallas-Lagos}}, \citenamefont {{Mezzacappa}}, \citenamefont {{Zanolin}}, \citenamefont {{Landfield}}, \citenamefont {{Lentz}}, \citenamefont {{Marronetti}}, \citenamefont {{Antelis}},\ and\ \citenamefont {{Moreno}}}]{Murphy24Dependence}%
  \BibitemOpen
  \bibfield  {author} {\bibinfo {author} {\bibfnamefont {R.~D.}\ \bibnamefont {{Murphy}}}, \bibinfo {author} {\bibfnamefont {A.}~\bibnamefont {{Casallas-Lagos}}}, \bibinfo {author} {\bibfnamefont {A.}~\bibnamefont {{Mezzacappa}}}, \bibinfo {author} {\bibfnamefont {M.}~\bibnamefont {{Zanolin}}}, \bibinfo {author} {\bibfnamefont {R.~E.}\ \bibnamefont {{Landfield}}}, \bibinfo {author} {\bibfnamefont {E.~J.}\ \bibnamefont {{Lentz}}}, \bibinfo {author} {\bibfnamefont {P.}~\bibnamefont {{Marronetti}}}, \bibinfo {author} {\bibfnamefont {J.~M.}\ \bibnamefont {{Antelis}}},\ and\ \bibinfo {author} {\bibfnamefont {C.}~\bibnamefont {{Moreno}}},\ }\bibfield  {title} {\bibinfo {title} {{Dependence of the reconstructed core-collapse supernova gravitational wave high-frequency feature on the nuclear equation of state in real interferometric data}},\ }\href {https://doi.org/10.1103/PhysRevD.110.083006} {\bibfield  {journal} {\bibinfo  {journal} {\prd}\ }\textbf {\bibinfo {volume} {110}},\ \bibinfo {eid} {083006} (\bibinfo
  {year} {2024})},\ \Eprint {https://arxiv.org/abs/2406.01784} {arXiv:2406.01784 [astro-ph.HE]} \BibitemShut {NoStop}%
\bibitem [{\citenamefont {{Abbott}}\ \emph {et~al.}(2018)\citenamefont {{Abbott}}, \citenamefont {{Abbott}}, \citenamefont {{Abbott}},\ and\ \citenamefont {{Acernese}}}]{abbott18gw170817}%
  \BibitemOpen
  \bibfield  {author} {\bibinfo {author} {\bibfnamefont {B.~P.}\ \bibnamefont {{Abbott}}}, \bibinfo {author} {\bibfnamefont {R.}~\bibnamefont {{Abbott}}}, \bibinfo {author} {\bibfnamefont {T.~D.}\ \bibnamefont {{Abbott}}},\ and\ \bibinfo {author} {\bibfnamefont {F.}~\bibnamefont {{Acernese}}} (\bibinfo {collaboration} {Ligo Scientific Collaboration and VIRGO Collaboration and KAGRA Collaboration}),\ }\bibfield  {title} {\bibinfo {title} {{GW170817: Measurements of Neutron Star Radii and Equation of State}},\ }\href {https://doi.org/10.1103/PhysRevLett.121.161101} {\bibfield  {journal} {\bibinfo  {journal} {\prl}\ }\textbf {\bibinfo {volume} {121}},\ \bibinfo {eid} {161101} (\bibinfo {year} {2018})},\ \Eprint {https://arxiv.org/abs/1805.11581} {arXiv:1805.11581 [gr-qc]} \BibitemShut {NoStop}%
\bibitem [{\citenamefont {Takami}\ \emph {et~al.}(2014)\citenamefont {Takami}, \citenamefont {Rezzolla},\ and\ \citenamefont {Baiotti}}]{Takami14Constraining}%
  \BibitemOpen
  \bibfield  {author} {\bibinfo {author} {\bibfnamefont {K.}~\bibnamefont {Takami}}, \bibinfo {author} {\bibfnamefont {L.}~\bibnamefont {Rezzolla}},\ and\ \bibinfo {author} {\bibfnamefont {L.}~\bibnamefont {Baiotti}},\ }\bibfield  {title} {\bibinfo {title} {Constraining the equation of state of neutron stars from binary mergers},\ }\href {https://doi.org/10.1103/PhysRevLett.113.091104} {\bibfield  {journal} {\bibinfo  {journal} {Phys. Rev. Lett.}\ }\textbf {\bibinfo {volume} {113}},\ \bibinfo {pages} {091104} (\bibinfo {year} {2014})}\BibitemShut {NoStop}%
\bibitem [{\citenamefont {{Radice}}\ \emph {et~al.}(2018)\citenamefont {{Radice}}, \citenamefont {{Perego}}, \citenamefont {{Bernuzzi}},\ and\ \citenamefont {{Zhang}}}]{Radice18Long}%
  \BibitemOpen
  \bibfield  {author} {\bibinfo {author} {\bibfnamefont {D.}~\bibnamefont {{Radice}}}, \bibinfo {author} {\bibfnamefont {A.}~\bibnamefont {{Perego}}}, \bibinfo {author} {\bibfnamefont {S.}~\bibnamefont {{Bernuzzi}}},\ and\ \bibinfo {author} {\bibfnamefont {B.}~\bibnamefont {{Zhang}}},\ }\bibfield  {title} {\bibinfo {title} {{Long-lived remnants from binary neutron star mergers}},\ }\href {https://doi.org/10.1093/mnras/sty2531} {\bibfield  {journal} {\bibinfo  {journal} {\mnras}\ }\textbf {\bibinfo {volume} {481}},\ \bibinfo {pages} {3670} (\bibinfo {year} {2018})},\ \Eprint {https://arxiv.org/abs/1803.10865} {arXiv:1803.10865 [astro-ph.HE]} \BibitemShut {NoStop}%
\bibitem [{\citenamefont {{Mitra}}\ \emph {et~al.}(2024)\citenamefont {{Mitra}}, \citenamefont {{Orel}}, \citenamefont {{Abylkairov}}, \citenamefont {{Shukirgaliyev}},\ and\ \citenamefont {{Abdikamalov}}}]{mitra24}%
  \BibitemOpen
  \bibfield  {author} {\bibinfo {author} {\bibfnamefont {A.}~\bibnamefont {{Mitra}}}, \bibinfo {author} {\bibfnamefont {D.}~\bibnamefont {{Orel}}}, \bibinfo {author} {\bibfnamefont {Y.~S.}\ \bibnamefont {{Abylkairov}}}, \bibinfo {author} {\bibfnamefont {B.}~\bibnamefont {{Shukirgaliyev}}},\ and\ \bibinfo {author} {\bibfnamefont {E.}~\bibnamefont {{Abdikamalov}}},\ }\bibfield  {title} {\bibinfo {title} {{Probing nuclear physics with supernova gravitational waves and machine learning}},\ }\href {https://doi.org/10.1093/mnras/stae714} {\bibfield  {journal} {\bibinfo  {journal} {MNRAS}\ }\textbf {\bibinfo {volume} {529}},\ \bibinfo {pages} {3582} (\bibinfo {year} {2024})},\ \Eprint {https://arxiv.org/abs/2310.15649} {arXiv:2310.15649 [astro-ph.HE]} \BibitemShut {NoStop}%
\bibitem [{\citenamefont {{Abylkairov}}\ \emph {et~al.}(2024)\citenamefont {{Abylkairov}}, \citenamefont {{Edwards}}, \citenamefont {{Orel}}, \citenamefont {{Mitra}}, \citenamefont {{Shukirgaliyev}},\ and\ \citenamefont {{Abdikamalov}}}]{Abylkairov24Evaluating}%
  \BibitemOpen
  \bibfield  {author} {\bibinfo {author} {\bibfnamefont {Y.~S.}\ \bibnamefont {{Abylkairov}}}, \bibinfo {author} {\bibfnamefont {M.~C.}\ \bibnamefont {{Edwards}}}, \bibinfo {author} {\bibfnamefont {D.}~\bibnamefont {{Orel}}}, \bibinfo {author} {\bibfnamefont {A.}~\bibnamefont {{Mitra}}}, \bibinfo {author} {\bibfnamefont {B.}~\bibnamefont {{Shukirgaliyev}}},\ and\ \bibinfo {author} {\bibfnamefont {E.}~\bibnamefont {{Abdikamalov}}},\ }\bibfield  {title} {\bibinfo {title} {{Evaluating machine learning models for supernova gravitational wave signal classification}},\ }\href {https://doi.org/10.1088/2632-2153/ada33a} {\bibfield  {journal} {\bibinfo  {journal} {Machine Learning: Science and Technology}\ }\textbf {\bibinfo {volume} {5}},\ \bibinfo {eid} {045077} (\bibinfo {year} {2024})},\ \Eprint {https://arxiv.org/abs/2409.14508} {arXiv:2409.14508 [astro-ph.HE]} \BibitemShut {NoStop}%
\bibitem [{\citenamefont {{Banik}}\ \emph {et~al.}(2014)\citenamefont {{Banik}}, \citenamefont {{Hempel}},\ and\ \citenamefont {{Bandyopadhyay}}}]{bhbeos}%
  \BibitemOpen
  \bibfield  {author} {\bibinfo {author} {\bibfnamefont {S.}~\bibnamefont {{Banik}}}, \bibinfo {author} {\bibfnamefont {M.}~\bibnamefont {{Hempel}}},\ and\ \bibinfo {author} {\bibfnamefont {D.}~\bibnamefont {{Bandyopadhyay}}},\ }\bibfield  {title} {\bibinfo {title} {{New Hyperon Equations of State for Supernovae and Neutron Stars in Density-dependent Hadron Field Theory}},\ }\href {https://doi.org/10.1088/0067-0049/214/2/22} {\bibfield  {journal} {\bibinfo  {journal} {\apjs}\ }\textbf {\bibinfo {volume} {214}},\ \bibinfo {eid} {22} (\bibinfo {year} {2014})},\ \Eprint {https://arxiv.org/abs/1404.6173} {arXiv:1404.6173 [astro-ph.HE]} \BibitemShut {NoStop}%
\bibitem [{\citenamefont {{Hempel}}\ and\ \citenamefont {{Schaffner-Bielich}}(2010)}]{hempel:10}%
  \BibitemOpen
  \bibfield  {author} {\bibinfo {author} {\bibfnamefont {M.}~\bibnamefont {{Hempel}}}\ and\ \bibinfo {author} {\bibfnamefont {J.}~\bibnamefont {{Schaffner-Bielich}}},\ }\bibfield  {title} {\bibinfo {title} {{A statistical model for a complete supernova equation of state}},\ }\href {https://doi.org/10.1016/j.nuclphysa.2010.02.010} {\bibfield  {journal} {\bibinfo  {journal} {Nucl. Phys. A}\ }\textbf {\bibinfo {volume} {837}},\ \bibinfo {pages} {210} (\bibinfo {year} {2010})},\ \Eprint {https://arxiv.org/abs/0911.4073} {arXiv:0911.4073 [nucl-th]} \BibitemShut {NoStop}%
\bibitem [{\citenamefont {{Hempel}}\ \emph {et~al.}(2012)\citenamefont {{Hempel}}, \citenamefont {{Fischer}}, \citenamefont {{Schaffner-Bielich}},\ and\ \citenamefont {{Liebend{\"o}rfer}}}]{hempel:12}%
  \BibitemOpen
  \bibfield  {author} {\bibinfo {author} {\bibfnamefont {M.}~\bibnamefont {{Hempel}}}, \bibinfo {author} {\bibfnamefont {T.}~\bibnamefont {{Fischer}}}, \bibinfo {author} {\bibfnamefont {J.}~\bibnamefont {{Schaffner-Bielich}}},\ and\ \bibinfo {author} {\bibfnamefont {M.}~\bibnamefont {{Liebend{\"o}rfer}}},\ }\bibfield  {title} {\bibinfo {title} {{New Equations of State in Simulations of Core-collapse Supernovae}},\ }\href {https://doi.org/10.1088/0004-637X/748/1/70} {\bibfield  {journal} {\bibinfo  {journal} {\apj}\ }\textbf {\bibinfo {volume} {748}},\ \bibinfo {eid} {70} (\bibinfo {year} {2012})},\ \Eprint {https://arxiv.org/abs/1108.0848} {arXiv:1108.0848 [astro-ph.HE]} \BibitemShut {NoStop}%
\bibitem [{\citenamefont {{Steiner}}\ \emph {et~al.}(2013)\citenamefont {{Steiner}}, \citenamefont {{Hempel}},\ and\ \citenamefont {{Fischer}}}]{steiner:13b}%
  \BibitemOpen
  \bibfield  {author} {\bibinfo {author} {\bibfnamefont {A.~W.}\ \bibnamefont {{Steiner}}}, \bibinfo {author} {\bibfnamefont {M.}~\bibnamefont {{Hempel}}},\ and\ \bibinfo {author} {\bibfnamefont {T.}~\bibnamefont {{Fischer}}},\ }\bibfield  {title} {\bibinfo {title} {{Core-collapse Supernova Equations of State Based on Neutron Star Observations}},\ }\href {https://doi.org/10.1088/0004-637X/774/1/17} {\bibfield  {journal} {\bibinfo  {journal} {\apj}\ }\textbf {\bibinfo {volume} {774}},\ \bibinfo {eid} {17} (\bibinfo {year} {2013})},\ \Eprint {https://arxiv.org/abs/1207.2184} {arXiv:1207.2184 [astro-ph.SR]} \BibitemShut {NoStop}%
\bibitem [{\citenamefont {{Shen}}\ \emph {et~al.}(2011{\natexlab{a}})\citenamefont {{Shen}}, \citenamefont {{Horowitz}},\ and\ \citenamefont {{O'Connor}}}]{gshen:11b}%
  \BibitemOpen
  \bibfield  {author} {\bibinfo {author} {\bibfnamefont {G.}~\bibnamefont {{Shen}}}, \bibinfo {author} {\bibfnamefont {C.~J.}\ \bibnamefont {{Horowitz}}},\ and\ \bibinfo {author} {\bibfnamefont {E.}~\bibnamefont {{O'Connor}}},\ }\bibfield  {title} {\bibinfo {title} {{Second relativistic mean field and virial equation of state for astrophysical simulations}},\ }\href {https://doi.org/10.1103/PhysRevC.83.065808} {\bibfield  {journal} {\bibinfo  {journal} {Phys. Rev. C}\ }\textbf {\bibinfo {volume} {83}},\ \bibinfo {eid} {065808} (\bibinfo {year} {2011}{\natexlab{a}})}\BibitemShut {NoStop}%
\bibitem [{\citenamefont {{Shen}}\ \emph {et~al.}(2011{\natexlab{b}})\citenamefont {{Shen}}, \citenamefont {{Horowitz}},\ and\ \citenamefont {{Teige}}}]{gshennl3}%
  \BibitemOpen
  \bibfield  {author} {\bibinfo {author} {\bibfnamefont {G.}~\bibnamefont {{Shen}}}, \bibinfo {author} {\bibfnamefont {C.~J.}\ \bibnamefont {{Horowitz}}},\ and\ \bibinfo {author} {\bibfnamefont {S.}~\bibnamefont {{Teige}}},\ }\bibfield  {title} {\bibinfo {title} {{New equation of state for astrophysical simulations}},\ }\href {https://doi.org/10.1103/PhysRevC.83.035802} {\bibfield  {journal} {\bibinfo  {journal} {\prc}\ }\textbf {\bibinfo {volume} {83}},\ \bibinfo {eid} {035802} (\bibinfo {year} {2011}{\natexlab{b}})},\ \Eprint {https://arxiv.org/abs/1101.3715} {arXiv:1101.3715 [astro-ph.SR]} \BibitemShut {NoStop}%
\bibitem [{\citenamefont {{Shen}}\ \emph {et~al.}(1998{\natexlab{a}})\citenamefont {{Shen}}, \citenamefont {{Toki}}, \citenamefont {{Oyamatsu}},\ and\ \citenamefont {{Sumiyoshi}}}]{hsheneos1}%
  \BibitemOpen
  \bibfield  {author} {\bibinfo {author} {\bibfnamefont {H.}~\bibnamefont {{Shen}}}, \bibinfo {author} {\bibfnamefont {H.}~\bibnamefont {{Toki}}}, \bibinfo {author} {\bibfnamefont {K.}~\bibnamefont {{Oyamatsu}}},\ and\ \bibinfo {author} {\bibfnamefont {K.}~\bibnamefont {{Sumiyoshi}}},\ }\bibfield  {title} {\bibinfo {title} {{Relativistic equation of state of nuclear matter for supernova and neutron star}},\ }\href {https://doi.org/10.1016/S0375-9474(98)00236-X} {\bibfield  {journal} {\bibinfo  {journal} {\nphysa}\ }\textbf {\bibinfo {volume} {637}},\ \bibinfo {pages} {435} (\bibinfo {year} {1998}{\natexlab{a}})},\ \Eprint {https://arxiv.org/abs/nucl-th/9805035} {arXiv:nucl-th/9805035 [nucl-th]} \BibitemShut {NoStop}%
\bibitem [{\citenamefont {{Shen}}\ \emph {et~al.}(1998{\natexlab{b}})\citenamefont {{Shen}}, \citenamefont {{Toki}}, \citenamefont {{Oyamatsu}},\ and\ \citenamefont {{Sumiyoshi}}}]{hsheneos2}%
  \BibitemOpen
  \bibfield  {author} {\bibinfo {author} {\bibfnamefont {H.}~\bibnamefont {{Shen}}}, \bibinfo {author} {\bibfnamefont {H.}~\bibnamefont {{Toki}}}, \bibinfo {author} {\bibfnamefont {K.}~\bibnamefont {{Oyamatsu}}},\ and\ \bibinfo {author} {\bibfnamefont {K.}~\bibnamefont {{Sumiyoshi}}},\ }\bibfield  {title} {\bibinfo {title} {{Relativistic Equation of State of Nuclear Matter for Supernova Explosion}},\ }\href {https://doi.org/10.1143/PTP.100.1013} {\bibfield  {journal} {\bibinfo  {journal} {Progress of Theoretical Physics}\ }\textbf {\bibinfo {volume} {100}},\ \bibinfo {pages} {1013} (\bibinfo {year} {1998}{\natexlab{b}})},\ \Eprint {https://arxiv.org/abs/nucl-th/9806095} {arXiv:nucl-th/9806095 [nucl-th]} \BibitemShut {NoStop}%
\bibitem [{\citenamefont {{Shen}}\ \emph {et~al.}(2011{\natexlab{c}})\citenamefont {{Shen}}, \citenamefont {{Toki}}, \citenamefont {{Oyamatsu}},\ and\ \citenamefont {{Sumiyoshi}}}]{hshenheos3}%
  \BibitemOpen
  \bibfield  {author} {\bibinfo {author} {\bibfnamefont {H.}~\bibnamefont {{Shen}}}, \bibinfo {author} {\bibfnamefont {H.}~\bibnamefont {{Toki}}}, \bibinfo {author} {\bibfnamefont {K.}~\bibnamefont {{Oyamatsu}}},\ and\ \bibinfo {author} {\bibfnamefont {K.}~\bibnamefont {{Sumiyoshi}}},\ }\bibfield  {title} {\bibinfo {title} {{Relativistic Equation of State for Core-collapse Supernova Simulations}},\ }\href {https://doi.org/10.1088/0067-0049/197/2/20} {\bibfield  {journal} {\bibinfo  {journal} {\apjs}\ }\textbf {\bibinfo {volume} {197}},\ \bibinfo {eid} {20} (\bibinfo {year} {2011}{\natexlab{c}})},\ \Eprint {https://arxiv.org/abs/1105.1666} {arXiv:1105.1666 [astro-ph.HE]} \BibitemShut {NoStop}%
\bibitem [{\citenamefont {{Lattimer}}\ and\ \citenamefont {{Swesty}}(1991)}]{lseos:91}%
  \BibitemOpen
  \bibfield  {author} {\bibinfo {author} {\bibfnamefont {J.~M.}\ \bibnamefont {{Lattimer}}}\ and\ \bibinfo {author} {\bibfnamefont {F.~D.}\ \bibnamefont {{Swesty}}},\ }\bibfield  {title} {\bibinfo {title} {{A generalized equation of state for hot, dense matter}},\ }\href {https://doi.org/10.1016/0375-9474(91)90452-C} {\bibfield  {journal} {\bibinfo  {journal} {Nucl. Phys. A}\ }\textbf {\bibinfo {volume} {535}},\ \bibinfo {pages} {331} (\bibinfo {year} {1991})}\BibitemShut {NoStop}%
\bibitem [{\citenamefont {{Dimmelmeier}}\ \emph {et~al.}(2002)\citenamefont {{Dimmelmeier}}, \citenamefont {{Font}},\ and\ \citenamefont {{M{\"u}ller}}}]{Dimmelmeier02a}%
  \BibitemOpen
  \bibfield  {author} {\bibinfo {author} {\bibfnamefont {H.}~\bibnamefont {{Dimmelmeier}}}, \bibinfo {author} {\bibfnamefont {J.~A.}\ \bibnamefont {{Font}}},\ and\ \bibinfo {author} {\bibfnamefont {E.}~\bibnamefont {{M{\"u}ller}}},\ }\bibfield  {title} {\bibinfo {title} {{Relativistic simulations of rotational core collapse I. Methods, initial models, and code tests}},\ }\href {https://doi.org/10.1051/0004-6361:20020563} {\bibfield  {journal} {\bibinfo  {journal} {A\&A}\ }\textbf {\bibinfo {volume} {388}},\ \bibinfo {pages} {917} (\bibinfo {year} {2002})},\ \Eprint {https://arxiv.org/abs/astro-ph/0204288} {arXiv:astro-ph/0204288 [astro-ph]} \BibitemShut {NoStop}%
\bibitem [{\citenamefont {{Dimmelmeier}}\ \emph {et~al.}(2005)\citenamefont {{Dimmelmeier}}, \citenamefont {{Novak}}, \citenamefont {{Font}}, \citenamefont {{Ib{\'a}{\~n}ez}},\ and\ \citenamefont {{M{\"u}ller}}}]{dimmelmeier:05MdM}%
  \BibitemOpen
  \bibfield  {author} {\bibinfo {author} {\bibfnamefont {H.}~\bibnamefont {{Dimmelmeier}}}, \bibinfo {author} {\bibfnamefont {J.}~\bibnamefont {{Novak}}}, \bibinfo {author} {\bibfnamefont {J.~A.}\ \bibnamefont {{Font}}}, \bibinfo {author} {\bibfnamefont {J.~M.}\ \bibnamefont {{Ib{\'a}{\~n}ez}}},\ and\ \bibinfo {author} {\bibfnamefont {E.}~\bibnamefont {{M{\"u}ller}}},\ }\bibfield  {title} {\bibinfo {title} {{Combining spectral and shock-capturing methods: A new numerical approach for 3D relativistic core collapse simulations}},\ }\href {https://doi.org/10.1103/PhysRevD.71.064023} {\bibfield  {journal} {\bibinfo  {journal} {Phys. Rev. D}\ }\textbf {\bibinfo {volume} {71}},\ \bibinfo {pages} {064023} (\bibinfo {year} {2005})},\ \Eprint {https://arxiv.org/abs/astro-ph/0407174} {astro-ph/0407174} \BibitemShut {NoStop}%
\bibitem [{\citenamefont {{Raynaud}}\ \emph {et~al.}(2022)\citenamefont {{Raynaud}}, \citenamefont {{Cerd{\'a}-Dur{\'a}n}},\ and\ \citenamefont {{Guilet}}}]{Raynaud22GW}%
  \BibitemOpen
  \bibfield  {author} {\bibinfo {author} {\bibfnamefont {R.}~\bibnamefont {{Raynaud}}}, \bibinfo {author} {\bibfnamefont {P.}~\bibnamefont {{Cerd{\'a}-Dur{\'a}n}}},\ and\ \bibinfo {author} {\bibfnamefont {J.}~\bibnamefont {{Guilet}}},\ }\bibfield  {title} {\bibinfo {title} {{Gravitational wave signature of proto-neutron star convection: I. MHD numerical simulations}},\ }\href {https://doi.org/10.1093/mnras/stab3109} {\bibfield  {journal} {\bibinfo  {journal} {\mnras}\ }\textbf {\bibinfo {volume} {509}},\ \bibinfo {pages} {3410} (\bibinfo {year} {2022})},\ \Eprint {https://arxiv.org/abs/2103.12445} {arXiv:2103.12445 [astro-ph.SR]} \BibitemShut {NoStop}%
\bibitem [{\citenamefont {Barsotti}\ \emph {et~al.}(2018)\citenamefont {Barsotti}, \citenamefont {McCuller}, \citenamefont {Evans},\ and\ \citenamefont {Fritschel}}]{barsotti2018a+}%
  \BibitemOpen
  \bibfield  {author} {\bibinfo {author} {\bibfnamefont {L.}~\bibnamefont {Barsotti}}, \bibinfo {author} {\bibfnamefont {L.}~\bibnamefont {McCuller}}, \bibinfo {author} {\bibfnamefont {M.}~\bibnamefont {Evans}},\ and\ \bibinfo {author} {\bibfnamefont {P.}~\bibnamefont {Fritschel}},\ }\bibfield  {title} {\bibinfo {title} {The a+ design curve},\ }\href@noop {} {\bibfield  {journal} {\bibinfo  {journal} {LIGO Document}\ }\textbf {\bibinfo {volume} {1800042}},\ \bibinfo {pages} {2018} (\bibinfo {year} {2018})}\BibitemShut {NoStop}%
\bibitem [{\citenamefont {{Abbott}}\ \emph {et~al.}(2017)\citenamefont {{Abbott}}, \citenamefont {{Abbott}}, \citenamefont {{Abbott}}, \citenamefont {{Abernathy}}, \citenamefont {{Ackley}}, \citenamefont {{Adams}}, \citenamefont {{Addesso}}, \citenamefont {{Adhikari}}, \citenamefont {{Adya}}, \citenamefont {{Affeldt}}, \citenamefont {{Aggarwal}}, \citenamefont {{Aguiar}}, \citenamefont {{Ain}}, \citenamefont {{Ajith}}, \citenamefont {{Allen}}, \citenamefont {{Altin}}, \citenamefont {{Anderson}}, \citenamefont {{Anderson}}, \citenamefont {{Arai}}, \citenamefont {{Araya}}, \citenamefont {{Arceneaux}}, \citenamefont {{Areeda}}, \citenamefont {{Arun}}, \citenamefont {{Ashton}}, \citenamefont {{Ast}}, \citenamefont {{Aston}}, \citenamefont {{Aufmuth}}, \citenamefont {{Aulbert}}, \citenamefont {{Babak}}, \citenamefont {{Baker}}, \citenamefont {{Ballmer}}, \citenamefont {{Barayoga}}, \citenamefont {{Barclay}}, \citenamefont {{Barish}}, \citenamefont {{Barker}}, \citenamefont {{Barr}}, \citenamefont {{Barsotti}},
  \citenamefont {{Bartlett}}, \citenamefont {{Bartos}}, \citenamefont {{Bassiri}}, \citenamefont {{Batch}}, \citenamefont {{Baune}}, \citenamefont {{Bell}}, \citenamefont {{Berger}}, \citenamefont {{Bergmann}}, \citenamefont {{Berry}}, \citenamefont {{Betzwieser}}, \citenamefont {{Bhagwat}}, \citenamefont {{Bhandare}}, \citenamefont {{Bilenko}}, \citenamefont {{Billingsley}}, \citenamefont {{Birch}}, \citenamefont {{Birney}}, \citenamefont {{Biscans}}, \citenamefont {{Bisht}}, \citenamefont {{Biwer}}, \citenamefont {{Blackburn}}, \citenamefont {{Blair}}, \citenamefont {{Blair}}, \citenamefont {{Blair}}, \citenamefont {{Bock}}, \citenamefont {{Bogan}}, \citenamefont {{Bohe}}, \citenamefont {{Bond}}, \citenamefont {{Bork}}, \citenamefont {{Bose}}, \citenamefont {{Brady}}, \citenamefont {{Braginsky}}, \citenamefont {{Brau}}, \citenamefont {{Brinkmann}}, \citenamefont {{Brockill}}, \citenamefont {{Broida}}, \citenamefont {{Brooks}}, \citenamefont {{Brown}}, \citenamefont {{Brown}}, \citenamefont {{Brown}},
  \citenamefont {{Brunett}}, \citenamefont {{Buchanan}}, \citenamefont {{Buikema}}, \citenamefont {{Buonanno}}, \citenamefont {{Byer}}, \citenamefont {{Cabero}}, \citenamefont {{Cadonati}}, \citenamefont {{Cahillane}}, \citenamefont {{Calder{\'o}n Bustillo}}, \citenamefont {{Callister}}, \citenamefont {{Camp}}, \citenamefont {{Cannon}}, \citenamefont {{Cao}}, \citenamefont {{Capano}}, \citenamefont {{Caride}}, \citenamefont {{Caudill}}, \citenamefont {{Cavagli{\`a}}}, \citenamefont {{Cepeda}}, \citenamefont {{Chamberlin}}, \citenamefont {{Chan}}, \citenamefont {{Chao}}, \citenamefont {{Charlton}}, \citenamefont {{Cheeseboro}}, \citenamefont {{Chen}}, \citenamefont {{Chen}}, \citenamefont {{Cheng}}, \citenamefont {{Cho}}, \citenamefont {{Cho}}, \citenamefont {{Chow}}, \citenamefont {{Christensen}}, \citenamefont {{Chu}}, \citenamefont {{Chung}}, \citenamefont {{Ciani}}, \citenamefont {{Clara}}, \citenamefont {{Clark}}, \citenamefont {{Collette}}, \citenamefont {{Cominsky}}, \citenamefont {{Constancio}},
  \citenamefont {{Cook}}, \citenamefont {{Corbitt}}, \citenamefont {{Cornish}}, \citenamefont {{Corsi}}, \citenamefont {{Costa}}, \citenamefont {{Coughlin}}, \citenamefont {{Coughlin}}, \citenamefont {{Countryman}}, \citenamefont {{Couvares}}, \citenamefont {{Cowan}}, \citenamefont {{Coward}}, \citenamefont {{Cowart}}, \citenamefont {{Coyne}}, \citenamefont {{Coyne}}, \citenamefont {{Craig}}, \citenamefont {{Creighton}}, \citenamefont {{Cripe}}, \citenamefont {{Crowder}}, \citenamefont {{Cumming}}, \citenamefont {{Cunningham}}, \citenamefont {{Dal Canton}}, \citenamefont {{Danilishin}}, \citenamefont {{Danzmann}}, \citenamefont {{Darman}}, \citenamefont {{Dasgupta}}, \citenamefont {{Da Silva Costa}}, \citenamefont {{Dave}}, \citenamefont {{Davies}}, \citenamefont {{Daw}}, \citenamefont {{De}}, \citenamefont {{DeBra}}, \citenamefont {{Del Pozzo}}, \citenamefont {{Denker}}, \citenamefont {{Dent}}, \citenamefont {{Dergachev}}, \citenamefont {{DeRosa}}, \citenamefont {{DeSalvo}}, \citenamefont {{Devine}},
  \citenamefont {{Dhurandhar}}, \citenamefont {{D{\'\i}az}}, \citenamefont {{Di Palma}}, \citenamefont {{Donovan}}, \citenamefont {{Dooley}}, \citenamefont {{Doravari}}, \citenamefont {{Douglas}}, \citenamefont {{Downes}}, \citenamefont {{Drago}}, \citenamefont {{Drever}}, \citenamefont {{Driggers}}, \citenamefont {{Dwyer}}, \citenamefont {{Edo}}, \citenamefont {{Edwards}}, \citenamefont {{Effler}}, \citenamefont {{Eggenstein}}, \citenamefont {{Ehrens}}, \citenamefont {{Eichholz}}, \citenamefont {{Eikenberry}}, \citenamefont {{Engels}}, \citenamefont {{Essick}}, \citenamefont {{Etzel}}, \citenamefont {{Evans}}, \citenamefont {{Evans}}, \citenamefont {{Everett}}, \citenamefont {{Factourovich}}, \citenamefont {{Fair}}, \citenamefont {{Fairhurst}}, \citenamefont {{Fan}}, \citenamefont {{Fang}}, \citenamefont {{Farr}}, \citenamefont {{Farr}}, \citenamefont {{Favata}}, \citenamefont {{Fays}}, \citenamefont {{Fehrmann}}, \citenamefont {{Fejer}}, \citenamefont {{Fenyvesi}}, \citenamefont {{Ferreira}}, \citenamefont
  {{Fisher}}, \citenamefont {{Fletcher}}, \citenamefont {{Frei}}, \citenamefont {{Freise}}, \citenamefont {{Frey}}, \citenamefont {{Fritschel}}, \citenamefont {{Frolov}}, \citenamefont {{Fulda}}, \citenamefont {{Fyffe}},\ and\ \citenamefont {{Gabbard}}}]{ET_CE_noise}%
  \BibitemOpen
  \bibfield  {author} {\bibinfo {author} {\bibfnamefont {B.~P.}\ \bibnamefont {{Abbott}}}, \bibinfo {author} {\bibfnamefont {R.}~\bibnamefont {{Abbott}}}, \bibinfo {author} {\bibfnamefont {T.~D.}\ \bibnamefont {{Abbott}}}, \bibinfo {author} {\bibfnamefont {M.~R.}\ \bibnamefont {{Abernathy}}}, \bibinfo {author} {\bibfnamefont {K.}~\bibnamefont {{Ackley}}}, \bibinfo {author} {\bibfnamefont {C.}~\bibnamefont {{Adams}}}, \bibinfo {author} {\bibfnamefont {P.}~\bibnamefont {{Addesso}}}, \bibinfo {author} {\bibfnamefont {R.~X.}\ \bibnamefont {{Adhikari}}}, \bibinfo {author} {\bibfnamefont {V.~B.}\ \bibnamefont {{Adya}}}, \bibinfo {author} {\bibfnamefont {C.}~\bibnamefont {{Affeldt}}}, \bibinfo {author} {\bibfnamefont {N.}~\bibnamefont {{Aggarwal}}}, \bibinfo {author} {\bibfnamefont {O.~D.}\ \bibnamefont {{Aguiar}}}, \bibinfo {author} {\bibfnamefont {A.}~\bibnamefont {{Ain}}}, \bibinfo {author} {\bibfnamefont {P.}~\bibnamefont {{Ajith}}}, \bibinfo {author} {\bibfnamefont {B.}~\bibnamefont {{Allen}}}, \bibinfo
  {author} {\bibfnamefont {P.~A.}\ \bibnamefont {{Altin}}}, \bibinfo {author} {\bibfnamefont {S.~B.}\ \bibnamefont {{Anderson}}}, \bibinfo {author} {\bibfnamefont {W.~G.}\ \bibnamefont {{Anderson}}}, \bibinfo {author} {\bibfnamefont {K.}~\bibnamefont {{Arai}}}, \bibinfo {author} {\bibfnamefont {M.~C.}\ \bibnamefont {{Araya}}}, \bibinfo {author} {\bibfnamefont {C.~C.}\ \bibnamefont {{Arceneaux}}}, \bibinfo {author} {\bibfnamefont {J.~S.}\ \bibnamefont {{Areeda}}}, \bibinfo {author} {\bibfnamefont {K.~G.}\ \bibnamefont {{Arun}}}, \bibinfo {author} {\bibfnamefont {G.}~\bibnamefont {{Ashton}}}, \bibinfo {author} {\bibfnamefont {M.}~\bibnamefont {{Ast}}}, \bibinfo {author} {\bibfnamefont {S.~M.}\ \bibnamefont {{Aston}}}, \bibinfo {author} {\bibfnamefont {P.}~\bibnamefont {{Aufmuth}}}, \bibinfo {author} {\bibfnamefont {C.}~\bibnamefont {{Aulbert}}}, \bibinfo {author} {\bibfnamefont {S.}~\bibnamefont {{Babak}}}, \bibinfo {author} {\bibfnamefont {P.~T.}\ \bibnamefont {{Baker}}}, \bibinfo {author} {\bibfnamefont
  {S.~W.}\ \bibnamefont {{Ballmer}}}, \bibinfo {author} {\bibfnamefont {J.~C.}\ \bibnamefont {{Barayoga}}}, \bibinfo {author} {\bibfnamefont {S.~E.}\ \bibnamefont {{Barclay}}}, \bibinfo {author} {\bibfnamefont {B.~C.}\ \bibnamefont {{Barish}}}, \bibinfo {author} {\bibfnamefont {D.}~\bibnamefont {{Barker}}}, \bibinfo {author} {\bibfnamefont {B.}~\bibnamefont {{Barr}}}, \bibinfo {author} {\bibfnamefont {L.}~\bibnamefont {{Barsotti}}}, \bibinfo {author} {\bibfnamefont {J.}~\bibnamefont {{Bartlett}}}, \bibinfo {author} {\bibfnamefont {I.}~\bibnamefont {{Bartos}}}, \bibinfo {author} {\bibfnamefont {R.}~\bibnamefont {{Bassiri}}}, \bibinfo {author} {\bibfnamefont {J.~C.}\ \bibnamefont {{Batch}}}, \bibinfo {author} {\bibfnamefont {C.}~\bibnamefont {{Baune}}}, \bibinfo {author} {\bibfnamefont {A.~S.}\ \bibnamefont {{Bell}}}, \bibinfo {author} {\bibfnamefont {B.~K.}\ \bibnamefont {{Berger}}}, \bibinfo {author} {\bibfnamefont {G.}~\bibnamefont {{Bergmann}}}, \bibinfo {author} {\bibfnamefont {C.~P.~L.}\ \bibnamefont
  {{Berry}}}, \bibinfo {author} {\bibfnamefont {J.}~\bibnamefont {{Betzwieser}}}, \bibinfo {author} {\bibfnamefont {S.}~\bibnamefont {{Bhagwat}}}, \bibinfo {author} {\bibfnamefont {R.}~\bibnamefont {{Bhandare}}}, \bibinfo {author} {\bibfnamefont {I.~A.}\ \bibnamefont {{Bilenko}}}, \bibinfo {author} {\bibfnamefont {G.}~\bibnamefont {{Billingsley}}}, \bibinfo {author} {\bibfnamefont {J.}~\bibnamefont {{Birch}}}, \bibinfo {author} {\bibfnamefont {R.}~\bibnamefont {{Birney}}}, \bibinfo {author} {\bibfnamefont {S.}~\bibnamefont {{Biscans}}}, \bibinfo {author} {\bibfnamefont {A.}~\bibnamefont {{Bisht}}}, \bibinfo {author} {\bibfnamefont {C.}~\bibnamefont {{Biwer}}}, \bibinfo {author} {\bibfnamefont {J.~K.}\ \bibnamefont {{Blackburn}}}, \bibinfo {author} {\bibfnamefont {C.~D.}\ \bibnamefont {{Blair}}}, \bibinfo {author} {\bibfnamefont {D.~G.}\ \bibnamefont {{Blair}}}, \bibinfo {author} {\bibfnamefont {R.~M.}\ \bibnamefont {{Blair}}}, \bibinfo {author} {\bibfnamefont {O.}~\bibnamefont {{Bock}}}, \bibinfo {author}
  {\bibfnamefont {C.}~\bibnamefont {{Bogan}}}, \bibinfo {author} {\bibfnamefont {A.}~\bibnamefont {{Bohe}}}, \bibinfo {author} {\bibfnamefont {C.}~\bibnamefont {{Bond}}}, \bibinfo {author} {\bibfnamefont {R.}~\bibnamefont {{Bork}}}, \bibinfo {author} {\bibfnamefont {S.}~\bibnamefont {{Bose}}}, \bibinfo {author} {\bibfnamefont {P.~R.}\ \bibnamefont {{Brady}}}, \bibinfo {author} {\bibfnamefont {V.~B.}\ \bibnamefont {{Braginsky}}}, \bibinfo {author} {\bibfnamefont {J.~E.}\ \bibnamefont {{Brau}}}, \bibinfo {author} {\bibfnamefont {M.}~\bibnamefont {{Brinkmann}}}, \bibinfo {author} {\bibfnamefont {P.}~\bibnamefont {{Brockill}}}, \bibinfo {author} {\bibfnamefont {J.~E.}\ \bibnamefont {{Broida}}}, \bibinfo {author} {\bibfnamefont {A.~F.}\ \bibnamefont {{Brooks}}}, \bibinfo {author} {\bibfnamefont {D.~A.}\ \bibnamefont {{Brown}}}, \bibinfo {author} {\bibfnamefont {D.~D.}\ \bibnamefont {{Brown}}}, \bibinfo {author} {\bibfnamefont {N.~M.}\ \bibnamefont {{Brown}}}, \bibinfo {author} {\bibfnamefont {S.}~\bibnamefont
  {{Brunett}}}, \bibinfo {author} {\bibfnamefont {C.~C.}\ \bibnamefont {{Buchanan}}}, \bibinfo {author} {\bibfnamefont {A.}~\bibnamefont {{Buikema}}}, \bibinfo {author} {\bibfnamefont {A.}~\bibnamefont {{Buonanno}}}, \bibinfo {author} {\bibfnamefont {R.~L.}\ \bibnamefont {{Byer}}}, \bibinfo {author} {\bibfnamefont {M.}~\bibnamefont {{Cabero}}}, \bibinfo {author} {\bibfnamefont {L.}~\bibnamefont {{Cadonati}}}, \bibinfo {author} {\bibfnamefont {C.}~\bibnamefont {{Cahillane}}}, \bibinfo {author} {\bibfnamefont {J.}~\bibnamefont {{Calder{\'o}n Bustillo}}}, \bibinfo {author} {\bibfnamefont {T.}~\bibnamefont {{Callister}}}, \bibinfo {author} {\bibfnamefont {J.~B.}\ \bibnamefont {{Camp}}}, \bibinfo {author} {\bibfnamefont {K.~C.}\ \bibnamefont {{Cannon}}}, \bibinfo {author} {\bibfnamefont {J.}~\bibnamefont {{Cao}}}, \bibinfo {author} {\bibfnamefont {C.~D.}\ \bibnamefont {{Capano}}}, \bibinfo {author} {\bibfnamefont {S.}~\bibnamefont {{Caride}}}, \bibinfo {author} {\bibfnamefont {S.}~\bibnamefont {{Caudill}}},
  \bibinfo {author} {\bibfnamefont {M.}~\bibnamefont {{Cavagli{\`a}}}}, \bibinfo {author} {\bibfnamefont {C.~B.}\ \bibnamefont {{Cepeda}}}, \bibinfo {author} {\bibfnamefont {S.~J.}\ \bibnamefont {{Chamberlin}}}, \bibinfo {author} {\bibfnamefont {M.}~\bibnamefont {{Chan}}}, \bibinfo {author} {\bibfnamefont {S.}~\bibnamefont {{Chao}}}, \bibinfo {author} {\bibfnamefont {P.}~\bibnamefont {{Charlton}}}, \bibinfo {author} {\bibfnamefont {B.~D.}\ \bibnamefont {{Cheeseboro}}}, \bibinfo {author} {\bibfnamefont {H.~Y.}\ \bibnamefont {{Chen}}}, \bibinfo {author} {\bibfnamefont {Y.}~\bibnamefont {{Chen}}}, \bibinfo {author} {\bibfnamefont {C.}~\bibnamefont {{Cheng}}}, \bibinfo {author} {\bibfnamefont {H.~S.}\ \bibnamefont {{Cho}}}, \bibinfo {author} {\bibfnamefont {M.}~\bibnamefont {{Cho}}}, \bibinfo {author} {\bibfnamefont {J.~H.}\ \bibnamefont {{Chow}}}, \bibinfo {author} {\bibfnamefont {N.}~\bibnamefont {{Christensen}}}, \bibinfo {author} {\bibfnamefont {Q.}~\bibnamefont {{Chu}}}, \bibinfo {author} {\bibfnamefont
  {S.}~\bibnamefont {{Chung}}}, \bibinfo {author} {\bibfnamefont {G.}~\bibnamefont {{Ciani}}}, \bibinfo {author} {\bibfnamefont {F.}~\bibnamefont {{Clara}}}, \bibinfo {author} {\bibfnamefont {J.~A.}\ \bibnamefont {{Clark}}}, \bibinfo {author} {\bibfnamefont {C.~G.}\ \bibnamefont {{Collette}}}, \bibinfo {author} {\bibfnamefont {L.}~\bibnamefont {{Cominsky}}}, \bibinfo {author} {\bibfnamefont {M.}~\bibnamefont {{Constancio}}, \bibfnamefont {Jr.}}, \bibinfo {author} {\bibfnamefont {D.}~\bibnamefont {{Cook}}}, \bibinfo {author} {\bibfnamefont {T.~R.}\ \bibnamefont {{Corbitt}}}, \bibinfo {author} {\bibfnamefont {N.}~\bibnamefont {{Cornish}}}, \bibinfo {author} {\bibfnamefont {A.}~\bibnamefont {{Corsi}}}, \bibinfo {author} {\bibfnamefont {C.~A.}\ \bibnamefont {{Costa}}}, \bibinfo {author} {\bibfnamefont {M.~W.}\ \bibnamefont {{Coughlin}}}, \bibinfo {author} {\bibfnamefont {S.~B.}\ \bibnamefont {{Coughlin}}}, \bibinfo {author} {\bibfnamefont {S.~T.}\ \bibnamefont {{Countryman}}}, \bibinfo {author} {\bibfnamefont
  {P.}~\bibnamefont {{Couvares}}}, \bibinfo {author} {\bibfnamefont {E.~E.}\ \bibnamefont {{Cowan}}}, \bibinfo {author} {\bibfnamefont {D.~M.}\ \bibnamefont {{Coward}}}, \bibinfo {author} {\bibfnamefont {M.~J.}\ \bibnamefont {{Cowart}}}, \bibinfo {author} {\bibfnamefont {D.~C.}\ \bibnamefont {{Coyne}}}, \bibinfo {author} {\bibfnamefont {R.}~\bibnamefont {{Coyne}}}, \bibinfo {author} {\bibfnamefont {K.}~\bibnamefont {{Craig}}}, \bibinfo {author} {\bibfnamefont {J.~D.~E.}\ \bibnamefont {{Creighton}}}, \bibinfo {author} {\bibfnamefont {J.}~\bibnamefont {{Cripe}}}, \bibinfo {author} {\bibfnamefont {S.~G.}\ \bibnamefont {{Crowder}}}, \bibinfo {author} {\bibfnamefont {A.}~\bibnamefont {{Cumming}}}, \bibinfo {author} {\bibfnamefont {L.}~\bibnamefont {{Cunningham}}}, \bibinfo {author} {\bibfnamefont {T.}~\bibnamefont {{Dal Canton}}}, \bibinfo {author} {\bibfnamefont {S.~L.}\ \bibnamefont {{Danilishin}}}, \bibinfo {author} {\bibfnamefont {K.}~\bibnamefont {{Danzmann}}}, \bibinfo {author} {\bibfnamefont {N.~S.}\
  \bibnamefont {{Darman}}}, \bibinfo {author} {\bibfnamefont {A.}~\bibnamefont {{Dasgupta}}}, \bibinfo {author} {\bibfnamefont {C.~F.}\ \bibnamefont {{Da Silva Costa}}}, \bibinfo {author} {\bibfnamefont {I.}~\bibnamefont {{Dave}}}, \bibinfo {author} {\bibfnamefont {G.~S.}\ \bibnamefont {{Davies}}}, \bibinfo {author} {\bibfnamefont {E.~J.}\ \bibnamefont {{Daw}}}, \bibinfo {author} {\bibfnamefont {S.}~\bibnamefont {{De}}}, \bibinfo {author} {\bibfnamefont {D.}~\bibnamefont {{DeBra}}}, \bibinfo {author} {\bibfnamefont {W.}~\bibnamefont {{Del Pozzo}}}, \bibinfo {author} {\bibfnamefont {T.}~\bibnamefont {{Denker}}}, \bibinfo {author} {\bibfnamefont {T.}~\bibnamefont {{Dent}}}, \bibinfo {author} {\bibfnamefont {V.}~\bibnamefont {{Dergachev}}}, \bibinfo {author} {\bibfnamefont {R.~T.}\ \bibnamefont {{DeRosa}}}, \bibinfo {author} {\bibfnamefont {R.}~\bibnamefont {{DeSalvo}}}, \bibinfo {author} {\bibfnamefont {R.~C.}\ \bibnamefont {{Devine}}}, \bibinfo {author} {\bibfnamefont {S.}~\bibnamefont {{Dhurandhar}}},
  \bibinfo {author} {\bibfnamefont {M.~C.}\ \bibnamefont {{D{\'\i}az}}}, \bibinfo {author} {\bibfnamefont {I.}~\bibnamefont {{Di Palma}}}, \bibinfo {author} {\bibfnamefont {F.}~\bibnamefont {{Donovan}}}, \bibinfo {author} {\bibfnamefont {K.~L.}\ \bibnamefont {{Dooley}}}, \bibinfo {author} {\bibfnamefont {S.}~\bibnamefont {{Doravari}}}, \bibinfo {author} {\bibfnamefont {R.}~\bibnamefont {{Douglas}}}, \bibinfo {author} {\bibfnamefont {T.~P.}\ \bibnamefont {{Downes}}}, \bibinfo {author} {\bibfnamefont {M.}~\bibnamefont {{Drago}}}, \bibinfo {author} {\bibfnamefont {R.~W.~P.}\ \bibnamefont {{Drever}}}, \bibinfo {author} {\bibfnamefont {J.~C.}\ \bibnamefont {{Driggers}}}, \bibinfo {author} {\bibfnamefont {S.~E.}\ \bibnamefont {{Dwyer}}}, \bibinfo {author} {\bibfnamefont {T.~B.}\ \bibnamefont {{Edo}}}, \bibinfo {author} {\bibfnamefont {M.~C.}\ \bibnamefont {{Edwards}}}, \bibinfo {author} {\bibfnamefont {A.}~\bibnamefont {{Effler}}}, \bibinfo {author} {\bibfnamefont {H.~B.}\ \bibnamefont {{Eggenstein}}}, \bibinfo
  {author} {\bibfnamefont {P.}~\bibnamefont {{Ehrens}}}, \bibinfo {author} {\bibfnamefont {J.}~\bibnamefont {{Eichholz}}}, \bibinfo {author} {\bibfnamefont {S.~S.}\ \bibnamefont {{Eikenberry}}}, \bibinfo {author} {\bibfnamefont {W.}~\bibnamefont {{Engels}}}, \bibinfo {author} {\bibfnamefont {R.~C.}\ \bibnamefont {{Essick}}}, \bibinfo {author} {\bibfnamefont {T.}~\bibnamefont {{Etzel}}}, \bibinfo {author} {\bibfnamefont {M.}~\bibnamefont {{Evans}}}, \bibinfo {author} {\bibfnamefont {T.~M.}\ \bibnamefont {{Evans}}}, \bibinfo {author} {\bibfnamefont {R.}~\bibnamefont {{Everett}}}, \bibinfo {author} {\bibfnamefont {M.}~\bibnamefont {{Factourovich}}}, \bibinfo {author} {\bibfnamefont {H.}~\bibnamefont {{Fair}}}, \bibinfo {author} {\bibfnamefont {S.}~\bibnamefont {{Fairhurst}}}, \bibinfo {author} {\bibfnamefont {X.}~\bibnamefont {{Fan}}}, \bibinfo {author} {\bibfnamefont {Q.}~\bibnamefont {{Fang}}}, \bibinfo {author} {\bibfnamefont {B.}~\bibnamefont {{Farr}}}, \bibinfo {author} {\bibfnamefont {W.~M.}\ \bibnamefont
  {{Farr}}}, \bibinfo {author} {\bibfnamefont {M.}~\bibnamefont {{Favata}}}, \bibinfo {author} {\bibfnamefont {M.}~\bibnamefont {{Fays}}}, \bibinfo {author} {\bibfnamefont {H.}~\bibnamefont {{Fehrmann}}}, \bibinfo {author} {\bibfnamefont {M.~M.}\ \bibnamefont {{Fejer}}}, \bibinfo {author} {\bibfnamefont {E.}~\bibnamefont {{Fenyvesi}}}, \bibinfo {author} {\bibfnamefont {E.~C.}\ \bibnamefont {{Ferreira}}}, \bibinfo {author} {\bibfnamefont {R.~P.}\ \bibnamefont {{Fisher}}}, \bibinfo {author} {\bibfnamefont {M.}~\bibnamefont {{Fletcher}}}, \bibinfo {author} {\bibfnamefont {Z.}~\bibnamefont {{Frei}}}, \bibinfo {author} {\bibfnamefont {A.}~\bibnamefont {{Freise}}}, \bibinfo {author} {\bibfnamefont {R.}~\bibnamefont {{Frey}}}, \bibinfo {author} {\bibfnamefont {P.}~\bibnamefont {{Fritschel}}}, \bibinfo {author} {\bibfnamefont {V.~V.}\ \bibnamefont {{Frolov}}}, \bibinfo {author} {\bibfnamefont {P.}~\bibnamefont {{Fulda}}}, \bibinfo {author} {\bibfnamefont {M.}~\bibnamefont {{Fyffe}}},\ and\ \bibinfo {author}
  {\bibfnamefont {H.~A.~G.}\ \bibnamefont {{Gabbard}}},\ }\bibfield  {title} {\bibinfo {title} {{Exploring the sensitivity of next generation gravitational wave detectors}},\ }\href {https://doi.org/10.1088/1361-6382/aa51f4} {\bibfield  {journal} {\bibinfo  {journal} {Classical and Quantum Gravity}\ }\textbf {\bibinfo {volume} {34}},\ \bibinfo {eid} {044001} (\bibinfo {year} {2017})},\ \Eprint {https://arxiv.org/abs/1607.08697} {arXiv:1607.08697 [astro-ph.IM]} \BibitemShut {NoStop}%
\bibitem [{\citenamefont {Flanagan}\ and\ \citenamefont {Hughes}(1998)}]{flanagan98}%
  \BibitemOpen
  \bibfield  {author} {\bibinfo {author} {\bibfnamefont {E.~E.}\ \bibnamefont {Flanagan}}\ and\ \bibinfo {author} {\bibfnamefont {S.~A.}\ \bibnamefont {Hughes}},\ }\bibfield  {title} {\bibinfo {title} {Measuring gravitational waves from binary black hole coalescences. i. signal to noise for inspiral, merger, and ringdown},\ }\href {https://doi.org/10.1103/PhysRevD.57.4535} {\bibfield  {journal} {\bibinfo  {journal} {Phys. Rev. D}\ }\textbf {\bibinfo {volume} {57}},\ \bibinfo {pages} {4535} (\bibinfo {year} {1998})}\BibitemShut {NoStop}%
\bibitem [{\citenamefont {Nitz}\ \emph {et~al.}(2024)\citenamefont {Nitz}, \citenamefont {Harry}, \citenamefont {Brown}, \citenamefont {Biwer}, \citenamefont {Willis}, \citenamefont {Canton}, \citenamefont {Capano}, \citenamefont {Dent}, \citenamefont {Pekowsky}, \citenamefont {Davies}, \citenamefont {De}, \citenamefont {Cabero}, \citenamefont {Wu}, \citenamefont {Williamson}, \citenamefont {Machenschalk}, \citenamefont {Macleod}, \citenamefont {Pannarale}, \citenamefont {Kumar}, \citenamefont {Reyes}, \citenamefont {dfinstad}, \citenamefont {Kumar}, \citenamefont {Tápai}, \citenamefont {Singer}, \citenamefont {Kumar}, \citenamefont {veronica villa}, \citenamefont {maxtrevor}, \citenamefont {Gadre}, \citenamefont {Khan}, \citenamefont {Fairhurst},\ and\ \citenamefont {Tolley}}]{pycbc}%
  \BibitemOpen
  \bibfield  {author} {\bibinfo {author} {\bibfnamefont {A.}~\bibnamefont {Nitz}}, \bibinfo {author} {\bibfnamefont {I.}~\bibnamefont {Harry}}, \bibinfo {author} {\bibfnamefont {D.}~\bibnamefont {Brown}}, \bibinfo {author} {\bibfnamefont {C.~M.}\ \bibnamefont {Biwer}}, \bibinfo {author} {\bibfnamefont {J.}~\bibnamefont {Willis}}, \bibinfo {author} {\bibfnamefont {T.~D.}\ \bibnamefont {Canton}}, \bibinfo {author} {\bibfnamefont {C.}~\bibnamefont {Capano}}, \bibinfo {author} {\bibfnamefont {T.}~\bibnamefont {Dent}}, \bibinfo {author} {\bibfnamefont {L.}~\bibnamefont {Pekowsky}}, \bibinfo {author} {\bibfnamefont {G.~S.~C.}\ \bibnamefont {Davies}}, \bibinfo {author} {\bibfnamefont {S.}~\bibnamefont {De}}, \bibinfo {author} {\bibfnamefont {M.}~\bibnamefont {Cabero}}, \bibinfo {author} {\bibfnamefont {S.}~\bibnamefont {Wu}}, \bibinfo {author} {\bibfnamefont {A.~R.}\ \bibnamefont {Williamson}}, \bibinfo {author} {\bibfnamefont {B.}~\bibnamefont {Machenschalk}}, \bibinfo {author} {\bibfnamefont {D.}~\bibnamefont
  {Macleod}}, \bibinfo {author} {\bibfnamefont {F.}~\bibnamefont {Pannarale}}, \bibinfo {author} {\bibfnamefont {P.}~\bibnamefont {Kumar}}, \bibinfo {author} {\bibfnamefont {S.}~\bibnamefont {Reyes}}, \bibinfo {author} {\bibnamefont {dfinstad}}, \bibinfo {author} {\bibfnamefont {S.}~\bibnamefont {Kumar}}, \bibinfo {author} {\bibfnamefont {M.}~\bibnamefont {Tápai}}, \bibinfo {author} {\bibfnamefont {L.}~\bibnamefont {Singer}}, \bibinfo {author} {\bibfnamefont {P.}~\bibnamefont {Kumar}}, \bibinfo {author} {\bibnamefont {veronica villa}}, \bibinfo {author} {\bibnamefont {maxtrevor}}, \bibinfo {author} {\bibfnamefont {B.~U.~V.}\ \bibnamefont {Gadre}}, \bibinfo {author} {\bibfnamefont {S.}~\bibnamefont {Khan}}, \bibinfo {author} {\bibfnamefont {S.}~\bibnamefont {Fairhurst}},\ and\ \bibinfo {author} {\bibfnamefont {A.}~\bibnamefont {Tolley}},\ }\href {https://doi.org/10.5281/zenodo.10473621} {\bibinfo {title} {gwastro/pycbc: v2.3.3 release of pycbc}} (\bibinfo {year} {2024})\BibitemShut {NoStop}%
\bibitem [{\citenamefont {{Ott}}\ \emph {et~al.}(2012)\citenamefont {{Ott}}, \citenamefont {{Abdikamalov}}, \citenamefont {{O'Connor}}, \citenamefont {{Reisswig}}, \citenamefont {{Haas}}, \citenamefont {{Kalmus}}, \citenamefont {{Drasco}}, \citenamefont {{Burrows}},\ and\ \citenamefont {{Schnetter}}}]{ott12correlated}%
  \BibitemOpen
  \bibfield  {author} {\bibinfo {author} {\bibfnamefont {C.~D.}\ \bibnamefont {{Ott}}}, \bibinfo {author} {\bibfnamefont {E.}~\bibnamefont {{Abdikamalov}}}, \bibinfo {author} {\bibfnamefont {E.}~\bibnamefont {{O'Connor}}}, \bibinfo {author} {\bibfnamefont {C.}~\bibnamefont {{Reisswig}}}, \bibinfo {author} {\bibfnamefont {R.}~\bibnamefont {{Haas}}}, \bibinfo {author} {\bibfnamefont {P.}~\bibnamefont {{Kalmus}}}, \bibinfo {author} {\bibfnamefont {S.}~\bibnamefont {{Drasco}}}, \bibinfo {author} {\bibfnamefont {A.}~\bibnamefont {{Burrows}}},\ and\ \bibinfo {author} {\bibfnamefont {E.}~\bibnamefont {{Schnetter}}},\ }\bibfield  {title} {\bibinfo {title} {{Correlated gravitational wave and neutrino signals from general-relativistic rapidly rotating iron core collapse}},\ }\href {https://doi.org/10.1103/PhysRevD.86.024026} {\bibfield  {journal} {\bibinfo  {journal} {\prd}\ }\textbf {\bibinfo {volume} {86}},\ \bibinfo {eid} {024026} (\bibinfo {year} {2012})},\ \Eprint {https://arxiv.org/abs/1204.0512} {arXiv:1204.0512
  [astro-ph.HE]} \BibitemShut {NoStop}%
\end{thebibliography}%

\end{document}